\documentclass{aa}
\usepackage[font=footnotesize]{caption}
\usepackage[utf8]{inputenc}
\usepackage{scrpage2,subfigure}
\usepackage{amssymb,amsmath,graphicx}
\usepackage{chngcntr}
\usepackage{natbib}
\usepackage{graphicx}
\usepackage{rotating}
\usepackage{color}
\usepackage{bm}
\usepackage{IEEEtrantools}

\abstract{} % optional Context
{The non-thermal spectra of jetted active galactic nuclei (AGN) show a variety of shapes and degree of curvature in their low- and high energy components. 
From some of the brightest {\it Fermi}-LAT blazars, prominent spectral breaks at a few GeV have been regularly detected, which is inconsistent with conventional cooling effects. 
We study the effects of continuous time-dependent injection of electrons into the jet with differing rates, durations, locations, and power-law spectral indices, and evaluate its impact on the 
ambient emitting particle spectrum that is observed at a given snapshot time in the framework of a leptonic blazar emission model. With this study, we provide a basis for analyzing
ambient electron spectra in terms of injection requirements, with implications for particle acceleration modes.
} % AIMS
{The emitting electron spectrum is calculated by Compton cooling the continuously injected electrons, where target photons are assumed to be provided by 
the accretion disk and broad line region (BLR). From this setup, we calculate the non-thermal photon spectra produced by inverse Compton scattering of these external target
radiation fields using the full Compton cross-section in the head-on approximation.} % Methods
{By means of a comprehensive parameter study we present the resulting ambient electron and photon spectra, and discuss the influence of each injection parameter individually.
We found that varying the injection parameters has a notable influence on the spectral shapes, which in turn can be used to set interesting constraints on the particle injection scenarios. 
By applying our model to the flare state spectral energy distribution (SED) of 3C~454.3, we confirm a previous suggestion that explained the observed spectral changes at a few GeV 
by a combination of the Compton-scattered disk and BLR radiation. We determine the required injection parameters for this scenario. We also show that this spectral turn-over can also be 
understood as Compton-scattered BLR radiation only, and provide the corresponding injection parameters. Here the spectral turn-over is explained by a corresponding break in the ambient electron 
spectrum. \textcolor{black}{ In a similar way, we also applied our model to the FSRQ PKS~1510-089, and present two possible model fits. Here,  the GeV-spectrum is either dominated by Compton-scattered accretion disk radiation or is a combination of Compton-scattered disk and 
BLR radiation. We provide the required injection parameters for these fits. 
In all four scenarios, we found that impulsive particle injection is disfavored.} 
} % Results
{The presented injection model that is embedded in a leptonic blazar emission model for external Compton-loss dominated jets of AGN
aims towards bridging jet emission with acceleration models using a phenomenological approach. 
Blazar spectral data can be analyzed with this model to constrain injection parameters, in addition to the conventional parameter 
values of steady-state emission models, if sufficient broad multifrequency coverage is provided.
} %optional Conclusion


\titlerunning{Shaping GeV-spectra of blazars}
\authorrunning{Hunger \& Reimer}

\begin{document}

\title{Shaping the GeV-spectra of bright blazars}
\author{Lars Hunger\inst{\ref{inst1},\ref{inst3}} \and  A. Reimer\inst{\ref{inst2},\ref{inst1}}}

\institute{Institut f\"ur Astro- und Teilchenphysik, Leopold-Franzens-Universit\"at Innsbruck, A-6020 Innsbruck, Austria\label{inst1}
\and
Institut f\"ur Theoretische Physik, Leopold-Franzens-Universit\"at Innsbruck, A-6020 Innsbruck, Austria\label{inst2}
\and
Brainlinks-Braintools, University of Freiburg, 79104 Freiburg im Brei{\ss}gau, Germany\label{inst3}}

\maketitle

\section{Introduction}

Jetted active Galactic Nuclei (AGN) comprise the most numerous variable source population in the $\gamma$-ray sky (e.g., \citealt{2013ApJ...771...57A}). 
Their jets are considered as the site of
intense broadband emission, with apparently random flaring events, shich cover the electromagnetic band from the radio up to GeV, or TeV energies. 
Variability timescales for this continuum emission range from months (radio) down to few minutes (TeV). Their broadband spectral energy 
distribution (SED) can be described as having two broad components, with the lower energy component usually attributed to synchrotron radiation
from a population of relativistic electrons in the magnetized emission region. The origin of the photons of the higher energy component is still 
under debate, and strongly depends on the relativistic particle content of the jet (e.g., \citealt{2013APh....43..103R}). In many cases, 
it is only by considering leptonic processes together with external target photon fields that can lead to snapshot or time-averaged SEDs that are in 
agreement with the corresponding multifrequency observations
(e.g., \citealt{2013ApJ...768...54B}). An often used approach for calculating these SEDs is based on an ad hoc assumption of the emitting electron 
spectrum as having log-parabolic shapes, (broken) power laws with possible exponential cutoffs, and  with
the minimum and maximum electron energies, energy of possible breaks and power law indices or curvature of this spectrum as free parameters 
(e.g., \citealt{2009ApJ...692...32D}, \citealt{2010ApJ...714L.303F}, \citealt{2014ApJ...782...82D}).

The observation of spectral breaks in the high energy photon spectra of bright flat spectrum radio quasars (FSRQ) was among the first findings
of the \textit{Fermi}-LAT in the extragalactic $\gamma$-ray sky (\citealt{2009ApJ...699..817A, 2010ApJ...710.1271A}, and confirmed subsequently in
\citealt{2011ApJ...743..171A}). Typically, bright FSRQs (and also some low frequency peaked-BL Lacs (LBLs) and intermediate frequency
peaked-BL Lacs (IBLs)) show $\gamma$-ray spectra that can be described phenomenologically either by 
broken power laws or log-parabolas with strong concave curvature \citep{2014MNRAS.441.3591H} between 1 and 10 GeV, which is too low to be caused by absorption in the extragalactic background light (e.g., \citealt{2010ApJ...723.1082A}), and with a 
power-law index change much larger than 
expected from cooling ($\Delta\Gamma=0.5$). 
For example, the turn-over in the GeV spectra of FSRQ 3C~454.3 is located at $\sim2.5$~GeV with a power law index change of 
about $\Delta\Gamma=1.1\pm 0.1$. The break energy does not seem to be correlated with the $\gamma$-ray luminosity. 
Several scenarios have been proposed to explain 
these breaks. These range from photon pair production in the broad-line region of the source (BLR; \citealt{2010ApJ...717L.118}), to a two-component $\gamma$-ray spectrum
(\citealt{2010ApJ...714L.303F}), or GeV breaks in the photon spectra owing to corresponding breaks in the electron spectra (\citealt{2010ApJ...710.1271A}).

Some recent studies, however, have cast doubt on the internal absorption scenario, which would predict well-defined break energies in the AGN's source frame.
\citet{2012ApJ...761....2H} did not find such universal values of the break energies. \citet{2015MNRAS.449.2901K} described the occurence of spectral breaks, which were investigated from a set of 40 bright LAT-blazars as ``random''. In fact they report a tendency for short duration flares to possess stronger curvature than those hat are integrated over a longer timescale.
The other two non-absorption scenarios mentioned above invoke ad hoc broken power-law particle spectra.
\citet{2009MNRAS.397..985G} show that Klein-Nishina or simple cooling effects cannot explain the strength and shape of the 
observed peaks
(see however, \citet{2013ApJ...771L...4C} who propose a high-energy cutoff at GeV energies from Klein-Nishina effects in scattering Ly$_{\alpha}$ photons).

Using the ambient particle spectrum as an input parameter, however, broadly ignores the build-up of this emitting particle spectrum that is impacted by the various 
particle energy losses, and gains through
particle energization mechanisms which, in general, requires a time-dependent treatment of the problem. As a consequence one can not gain much information
on the particle acceleration mechanisms at work in these environments from modeling broadband SEDs this way.

In the present work we attempt to improve on this situation by studying the impact of various injection scenarios of relativistic particles on the resulting
emitting particle and photon spectrum. Specifically, we calculate the ambient particle spectrum that results from injection during a finite time range in the past, and viewed at a given (snapshot) time. Comparing to the broken power-law, emitting-particle spectra, which are required in broadband modeling of blazar 
snapshot SEDs, will then help us draw inferences on the required particle acceleration mechanisms.

In this sense, our approach differs from works which directly implement a specific acceleration mechanism into an emission code (e.g., \citealt{2014JHEAp...1...63D}).

We restrict ourselves to leptonic emission processes in blazar jets here where external photon fields, such as accretion disk radiation and radiation partly re-scattered 
at the BLR, present the dominant target for particle-photon interactions. This setup is expected to be suitable for application to
those radio-loud jetted AGN, which show a strong accretion disk radiation field, such as FSRQs and LBLs. 
The analog procedure, including hadronic interactions as well, is more complex and will be postponed to a subsequent paper.

We aim to explain the observed breaking spectra at GeV-energies, and hence focus here on steep spectrum LAT-AGN. This is unlike \citet{2014ApJ...790...45P}, whose study on spectral shapes is based on a data set in which those periods are exclusively considered where the sources emitted significant $>10$ GeV photons and are therefore associated with hard $\gamma$-ray spectrum-flaring events. 
\citet{2014JHEAp...1...63D} complemented their time-dependent leptonic blazar emission code with an implementation of a Fermi-II acceleration scenario, in addition to time-dependent particle ``pick-up'', with the goal of identifying observational signatures for the origin of flares. Hence they studied the effect of a perturbation of various input parameters that could potentially be associated with causing outbursts on top of a steady-state situation. They explored observational signatures such as correlations between lightcurves at different frequencies, possible time lags, etc. Our present work attempts to explore the impact of a specific injection history on the resulting high-energy spectrum thereby offering an alternative explanation for the often observed strong breaks in bright FSRQ spectra at GeV-energies.

%somehow this whole thing throws an error when I use just textbf

Section~2 of this paper describes the emission model used here (an external inverse Compton emission model that broadly 
follows \citealt{1993ApJ...416..458D}, \citealt{2009ApJ...692...32D}), 
and the calculation of the emitting electron spectrum, which is the result of radiative losses that impact upon an
electron population that has been continuously injected while propagating along the jet.
Such continuously injected electron spectra, like the one we use in this work, were first proposed by \citet{2000MNRAS.312..177M}.  
We present a detailed parameter study in Sect.~3, and summarize and discuss the results of our work in Sect.~4.

\section{Model}
\subsection{Basic model}
\label{sec:Basicmodel}
\begin{figure}[htbp]
 	\centering
 	\includegraphics[width=0.49\textwidth]{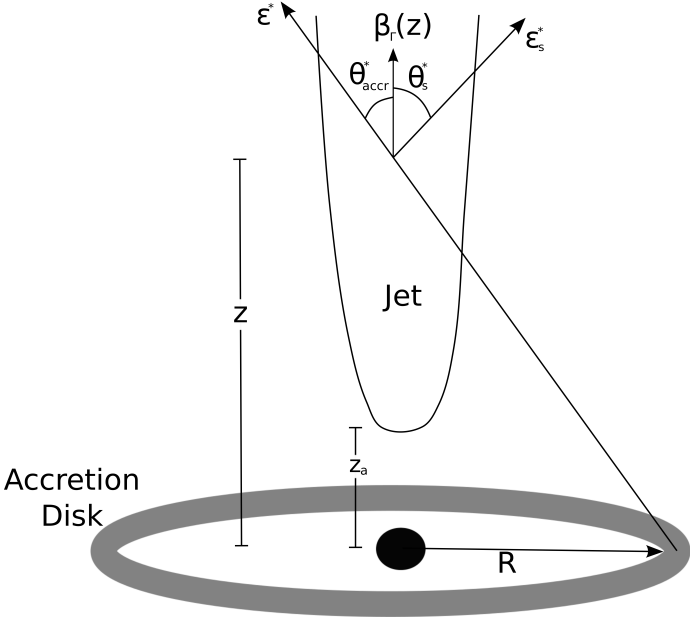}
 	\caption{Geometry of the blazar jet model considered in this work (Figure adapted from \citealt{1993ApJ...416..458D}).}
 	\label{fig:basemodel}
 \end{figure}

We consider the AGN to be powered by accretion onto a supermassive Schwarzschild black hole of mass $M_{\rm BH}$.
The central black hole is surrounded by a Shakura-Sunyaev accretion disk (\citealt{1973A&A....24..337S}).
Accretion disk photons, with energy $\varepsilon^*$, enter the jet under the angle $\theta^*_{accr}$. These photons are scattered into the angle $\theta_s^*$ 
with scattered photon energy $\varepsilon_s^*$, by the jet electrons.   
The plasma outflow of the jet starts at a distance $z_0$ from the central black hole and flows along the symmetry axis of the system with bulk Lorentz factor $\Gamma$ and the velocity $\beta_{\Gamma} c$,
where $\beta_{\Gamma}=\sqrt{1-1/\Gamma^2}$. 
\textbf{Figure~\ref{fig:basemodel}} illustrates the basic model used in this work.
Other than in the model of \citet{1993ApJ...416..458D}, electrons are injected continuously along the jet axis between heights $z_a$ and $z_b$ with a time-dependent injection rate $\propto z(t)^{-\alpha}$, with $\alpha$ being a free parameter.
We consider cooling of the the injected electrons as being dominantly the result of inverse Compton losses on external target 
radiation fields.
Two target photon fields are considered here: the aforementioned accretion disk radiation, which generally enters the jet under small angles $\theta^*_{accr}$, and the accretion disk radiation that has been backscattered by the BLR. Most of this backscattered radiation enters the jet under angles $\theta^*>90^{\circ}$, which leads to a different scattering behavior.
Because the two target photon fields have different angular distributions in the jet frame, the inverse Compton scattered photon distributions that are produced by each of the target photon fields 
shows correspondingly different spectral shapes.
The BLR model used here largely follows \citet{2009ApJ...692...32D} and is described in Sect.~\ref{ssec:BLRphotdens}. 

In the following, quantities with asterisks are in the rest frame of the accretion disk.
All energies are in units of the electron rest mass $m_e c^2$.

\subsection{Compton cooling rates}
\label{sec:ecalc}
We consider Compton cooling of electrons in the comoving frame of the jet. This radiative cooling rate is dependent on the 
differential Compton photon production rate,  

\begin{IEEEeqnarray}{rCl}  
\dot{n}(\varepsilon_s,\Omega_s,z) & = & c\int^\infty_0 d\varepsilon \oint d\Omega  \int^\infty_1 d\gamma \oint d\Omega_e (1-\beta \cdot \cos (\psi)) \nonumber\\
&& \cdot n_{\rm{ph}}(\varepsilon,\Omega,z) \cdot n_e(\gamma,\Omega_e) \frac{d\sigma_C}{d\varepsilon_s d\Omega_s},
\label{diffphotoprod1}
\end{IEEEeqnarray}

which is dependent on
\begin{itemize}
\item $n_{\rm{ph}}(\varepsilon,\Omega,z)$, the photon density of the target photon field at solid angle ($\Omega$), energy ($\varepsilon$) and position ($z$) along the jet axis,
\item $n_e(\gamma,\Omega_e)$, the electron density at electron Lorentz factor $\gamma$ and the solid angle ($\Omega_e$),
\item the differential Compton scattering cross section $\frac{d\sigma_C}{d\varepsilon_s d\Omega_s}$.
 \end{itemize}  
Here, $\cos(\psi)=\mu \mu_e + \sqrt{(1-\mu^2)}\sqrt{1-\mu_e^2}\cos(\phi-\phi_e)$ describes the three-dimensional collision angle.

We assume that the electron density is isotropic in the comoving frame of the jet, $n_e(\gamma,\Omega_e)=\frac{n_e(\gamma)}{4 \pi}$.
To calculate the energy loss rate, we set $n_e(\gamma)=\delta(\gamma - \bar{\gamma})$, where $\delta()$ denotes the Delta distribution.
By only considering relativistic electrons, the assumption $\gamma \gg 1$ is justified. Furthermore, we use the Thomson approximation for the differential cross-section with the $\delta$-approximation in the head-on case:
\begin{equation}  
\frac{d\sigma}{d\varepsilon_s d\Omega_s}=\sigma_T \delta[\varepsilon_s-\gamma^2\varepsilon(1-\beta \cos (\psi))]\delta(\Omega_s-\Omega_e)
\label{Thomsoncross}
\end{equation}

where $\sigma_T$ denotes the Thomson cross-section. This is valid as long as $\varepsilon' \equiv \gamma \varepsilon (1 - \cos(\psi)) \leq 1$.
For $\varepsilon' > 1$ (Klein-Nishina regime) we assume that the cross-section is zero when calculating the cooling rates, since the Klein-Nishina cross-section declines rapidly at high energies.

With these approximations, the integrations over $\Omega_e$ and $\varepsilon$ are trivial.
To achieve further simplification, we average the azimuthally symmetric problem over $\phi$ implying
%\begin{equation}
$\langle \cos(\psi) \rangle=\mu \mu_s$ .
%\label{meanphi}
%\end{equation}

The integrations then lead to

\begin{equation}  
\dot{n}(\varepsilon_s,\mu_s,z)=\frac{c \sigma_T}{4\pi} \int^1_{-1} d\mu \, \gamma^{-2}\cdot n_{\rm{ph}}\left(\frac{\varepsilon_s}{\gamma^2 (1-\beta \cos(\psi))},\mu,z\right).
\label{diffphotoprod4}
\end{equation}

To make progress in calculating the scattered differential photon production rate the two differential target photon densities $n_{\rm{ph}}$ (in the comoving frame) are required.
Their calculations are shown in the following two sections.

\subsubsection{BLR target photon density}
\label{ssec:BLRphotdens}
\begin{figure}[htbp]
 	\centering
 	\includegraphics[width=0.49\textwidth]{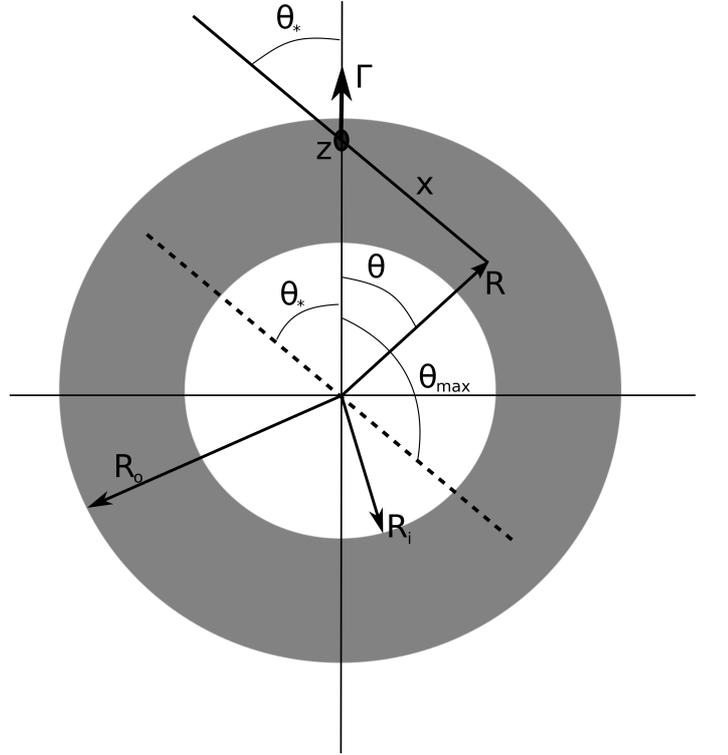}
 	\caption{Our BLR model consists of a spherically symmetric shell of gas between $R_i$ and $R_o$ and a density gradient $\zeta$ with $n_e(R)\propto R^{\zeta}$
	(Figure adapted from \citealt{2009ApJ...692...32D}).}.
 	\label{fig:BLRmodel}
 \end{figure}

\textbf{Figure~\ref{fig:BLRmodel}} sketches the BLR model used here, mainly following \citet{2009ApJ...692...32D}, where the BLR target photon field is the result of reprocessing the accretion disk radiation and with any direct BLR line emissivity being neglected. For the purpose of calculating the target photon field re-scattered at the BLR (BLR target photon field), we consider the central source as an 
isotropically emitting point source. 

An anisotropic central source would slightly increase the energy density caused by the BLR in the jet-frame with distance from the illuminating disk  (\citealt{1996MNRAS.280...67G}).

The central source radiation is isotropically Thomson-scattered by material of the BLR, back to a distance $z$ (above the black hole) on the jet axis.
We consider a spherically symmetric shell of thin gas for the scattering BLR material with a density gradient inside this shell (see \textbf{Fig.~2}). Taking into account a clumpy BLR, instead, would require the introduction of further free parameters (cloud radius distribution) while the expected impact on the distribution of the reprocessed radiation likely stays small for not too large cloud sizes: one would expect additional inhomogenities to some extent on top of an R-dependence from the BLR density gradient.

The calculations here are all carried out in the rest frame of the BLR.
\newline

The Thomson-scattered photon density is obtained by

\begin{equation}  
n^{\rm{BLR}}_{\rm{ph}}(\varepsilon_*;z)=\int dV \frac{\dot{n}(\varepsilon_*; \vec{R})}{4 \pi x^2 c}.
\label{BLRtarphot1}
\end{equation}

Here $x^2=R^2+z^2-2zR\cos(\theta)$, and $\dot{n}(\varepsilon_*;\vec{R})$, the differential BLR-scattered photon density rate, is given by

\begin{equation}  
\dot{n}(\varepsilon_*; \vec{R})=\frac{\dot{N}_{\rm{ph}}(\varepsilon_*)\sigma_T n_e^{BLR}(R)}{4\pi R^2}, 
\label{diffBLRprodrate}
\end{equation}

where
\begin{itemize}
\item $\dot{N}_{\rm{ph}}(\varepsilon_*)$ is the central source photon production rate, 
\item $\sigma_T n_e^{BLR}(R)$ gives the fraction of the incoming flux that is scattered,
\item $\frac{1}{4\pi R^2}$ is the geometric dilution from an isotropic source.
\end{itemize}  

With spherical coordinates and Eq.(\ref{diffBLRprodrate}), Eq.(\ref{BLRtarphot1}) we find:

\begin{equation}  
n^{\rm{BLR}}_{\rm{ph}}(\varepsilon_*;z)=\frac{\sigma_T \dot{N}_{\rm{ph}}(\varepsilon_*)}{8\pi c}
 \int_{-1}^1 d\mu \int_0^{\infty} dR \frac{n_e(R)}{x^2}.
\label{BLRtarphot2}
\end{equation}

Since inverse Compton scattering is angle dependent, we need the angle-dependent target photon spectrum.
With $\delta(\mu_*-\bar{\mu}_*)$, $\mu_*=\cos \ \theta_*$ and by changing the variables to $g=R/z$ we get

\begin{IEEEeqnarray}{rCl}  
n^{\rm{BLR}}_{\rm{ph}}(\varepsilon_*;z) & = & \frac{\sigma_T \dot{N}_{\rm{ph}}(\varepsilon_*)}{8\pi c z} \\\nonumber
&& \cdot \int_{-\mu^*}^1 d\mu \int_0^{\infty} dg \frac{n_e(gz) \delta[\mu_*-\bar{\mu}_*(\mu,g)]}{g^2+1-2g\mu}.
\label{BLRtarphotdelta}
\end{IEEEeqnarray}

We use the law of sines $R^2(1-\mu^2)=x^2(1-\mu_*^2)$ to calculate the angle $\mu_*$ under which the scattered photons reach the point $z$:

\begin{equation}  
{\bar{\mu}}_*(\mu,g)=\frac{\pm (1-g\mu)}{\sqrt{(1+g^2-2g\mu)}}. 
\label{scatangle}
\end{equation}

$\bar{\mu}_*(\mu,g)$ defines a line. If the scattering process takes place on that line, a photon emitted into the angle 
$\mu$ arrives with angle $\mu_*$ at point $z$. 
Formally, there are two solutions with the positive one being the physically relevant one for the already $\phi$-integrated quantity ${\bar{\mu}}_*(\mu,g)$. 
%\textcolor{red}{***We ignore the negative one, since it corresponds to a
%rotation of $\pi$ around $\phi$ and we already integrated over $\phi$.***}

Carrying out the integration we find:

\begin{equation}  
n^{\rm{BLR}}_{\rm{ph}}(\varepsilon_*,\mu_*;z)=\frac{\sigma_T \dot{N}_{\rm{ph}}(\varepsilon_*)}{8\pi c} \frac{\Pi(\mu_*,z)}{z},
\label{BLRtarphot3}
\end{equation}

where
$$\Pi(\mu_*,z)\equiv \int^1_{-\mu_*} d\mu \ n_e(\bar{g}z) \frac{\sqrt{1+\bar{g}^2-2\bar{g}\mu}}{\bar{g}(1-\mu^2)}$$

and
$$\bar{g}=\bar{g}(\mu,\mu_*) \equiv \frac{-\mu(1-\mu_*^2)+\mu_* \sqrt{(1-\mu^2)(1-\mu_*^2)}}{\mu_*^2-\mu^2}. $$

Here $\bar{g}$ describes the radius $R$ at which the photons, emitted into the angle $\mu$, have to be scattered to end up in $\mu_*$.
$n_e(\bar{g}z)$ is the electron density at the scattering point. This means $\Pi(\mu_*,z)$ gives a measure for the total number of photons that end up 
in the angle element $\mu_*$ at point $z$ along the jet axis.

\begin{figure}[htbp]
 	\centering
 	\includegraphics[width=0.49\textwidth]{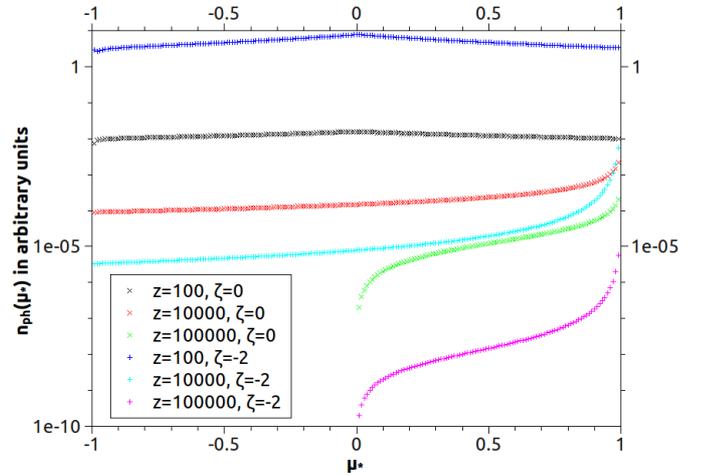}
 	\caption{Angle dependence of $n_{\rm{ph}}^{BLR}$ for a BLR extending from $10^2$ to $10^5$ $R_g$ for different values of $z$ (in $R_g$) and $\zeta$, in agreement with \citealt{2009ApJ...692...32D}.}
 	\label{fig:BLRphots}
 \end{figure}

$\Pi$ cannot be solved analytically, but numerically.
The results of these calculations, which we find in agreement with \citet{2009ApJ...692...32D}, are shown in \textbf{Fig. \ref{fig:BLRphots}}. The figure shows the angular distribution of the BLR photons for different positions $z$.
For $z\geq R_o$ there are no incoming photons from the front ($\mu_*<0$) since there is no scattering BLR material in front of $z$.

The central source photon-production rate $\dot{N}_{\rm{ph}}(\varepsilon_*)$ is calculated
by assuming that this photon source emits a monochromatic spectrum with luminosity $L_0$:

\begin{equation}  
\dot{N}_{\rm{ph}}(\varepsilon_*)=\frac{L_0 \delta(\varepsilon_*-\varepsilon_{0*})}{m_e c^2 \varepsilon_*}.
\label{Nph}
\end{equation}
Here $ L_0 = L_{\rm{edd}} l_{\rm{edd}} $, with $L_{\rm{edd}} $ the Eddington luminosity of the black hole.  \citet{2008MNRAS.386..945T} have shown that the BLR radiation, simulated with the photoionization code CLOUDY, as the target field for inverse Compton scattering can be adequately approximated by a blackbody with a peak at 1.5 times the Ly$_\alpha$-frequency , when considering the IC spectrum above a few keV.
Afterwards, we approximate this blackbody with a monochromatic spectrum.

The resulting BLR target photon density is

\begin{equation}  
n^{\rm{BLR}}_{\rm{ph}}(\varepsilon_*,\mu_*;z)=\frac{\sigma_T L_{\rm{edd}} l_{\rm{edd}} \delta(\varepsilon_*-\varepsilon_{0*})}{8\pi m_e c^3 \varepsilon_*} \frac{\Pi(\mu_*,z)}{z}.
\label{BLRtarphotfinal}
\end{equation}   
To transform this into the comoving frame we use $\varepsilon_*=\Gamma\varepsilon(1+\beta_{\Gamma}\mu)$ and $\mu_*=\frac{(\mu+\beta_{\Gamma})}{1+\beta_{\Gamma} \mu}$, to get

\begin{IEEEeqnarray}{rCl} 
n^{\rm{BLR}}_{\rm{ph}}(\varepsilon,\mu;z) & = & \frac{\sigma_T L_{\rm{edd}} l_{\rm{edd}} \delta(\Gamma (1+\beta_{\Gamma}\mu)(\varepsilon-\varepsilon_{0}))}{8\pi m_e c^3 \varepsilon\Gamma (1+\beta_{\Gamma}\mu)} \nonumber\\ 
&& \cdot  \frac{\Pi(\frac{(\mu+\beta_{\Gamma})}{1+\beta_{\Gamma} \mu},z)}{z}.
\label{BLRtarphotfinalBF}
\end{IEEEeqnarray}

\subsubsection{Accretion disk target photon density}
To calculate the accretion disk photon density in the comoving jet-frame, we follow the procedure outlined in \citet{1993ApJ...416..458D}.
The accretion disk target photon density at the point $z$ is given by

\begin{equation}  
n^{\rm{SSD}}_{\rm{ph}}(\varepsilon_*,\mu_*,R;z)=\frac{\dot{N}_{\rm{ph}}(\varepsilon_*,R)}{4 \pi x^2 c} \frac{\delta(\mu_*-\bar{\mu}_{*,accr})}{2 \pi},
\label{SSDtarphot1}
\end{equation} 
 
with 
\begin{equation}
 \dot{N}_{\rm{ph}}(\varepsilon_*,R)=\frac{2\pi R F(\varepsilon_*,R)}{\varepsilon_*}
\label{NphSSD}
\end{equation}

and the surface energy flux for a Shakura-Sunyaev disk is
\begin{equation}
 F(R)=\frac{3GM\dot{M}}{8 \pi m_e c^2 R^3} I(R).
\label{FluxF}
\end{equation}

Here $I(R)=1-\sqrt{R_i/R}$, $\dot{M}$ is the mass accretion rate and $R_i=6R_g$ for a Schwarzschild black hole with $R_g=GM/c^2$ the gravitational radius. 
In gravitational units, $\tilde R=R/R_g$, Eq.~(\ref{FluxF}) becomes
\begin{equation}
 F(\tilde{R})=\frac{\Phi_F}{\tilde{R}^3} I(\tilde{R}),
\label{FluxF2}
\end{equation}

with $\Phi_F=8.43 \cdot 10^{24} \Psi \mathrm{cm}^{-2}\mathrm{s}^{-1}$ and $\Psi=l_{\rm{edd}}/ \epsilon_f M_8$
where $M_8$ is the black hole mass in $10^8$ solar masses \color{black}.

We use a monochromatic approximation for the blackbody radiation that is emitted locally at a specific radius $\tilde{R}$ of the accretion disk.
The mean photon energy emitted at a radius $\tilde{R}$ is then given by
\begin{equation}
\bar{\varepsilon}^*=2.69 \cdot 10^{-4}\tilde{R}^{-3/4}\Psi^{1/4}I(\tilde{R})^{1/4}.
\label{Meaneenergy}
\end{equation}

By combining Eq.~(\ref{NphSSD}), ~(\ref{FluxF2}) and, ~(\ref{Meaneenergy}) we get
\begin{equation}
 \dot{N}_{\rm{ph}}(\varepsilon^*,\tilde{R})=\frac{ 2\pi R_g^2\Phi_F I(\tilde{R})}{\bar{\varepsilon}^*\tilde{R}^2} \delta(\varepsilon^*-\bar{\varepsilon}^*).
\label{NphSSDfin}
\end{equation}

After substituting $\dot{N}_{\rm{ph}}$ in Eq.(\ref{SSDtarphot1}) with Eq.(\ref{NphSSDfin}) and transforming the result into the comoving frame we find

\begin{IEEEeqnarray}{rCl} 
n^{\rm{SSD}}_{\rm{ph}}(\varepsilon,\mu,\tilde{R};z) & = & \frac{R_g^2 \Phi_F I(\tilde{R})}{4\pi x^2 c  \tilde{R}^2 \bar{\varepsilon_*}} \delta(\Gamma(1+\beta_{\Gamma}\mu)(\varepsilon-\bar{\varepsilon}))  \nonumber\\
&& \cdot \delta(\mu-\frac{\bar{\mu}_{*,accr}-\beta_{\Gamma}}{1-\beta_{\Gamma}\bar{\mu}_{*,accr}}).
\label{NphSSDfin2}
\end{IEEEeqnarray}

\subsection{Energy loss and equation of motion}

\subsubsection{Energy loss in the BLR radiation field}
With the differential target photon density, we now calculate the scattered photon density.
We use Eq.(\ref{BLRtarphotfinalBF}) together with Eq.~(\ref{diffphotoprod4}) to calculate $\dot{n}^{\rm{BLR}}(\varepsilon_s,\mu_s,z)$.
By integrating over all scattering angles $\mathbf{\mu_s}$, and all scattering energies $\varepsilon_s$, we get the energy-loss rate of a single electron
of given $\gamma$:

\begin{equation}  
-\dot{\gamma}(\gamma,z)=\int^{1}_{-1} d\mu_s \int^\infty_0 d\varepsilon_s \; \varepsilon_s  \cdot \dot{n}^{\rm{BLR}}(\varepsilon_s,\mu_s,z).
\label{Energyloss}
\end{equation} 

After integrating over $\varepsilon_s$ we get:

\begin{IEEEeqnarray}{rCl}  
-\dot{\gamma}(\gamma,z) &= & \frac{{\sigma_T}^2 L_{\rm{edd}} l_{\rm{edd}} \cdot \gamma^2}{16 \pi \cdot m_e c^2 \Gamma^2} \int^{1}_{-1} d\mu_s \int^{1}_{-1} d\mu \; \frac{(1-\mu \mu_s)^2}{(1+\beta_{\Gamma} \mu)^2} \nonumber\\
&& \cdot \frac{\Pi(\frac{\mu+\beta_{\Gamma}}{1+\beta_{\Gamma} \mu},z)}{z}.
\label{lossrateBLRfinal}
\end{IEEEeqnarray}

\begin{figure*} %[!h]
        \centering
        \subfigure[ Radiative electron energy loss rates for varying $z$ for electron energy $\gamma=500$ and BLR1 with $\tau_{\rm{BLR}}=0.01$.]{\includegraphics[width=0.49\textwidth]{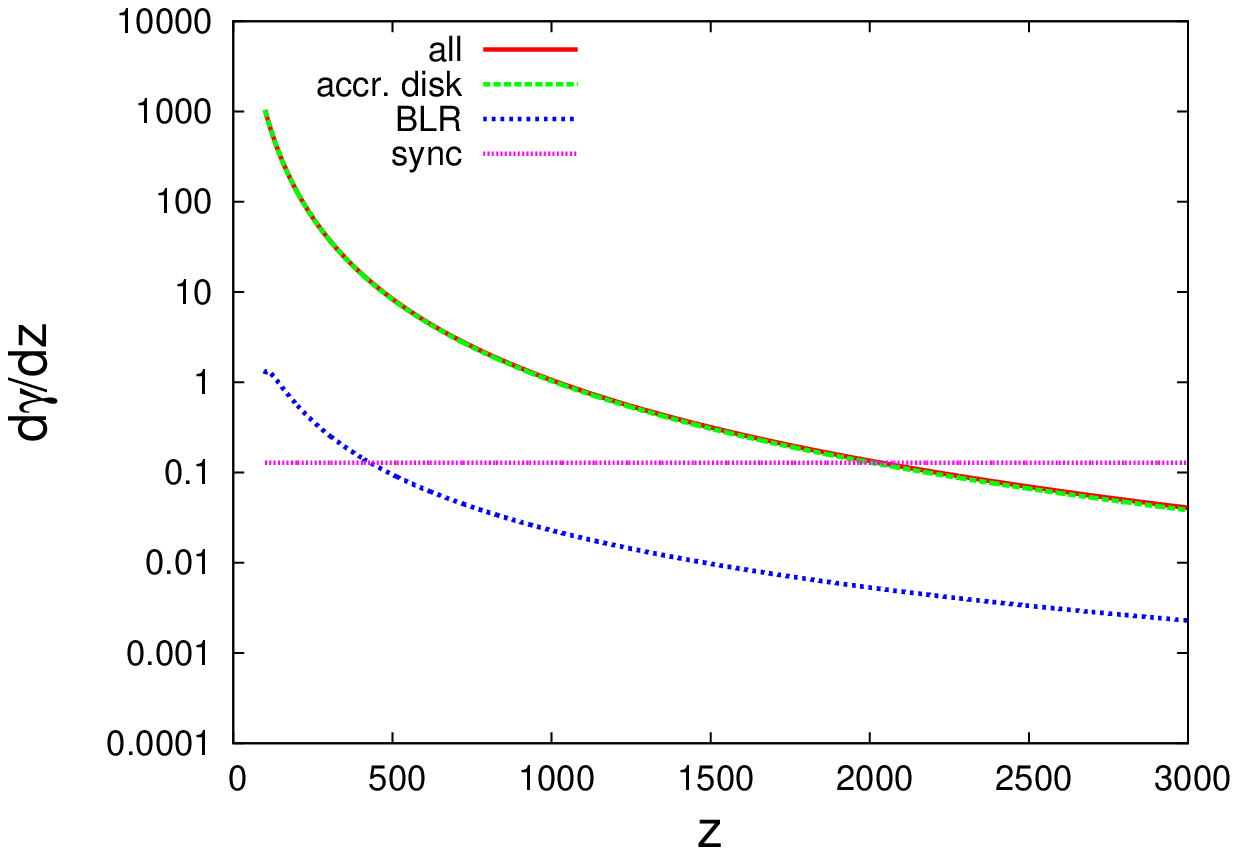}}\hfill
        \subfigure[ Radiative electron energy loss rates for varying $\gamma$ at $z=100$ and BLR1 with $\tau_{\rm{BLR}}=0.01$. BLR shows the BLR energy loss used in our calculations, here we set the energy loss in the Klein-Nishina regime to 0.]{\includegraphics[width=0.49\textwidth]{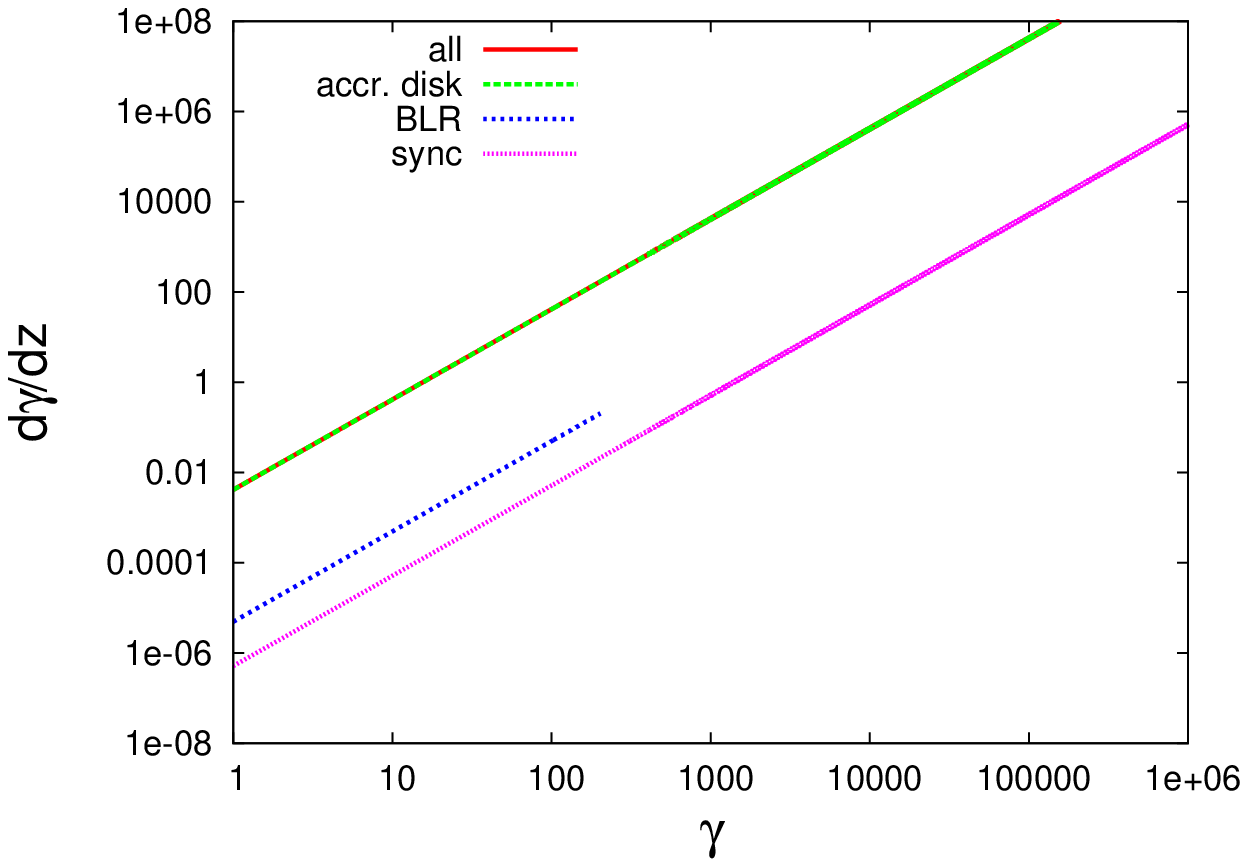}}\\
	\caption{  Comparison of radiative electron energy-loss rates. Here we choose BLR1 at $z=100$ as an example, while noting that the qualitative picture, with the negligible BLR losses, is the same in all the cases we considered in our study.
	We compare with synchrotron loss rates for a $B=4.5$G magnetic field. }
\label{fig:BLR_losses_negligible}
\end{figure*}

Because the electron energy loss rate depends on $(1-\mu\mu_s)^2/(1+\beta_{\rm{\Gamma}}\mu)^2$ photons with incident angles $\mu < 0$ contribute more strongly to the electron energy loss rate 
than photons with incident angles $\mu >0$. The small number of reprocessed BLR photons causes the energy loss due to the BLR to be negligible for the cases that we considered here (see \textbf{Fig.\ref{fig:BLR_losses_negligible}}).

\subsubsection{Energy loss in the accretion disk photon field}

The procedure is analogous to the BLR case: We use Eq.(\ref{diffphotoprod4}) to calculate the differential photon production rate
for one electron:
\begin{IEEEeqnarray}{rCl}
 \dot{n}^{\rm{SSD}}(\varepsilon_s,\mu_s,R,z) & = & \frac{\sigma_T \Phi_F I(\tilde{R})}{8\pi x^2 R_g^2  \tilde{R}^2 \bar{\varepsilon_*}} \\
 && \cdot \delta(\Gamma(1+\beta_{\Gamma}\bar{\mu})(\frac{\varepsilon_s}{\gamma^2(1-\bar{\mu}\mu_s)}-\bar{\varepsilon})) \nonumber.
\end{IEEEeqnarray}

This leads to the energy-loss rate induced by the accretion disk target photons of

\begin{IEEEeqnarray}{rCl} 
-\dot{\gamma}(\gamma,z) & = & \frac{\sigma_T \Phi_F }{8\pi \gamma^2} \int_{R_i}^{\infty} dR \tilde{R}^{-2} \tilde{x}^2 \int_0^{\infty} d\varepsilon_s \ 
\varepsilon_s \frac{I(\tilde{R})}{\varepsilon^*} \nonumber\\ 
&& \cdot \delta(\Gamma(1+\beta_{\Gamma}\bar{\mu})(\frac{\varepsilon_s}{\gamma^2(1-\bar{\mu}\mu_s)}-\bar{\varepsilon}).
\label{lossrateSSfinal}
\end{IEEEeqnarray}

The very good approximation for the solution of Eq.~(\ref{lossrateSSfinal}) in the near-field regime 
($z<0.1\Gamma^4(1+\beta_\Gamma)^4$) is given by \citet{1993ApJ...416..458D}, which 
we use in the following.

\subsubsection{Thomson scattering criterion}

% To calculate the electron spectrum, we need to consider the energy-loss rates due to both target photon fields.
% The overall energy-loss rate is the sum of the loss rates due to BLR and accretion disk target photons.
% So the equation of motion is: 
% \begin{IEEEeqnarray}{rCl}
%  -\frac{d\gamma}{d\tilde{z}} & =& 34.5 \frac{l_{\rm{edd}}}{\epsilon_f} \frac{(1+\beta_{\Gamma})^2 \Gamma}{\beta_{\Gamma}} \frac{\gamma^2}{\tilde{z}^3}
% +\frac{{\sigma_T}^2 L_{\rm{edd}} l_{\rm{edd}} \epsilon_f \cdot \gamma^2}{16 \pi R_g \cdot m_e c^3 \Gamma^3 \beta_{\Gamma}} \nonumber\\ 
% && \cdot \int^{1}_{-1} d\mu_s \int^{1}_{-1} d\mu \; \frac{(1-\mu \mu_s)^2}{(1+\beta_{\Gamma} \mu)^2}
% \frac{\Pi(\frac{\mu+\beta_{\Gamma}}{1+\beta_{\Gamma} \mu},\tilde{z})}{\tilde{z}}
% \label{EOMfinal}
% \end{IEEEeqnarray}

The energy loss rates are calculated in the Thomson regime, and are therefore only valid for $\varepsilon' \equiv \gamma \varepsilon (1 - \cos(\psi)) \leq 1$ while, in the Klein-Nishina regime, the cross section is set to zero. 
After transforming the target photon energy $\varepsilon$ into the rest frame of the BLR, we obtain the Thomson limit criterion:

\begin{equation}
\gamma \Gamma \varepsilon_* (1-\beta_{\Gamma} \mu_*)(1-\beta \cos (\psi))<1.
 \label{Thomsoncrit}
\end{equation}

Again, we use the monochromatic approximation for both external target photon fields to confine the range of $\gamma$, where
both energy loss rates are in the Thomson regime. Since $1-\beta \cos (\psi)$ can never be larger than 2, the scattering will always be in the Thomson regime if the 
following relation is fulfilled:
\begin{equation}
 2\gamma \Gamma \varepsilon_* (1-\beta_{\Gamma} \mu_*)<1 .
\end{equation}
% Now we have to distinguish between the two target photon fields.
The accretion disk photons will mainly enter the jet from behind, i.e. $\mu_* \approx 1$. In this case, the Thomson criterion for scattering the accretion disk target photons
is:
\begin{equation}
 \gamma < \frac{1}{2\Gamma \varepsilon_* (1-\beta_{\Gamma}) }.
\end{equation}

The BLR photons that are most important for the problem will enter the jet from the front, $\mu_* \approx -1$. In this case the Thomson criterion changes to:
\begin{equation}
 \gamma < \frac{1}{4\Gamma \varepsilon_*}.
\end{equation}

For larger $\gamma$, the rate of inverse Compton scattering in the BLR reduces to the Klein-Nishina regime, which we neglect in the following.
We can neglect the energy loss in the Klein-Nishina regime for the BLR target photons. This neglect is justified since the energy loss owing to the BLR is negligible with respect to the energy loss caused by the accretion disk target photons in the cases that are considered in this work (see \textbf{Fig.\ref{fig:BLR_losses_negligible}}).
In the cases we studied, the accretion disk photons are usually scattered in the Thomson regime.

\subsubsection{Omission criterium for synchrotron energy losses}

In particular, this work targets accretion radiation-strong blazars such as FSRQs. Here synchrotron energy losses of the charged particles are typically weak when compared to
Compton losses (see also \textbf{Fig.\ref{fig:BLR_losses_negligible}}). In the Thomson limit, synchrotron losses can therefore be neglected when, in the jet-frame, the magnetic field energy density $u'_B=\frac{B^2}{8\pi}$ is much lower than
the sum of all target photon fields, i.e.,:
\begin{equation}
u'_B \ll u'_{BLR} + u'_{acc}
\end{equation}
The jet-frame accretion disk energy density in the near field regime for a Schwarzschild black hole reads
$$
u'_{acc} = \frac{0.207R_g l_{edd}L_{edd}}{\pi c z^3 \Gamma^2}
$$
(see \citealt{2009herb.book.....D}), while the jet-frame BLR radiation field is given by
$$
u'_{BLR} = u_{BLR}\Gamma^2 (1+\frac{1}{3}\beta_{\Gamma}^2 )
$$
where the galaxy-frame BLR energy density $u_{BLR}$ can be evaluated following \citet{2010ApJ...714L.303F}.
In the case of negligible BLR Compton energy losses, Eq.~28 reduces to:
$$
z \ll 380R_g [\frac{1+\beta_{\Gamma})\Gamma}{B_G}]^{2/3} (\frac{l_{edd}}{\epsilon_f M_8})^{1/3}
$$
where $l_{edd}L_{edd}=\epsilon_f \dot M c^2$ (see \citealt{1993ApJ...416..458D}).

\subsection{The emitting electron spectrum}

The equation of motion of the relativistically outflowing jet with constant velocity $\beta_{\Gamma} c$ is 
\begin{equation}
 z(t^*)=z_i + \beta_{\Gamma}c t^* .
\label{BlobEoM}
\end{equation}
    
The time interval in the rest frame of the disk is related to the comoving frame by $\delta t^*=\Gamma \delta t$.
By combining Eq.~(\ref{BlobEoM}) with Eq.~(\ref{lossrateBLRfinal}) and using the solution of Eq.~(\ref{lossrateSSfinal}) in the near-field regime we get
% the time reversed equation of motion
%for one electron (the time reversal is a trivial multiplication with -1): 

% Eq.(\ref{EOMfinal}) describes how much energy a single electron looses while
% propagating an infinitesimal distance $d$z. For our purpose we need the time-reverse of Eq.(\ref{EOMfinal}) which is simply done by changing the sign of the 
% whole equation.  
%   
% 
\begin{IEEEeqnarray}{rCl}
 \frac{d\gamma}{d\tilde{z}} & = & 34.5 \frac{l_{\rm{edd}}}{\epsilon_f} \frac{(1+\beta_{\Gamma})^2 \Gamma}{\beta_{\Gamma}} \frac{\gamma^2}{\tilde{z}^3} + \frac{{\sigma_T}^2 L_{\rm{edd}} l_{\rm{edd}}\cdot \gamma^2}{16 \pi \cdot m_e c^3 \Gamma^3 \beta_{\Gamma}} \int^{1}_{-1} d\mu_s \nonumber\\
&& \cdot \int^{1}_{-1} d\mu \; \frac{(1-\mu \mu_s)^2}{(1+\beta_{\Gamma} \mu)^2}
\frac{\Pi(\frac{\mu+\beta_{\Gamma}}{1+\beta_{\Gamma} \mu},\tilde{z})}{\tilde{z}},
\label{EOMfinalbw}
\end{IEEEeqnarray}

which describes the change of electron energy over a distance $dz$.
We note that, while in the model of \citet{1993ApJ...416..458D} only Compton energy losses in the disk radiation field are discussed, here we also take energy losses from Compton scattering in the BLR radiation field into account.
The initial value problem for calculating the electron spectrum consists of Eq.~(\ref{EOMfinalbw}), the point $z$, the injection point $z_i$,
and the Lorentz factor $\gamma$ that the electrons possess, after having propagated from $z_i$ to $z$.
This can only be solved numerically. For all  $\gamma > \frac{1}{4\Gamma \varepsilon_*}$, the BLR component does not play a role.
In this case  Eq.~(\ref{EOMfinalbw}) simplifies to

\begin{equation}
 \frac{d\gamma}{d\tilde{z}}=34.5 \frac{l_{\rm{edd}}}{\epsilon_f} \frac{(1+\beta_{\Gamma})^2 \Gamma}{\beta_{\Gamma}} \frac{\gamma^2}{\tilde{z}^3}\equiv K_n \frac{\gamma^2}{\tilde{z}^3},
\label{EOMSSDfinalbw}
\end{equation}

which has an analytic solution:

\begin{equation}
\gamma^{-1}(z) = \gamma_i^{-1}+0.5K_n(z_i^{-2}-z^{-2}) 
\label{solutionEOM}
\end{equation}

We consider the continuous injection of an electron power-law spectrum, with spectral index $s$, into a moving plasma blob along the jet at an injection rate 
 that depends on the height above the disk and with an injection index $\alpha$.
The electron spectrum injected at height $z_i$ can be described by $N_e(\gamma_i)=Q_0 \gamma_i^{-s}H[\gamma_i - \gamma_1] H[\gamma_2-\gamma_i]$, 
with $Q_0$ the normalization parameter. 
Using Eq.~\ref{solutionEOM}, the cooled injected electron spectrum then has the form 
\begin{equation}
N_e(\gamma,z,z_i)=Q_0 \cdot \gamma^{-2} \left[\gamma^{-1}-0.5K_n(z_i^{-2}-z^{-2})\right]^{s-2}
\label{cooled_inj_espec}
\end{equation}
(see also \citealt{1993ApJ...416..458D}).

When continuously injecting power-law electron spectra into the moving plasma blob with a given injection rate $\propto z_i^{-\alpha}$, the
resulting emitting electron spectrum $N_e(\gamma,z)$ is a superposition of the cooling electron spectra that have been injected and weighted by the 
corresponding injection rate. 
This is described by

\begin{equation}
N_e(\gamma,z)=\int_{z_a}^{z_b} d z_i {z_i}^{-\alpha} N_e(\gamma,z,z_i) H[\gamma - \tilde{\gamma}_1(z)] H[\tilde{\gamma}_2(z)-\gamma].
\label{espec}
\end{equation}
where
\begin{equation}
\tilde\gamma_j^{-1}(z) = \gamma_j(z)^{-1}+0.5K_n(z_i^{-2}-z^{-2}), j=1,2 
\label{coolinglimits}
\end{equation}
describe the limits of the cooling electron spectrum (see also \citealt{2000MNRAS.312..177M}).

An electron that was injected with an initial energy $\gamma_i$ cooled down to 
$\tilde{\gamma}$ after having propagated a distance $z_{\rm{travel}}=z-z_i$. Accordingly, $\tilde{\gamma}_1$ corresponds to the energy that an electron injected at point $z_i$ with the energy $\gamma_1$ possesses after having propagated from $z_i$ to $z$. 
The two Heaviside functions $H[\gamma - \tilde{\gamma}_1(z)]$ and $H[\tilde{\gamma}_2(z)-\gamma]$ prevent all 
electrons that are injected outside of the contributing energy range from adding to $N_e(\gamma,z)$.

\subsection{Calculating the photon spectrum}

\label{sec:Photocalc}
Having determined the emitting electron spectrum, we now proceed to calculate the scattered photon spectrum of both the accretion disk target photon field
and the BLR target photon field.
For this purpose, we follow \citet{2009ApJ...692...32D}.
The $\nu F_{\nu}$ spectrum is given by $$ f^C_{\varepsilon_{obs}}=\frac{\varepsilon_s^* L_C (\varepsilon_s^*,\Omega_s^*)}{d_L^2},$$
where $d_L$ is the distance to the source.

Here $\varepsilon_s^* \equiv (1+z_r)\varepsilon_{\rm obs}$ with $z_r$ the redshift of the photons that are emitted by the source.
The $\nu L_{\nu}$ spectrum is related to the photon production rate via
\begin{equation} 
\varepsilon_s^* L_C (\varepsilon_s^*,\Omega_s,z)=V_{b}m_e c^2 \varepsilon_s^{*2} \dot{n}(\varepsilon_s^*,\Omega_s^*,z)
\label{nu_fnu_start}
\end{equation}

where $V_b=(4/3) \pi R_b^3$ is the (observer frame) volume of the emitting plasma blob with a radius $R_b$.
We now calculate the photon scattering in the rest frame of the accretion disk.  
Eq.~(\ref{diffphotoprod1}) in the rest frame of the accretion disk reads

\begin{IEEEeqnarray}{rCl}
\dot{n}(\varepsilon_s^*,\Omega_s^*,z) &=& c\int^\infty_0 d\varepsilon^* \oint d\Omega^*  \int^\infty_1 d\gamma^* \oint (1-\beta \cdot \cos (\psi))  \nonumber\\ 
&&  \cdot n_{\rm{ph}}(\varepsilon^*,\Omega^*,z) \cdot n_e^*(\gamma^*,\Omega_s^*) \frac{d\sigma_C}{d\varepsilon_s},
\label{diffphotoprodRF}
\end{IEEEeqnarray}
where, again, the head-on approximation has been used, which implies $\Omega_e=\Omega_s$.

To calculate the scattered photon spectrum, we use the full Compton cross-section
with the scattering treated in the head-on approximation:
\begin{equation}
\frac{d\sigma_C}{d \varepsilon_s} \approx \frac{\pi r_e^2}{\gamma\bar{\varepsilon}} \ \Xi \  H(\varepsilon_s^*;\frac{\bar{\varepsilon}}{2\gamma}
, \frac{2 \gamma \bar{\varepsilon}}{1+2\bar{\varepsilon}}). 
 \label{crossection}
\end{equation}

$\bar{\varepsilon} \equiv \gamma \varepsilon^* (1-\beta \mu^* \mu_s^*)$ is the invariant collision energy after having averaged over $\phi$ and
\begin{equation}
\Xi \equiv y+y^{-1}- \frac{2\varepsilon_s^*}{\gamma \bar{\varepsilon}y}+(\frac{\varepsilon_s}{\gamma \bar{\varepsilon}y})^2 
\label{Xi},
\end{equation}

with $$y \equiv 1- \frac{\varepsilon_s^*}{\gamma}.$$

The Heaviside function gives the integration limits:
\begin{equation}
 \gamma_{low}=\frac{\varepsilon_s^*}{2} \left(1+\sqrt{1+\frac{2}{\varepsilon^*\varepsilon_s^* (1-\beta \mu^* \mu_s^*)}}\right)
\label{ylow}
\end{equation}
is the lowest $\gamma$ value at which an electron can still transfer enough energy to a photon so that it reaches the energy $\varepsilon_s^*$
after the scattering.   

\begin{equation}
 \varepsilon_{hi}^*=\frac{2\varepsilon_s}{1-\beta \mu^* \mu_s^*},
\label{ehi}
\end{equation}
is the highest photon energy, so that the smallest energy transfer still yields a scattered photon with $\varepsilon_s^*$. 

%Since the scattering is calculated in the rest frame of the accretion disk, we have to 
After transforming the emitting electron spectrum (calculated in the comoving frame) into the rest frame of the accretion disk by
 \begin{equation}
  n_e^*(\gamma^*,\Omega_s^*)=\delta_D^3 \frac{n_e(\frac{\gamma^*}{\delta_D})}{4 \pi},
\label{Netransform}
 \end{equation}

we combine Eq.~(\ref{nu_fnu_start}) with Eqs.~(\ref{diffphotoprodRF}),~(\ref{crossection}),~(\ref{Xi}) and ~(\ref{Netransform}) to arrive at

\begin{IEEEeqnarray}{rCl}
\label{diffphotoprodRFfinal}
 \varepsilon_s^* L_C (\varepsilon_s^*,\Omega_s,z) & = & \frac{c r_e^2\pi}{2} V_b \varepsilon_s^{*2} \delta_D^3 m_e c^2 \int_{-1}^{1} d\mu^*\\ 
&&\cdot  \int_0^{\varepsilon^*_{hi}} d\varepsilon^*
  \frac{n_{\rm{ph}}(\varepsilon^*,\mu^*,z)}{\varepsilon^{*}} \int_{\gamma_{low}}^{\infty} d\gamma^* \frac{n_e(\frac{\gamma^*}{\delta_D})}{\gamma^{*2}} \Xi \nonumber 
\end{IEEEeqnarray}
%\begin{IEEEeqnarray}{rCl}
%\label{diffphotoprodRFfinal}
% \varepsilon_s^* L_C (\varepsilon_s^*,\Omega_s) & = & \frac{c r_e^2}{4} V_b \varepsilon_s^{*2} \delta_D^3 m_e c^2 \\ 
%&&\cdot \int_{-1}^{1} d\mu^* \int_0^{\varepsilon^*_{hi}} d\varepsilon^*
%  \frac{n_{\rm{ph}}(\varepsilon^*,\mu^*)}{\varepsilon^{*}} \int_{\gamma_{low}}^{\infty} d\gamma^* \frac{n_e(\frac{\gamma^*}{\delta_D})}{\gamma^{*2}} \Xi \nonumber 
%\end{IEEEeqnarray}

for the $\nu L_{\nu}$ spectrum. 

The final photon spectrum that is due to the BLR target photons is calculated by using Eq.~(\ref{BLRtarphotfinal}) in conjunction with Eq.~(\ref{diffphotoprodRFfinal}).

To calculate the target photon density $n_{\rm{ph}}(\varepsilon^*,\mu^*)$ that is produced by the accretion disk photons, we integrate over the whole disk first.
We substitute $\dot{N}_{\rm{ph}}$ in Eq.~(\ref{SSDtarphot1}) with Eq.~(\ref{NphSSDfin}) and get

\begin{IEEEeqnarray}{rCl}
&n_{\rm ph}&^{\rm SSD}(\epsilon_*,\mu_*,z) =   \\
&&\int^{R_{max}}_{R_{min}}dR ~\frac{R_g^2 \cdot \Phi_F I(\tilde{R})}{\bar{\varepsilon}^* \tilde{R}^2 4 \pi x^2 c} \delta(\mu^*-\bar{\mu}_{*,{\rm accr}}) \delta(\varepsilon^*-\bar{\varepsilon}^*). \nonumber
 \label{SSDphotRunabh}
\end{IEEEeqnarray}

From $\bar{\mu}_{*,{\rm accr}}=(1+R^2/z^2)^{-1/2}$ we introduce $$a=\mu^*-\frac{1}{\sqrt{1+\frac{R^2}{z^2}}},$$ with
 $$dr=da \cdot \frac{z^2 (\frac{R^2}{z^2}+1)^{\frac{3}{2}}}{R},$$ and  
$$R(\mu^*)=\sqrt{z^2\cdot(\frac{1}{\mu^{*2}}-1)}.$$

Scaling the length scales $z=R_g \tilde{z}$ and $R=R_g \tilde{R}$ in terms of the gravitational radius $R_g$ we obtain
\begin{equation}
 n_{\rm{ph}}(\varepsilon^*,\mu^*,z)=\frac{\tilde{z}^2(1/\mu^*) R_g \Phi_F I(\tilde{R})}{4 \pi \sqrt{\tilde{z}^2\cdot(\frac{1}{\mu^{*2}}-1)}^3 \tilde{x}^2 \bar{\varepsilon}^* c} \delta (\varepsilon^*-\bar{\varepsilon}^*).
\end{equation}

This can be used in conjunction  with Eq.~(\ref{diffphotoprodRFfinal}) to find the scattered photon spectrum that is produced by the accretion disk target photon field.
The total photon spectrum produced is the sum of the photon spectra from both target photon fields.

We note that a further leptonic high-energy radiation process which contributes to the broadband blazar SED is synchrotron-self Compton (SSC) scattering. In the case of quasars with strong accretion disks, as considered in this work, the radiative output from this process has been found to typically dominate in a rather small energy range at X-rays (see e.g., \citealt{2000AIPC..515...31B}, \citealt{2009ApJ...692...32D}). Hence its contribution to the overall shape of the SED is markedly limited. We therefore do not include this process in our model, noting that our studies and the predictions given in the following should be considered with care at X-ray energies.

\section{Results}
\label{sec:Results}
We present the electron and photon spectra following the calculations outlined in Sect.~\ref{sec:ecalc} and Sect.~\ref{sec:Photocalc}.
For our calculations we set the black hole mass to $10^8 M_{\odot}$ (noting that the spectral shape only depends weakly on the black hole mass), the bulk Lorentz factor to $\Gamma=25$,
a fraction of $l_{\rm{edd}}=0.1$ of the Eddington luminosity for the central source and use a value of $\epsilon_{*0}=25(m_e c^2)^{-1}$eV as the characteristic photon energy of the BLR (monochromatic) spectrum. Here, synchrotron energy losses can be neglected for field values smaller than $\sim 4.5(\frac{z}{2000R_g})^{-3/2}$G.

In the following, we focus on photon spectra up to $\sim 10$GeV. Hence, $\gamma$-ray absorption during propagation through the diffuse extragalactic background light
can be neglected here (e.g., \citealt{2010ApJ...723.1082A}).
We also neglect internal $\gamma$-ray absorption in the external radiation fields, which could potentially affect
the spectral shape beyond $\sim 50$GeV for typical bright {\it Fermi}-LAT FSRQs (see Sect.~\ref{sec:Data_fits}), and focus our spectral considerations
 below this energy. We also note that most {\it Fermi}-LAT FSRQs were not detected beyond such energies owing to their
typically strong spectral decline at a few GeV (e.g., \citealt{2011ApJ...743..171A}), although few notable exceptions 
(e.g., 3C~279 (\citealt{2008Sci...320.1752M}), 4C +21.35 (\citealt{2011ApJ...730L...8A}), PKS~1510-089 (\citealt{2013A&A...554A.107H})) exist.

Below, the photon spectra are shown in the observer frame using a source redshift of $z_r=1$. All distances are given in terms of the gravitational radius $R_g$.

In our parameter study, we discuss the impact of varying selected parameters, with respect to a reference model, on the resulting particle and photon spectra.
The varied parameters are:
\begin{itemize}
 \item The height $z$ above the disk where the emission region is located at observing time $t_{\rm obs}$;
 \item The injection-rate index $\alpha$ at the position $z$;
 \item The starting and end points $z_a<z$ and $z_b>z_a$ of the particle injection; we set $z=z_b$ unless specified otherwise;
 \item The spectral index $s$ and the high (low) energy cutoff, $\gamma_2$ ($\gamma_1$), of the injected power-law electron spectrum $N_e^0(\gamma)=Q_0 \gamma^{-s}H(\gamma-\gamma_1)H(\gamma_2-\gamma)$;
 \item The geometry of the BLR: 
      \begin{itemize}
       \item $R_i$, $R_o$, inner and outer border of the BLR;
	\item $\zeta$ describes the density decline inside the BLR, $n_e^{\rm BLR} \propto R^{-\zeta}$;
	\item the BLR optical depth $\tau_{\rm{BLR}}=\sigma_T\int_{R_i}^{R_0}dR n_e^{\rm BLR}(R)$;
      \end{itemize}
\end{itemize}
The free parameters of the BLR are combined to result in four different BLR configurations BLR1-BLR4 (see \textbf{Table \ref{tab:BLR2s}} that are used to
explore their effect on the emitting electron and photon spectrum.
For the reference model, we use the following parameters: $s=3, \alpha=2, z_a=200, z=z_b=2000, \gamma_1=1000, \gamma_2=5\cdot 10^5, \tau_{\rm{BLR}}=0.01$, and BLR1 as the BLR geometry.
We use a dimensionless injection energy parameter $\Upsilon_{\rm inj}= E_{\rm inj}/E_{\rm inj}^{(ref)}$ to provide the total injected particle energy $E_{\rm inj}$ in each model in terms of the total injected particle energy of the reference model, $E_{\rm inj}^{(ref)}$.

\subsection{The BLR geometry}
\label{ssec:BLR}
\begin{figure*} % [!h]
        \centering
	\subfigure[Composite IC photon spectrum for the case of BLR1.]{\includegraphics[width=0.49\textwidth]{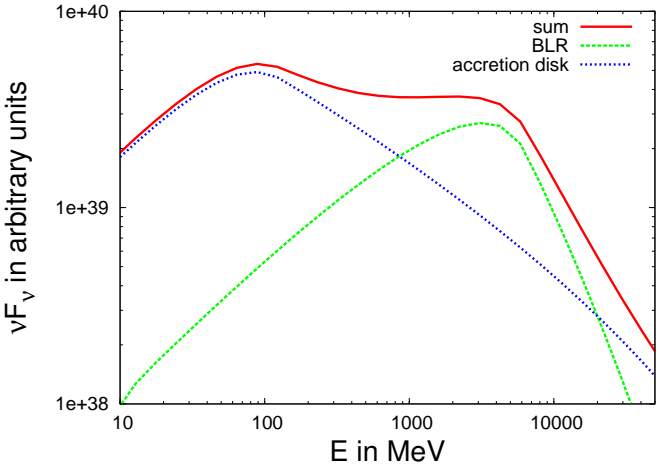}}
     	\subfigure[Resulting IC photon spectra when varying the BLR geometries. The inlay shows the underlying ambient electron spectrum.]{\includegraphics[width=0.49\textwidth]{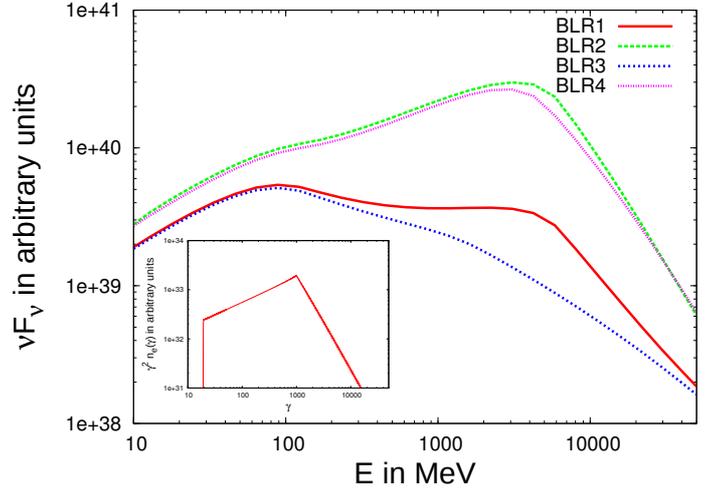}}
	\caption{IC photon spectra (and ambient electron spectrum as an inlay in (b) for the reference model) for different BLR geometries used.% Since the electron spectra is the same for all BLR geometries the total injected electron energy is $E_{inj,ref}=1$ for the cases shown here.  
} 
\label{fig:BLRgeometry}
\end{figure*}

\begin{table}[!htb]

 \centering
  \begin{tabular}{c|c|c|c|c}
	 & BLR1 & BLR2 & BLR3 & BLR4 \\ \hline
  $R_i$  & 100	& 1000 & 100  & 1000 \\
  $R_o$	 &10000	& 10000& 1000 & 2000 \\
  $\zeta$& -2	& -2   & 0    & 0    \\
  \end{tabular}
\caption{List of BLR geometries used. $R_i$, $R_o$ denote the inner and outer boundary of the BLR, $\zeta$ describes the gas-density decline in the BLR. All distances are in gravitational units.}
\label{tab:BLR2s}
\end{table}

We first study the influence of different BLR geometries (as listed in \textbf{Table~\ref{tab:BLR2s}}) upon the inverse Compton (IC) photon spectrum. 
Geometries BLR1 and BLR2 follow a $\zeta = -2$ density distribution of the BLR gas. Here most of its density is located close to the 
inner BLR boundary.
BLR3 and BLR4 follow a constant density distribution between the inner and the outer radius of the BLR.
The BLR geometry has no significant influence on the shape of the electron spectrum, a consequence of the
electron energy-loss rates in the BLR being rather low in general (see Sect.~\ref{ssec:BLRphotdens}). 
%That is important mention that ! more Figures ? The energy-loss figure ? 

\textbf{Figure~\ref{fig:BLRgeometry}a} shows the total IC photon spectrum being composed of two components:
The lower energy one is produced by IC scattering the accretion disk target photons, the second component, with a peak at 
around 3 GeVs, is produced by scattering the BLR target photons.
Using different BLR configurations with respect to the emission region leads to quite 
different IC spectra, as we present in \textbf{Fig.~\ref{fig:BLRgeometry}b}.

The BLR3 gas is distributed in a rather thin shell. With the emitting region located at $z=2000$ at the time of observation, 
i.e., beyond the outer boundary of the BLR3, the 
BLR photons enter the blob from behind 
with a collision angle similar to the accretion disk photons.
The BLR3 contribution is drowned out by the much more numerous accretion disk photons. Hence, \textbf{Fig.~\ref{fig:BLRgeometry}b} shows 
that the total IC spectrum for the BLR3 case
is nearly indistinguishable from the dominating accretion disk IC component (\textbf{Fig.~\ref{fig:BLRgeometry}a}).

In the BLR1 configuration the emission region is located within the BLR with
parts of the BLR1 gas being in front of this region. 
Target photons from this front part experience a higher energy transfer than the photons from the accretion disk, therefore they are scattered to higher energies.
Hence at high energies, the BLR IC component starts to poke through the accretion disk IC component (\textbf{Fig.~\ref{fig:BLRgeometry}b}).
This behaviour becomes most prominent in the BLR2 configuration, where an even larger fraction of the BLR gas lies in front of the blob.
Here again, the high energy BLR peak is produced by target photons that are backscattered by the BLR gas in front of the emission region.

The resulting IC spectrum from the BLR4 configuration, also shown in \textbf{Fig.~\ref{fig:BLRgeometry}b}, displays a shifted BLR peak position with respect to 
 BLR1 and BLR2.
In the BLR4 configuration, the emission region is located on the outer edge of the BLR. No target photons enter the blob from the front, 
but a large number of target photons enter from the sides. 
The maximum energy transfer for these photons is smaller than for photons that enter the blob from the front, thereby
shifting the peak position of the IC component to smaller energies.

In summary, if the ambient electron spectrum is unambiguously
determined, the relative position of the emission region with respect to the BLR geometry influences the relative strength of BLR and accretion disk component,
and may shift the peak energy of the BLR IC component.

\subsection{The luminosity of the accretion disk}
\label{ssec:led}

\begin{figure*} %[!h]
        \centering
        \subfigure[Emitting electron spectra for various $l_{\rm{edd}}$.]{\includegraphics[width=0.49\textwidth]{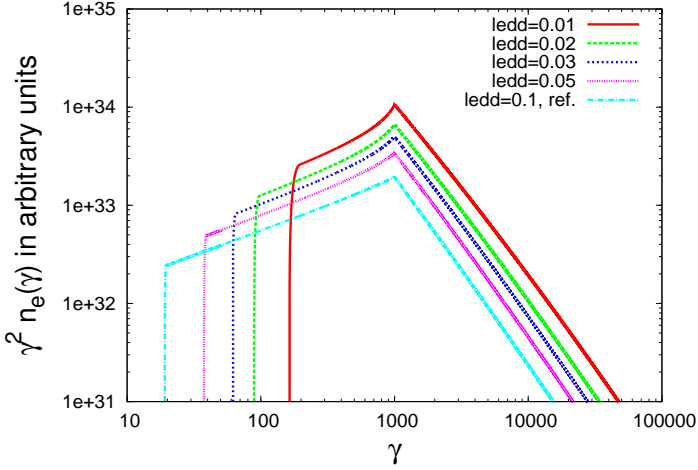}}\hfill
        \subfigure[Corresponding photon spectra to (a).]{\includegraphics[width=0.49\textwidth]{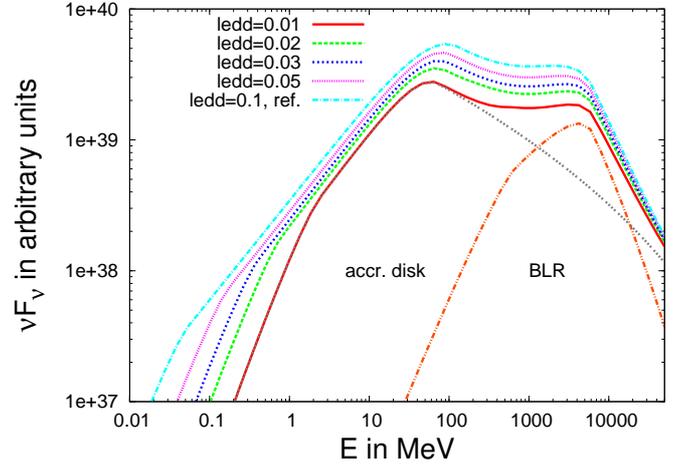}}\\
	\caption{Emitting electron and resulting photon spectra for various $l_{\rm{edd}}$. %The total injected electron energy is $E_{inj,ref}=1$ for all values of $l_{\rm{edd}}$
%since the injection behavior does not change in this example. 
Where $l_{\rm{edd}}=0.01$ we also show the photon components produced by both target photon fields separately.} 
\label{fig:ledd}
\end{figure*}

The accretion disk luminosity, here measured in terms of a fraction $l_{\rm edd}$ of the Eddington luminosity, directly influences the density of the target 
radiation field along the jet.
Higher accretion disk luminosities lead to increased particle cooling; the ambient electron spectrum reaches to lower minimum electron energies (see 
\textbf{Fig.~\ref{fig:ledd}a}). Consequently, the corresponding IC scattered photon spectra also reach lower energies as shown in \textbf{Fig.~\ref{fig:ledd}b}, 
as do the corresponding synchrotron radiation spectra.

\subsection{The injected particle spectrum}
\label{ssec:s}
The injected electron spectrum, a simple power law with spectral index $s$, is bracketed by the low-energy and high-energy cutoff particle Lorentz factors
 $\gamma_1$ and $\gamma_2$. 
The effects of varying $\gamma_1$, $\gamma_2$ and $s$ with respect to the reference model are explored below.

\begin{figure*} %[!h]
        \centering
        \subfigure[Emitting electron spectra when varying $\gamma_1$.]{\includegraphics[width=0.49\textwidth]{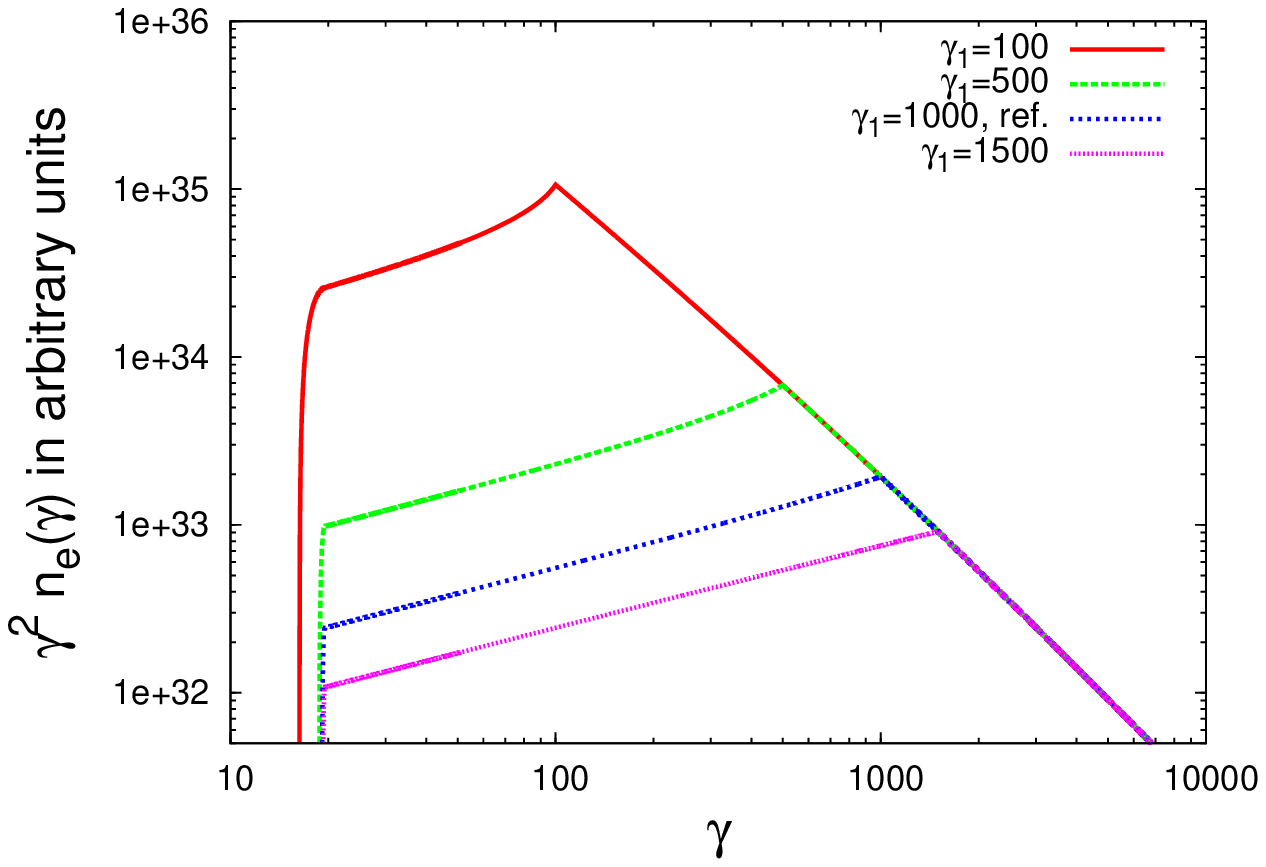}}\hfill
       	\subfigure[Corresponding photon spectra to (a).]{\includegraphics[width=0.49\textwidth]{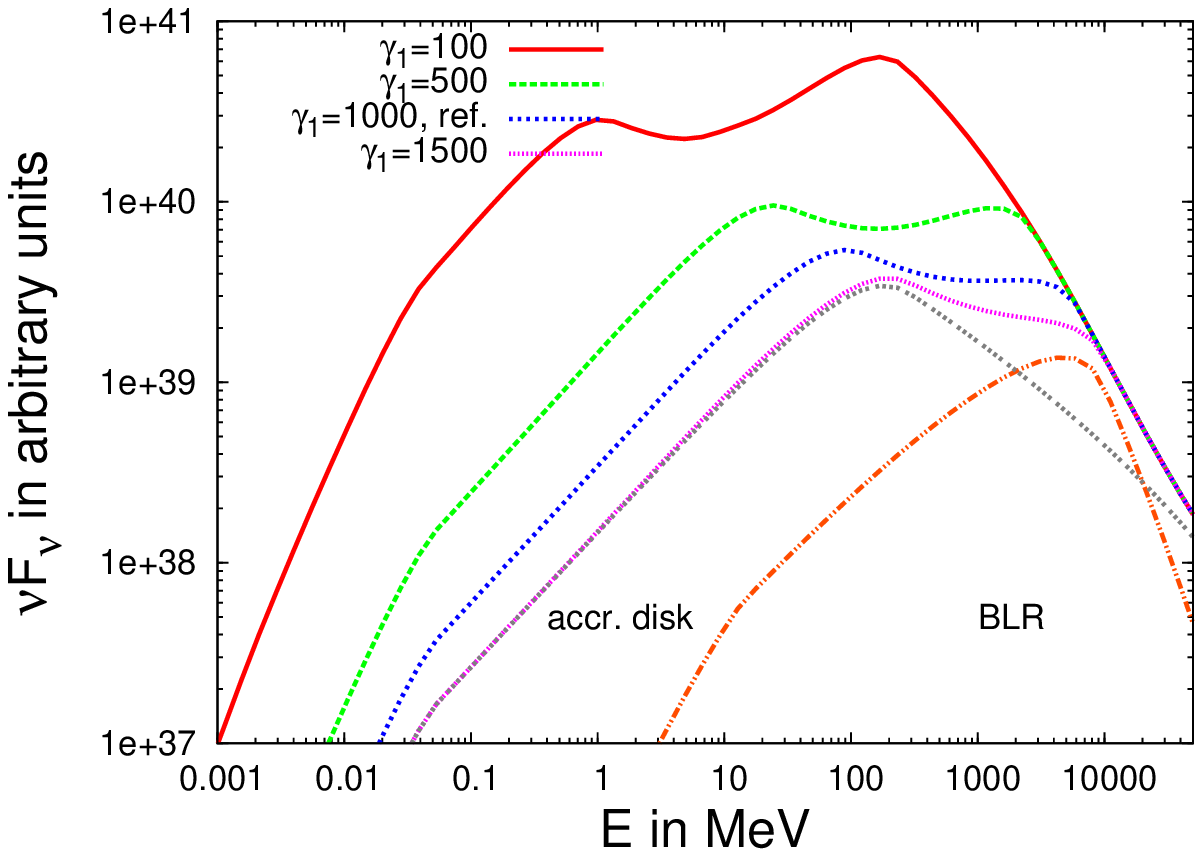}}
	\caption{Emitting electron spectra and corresponding photon spectra for $\gamma_1$ that varies with respect to the reference model. 
The total injected particle energies are strongly determined by $\gamma_1$. For $\gamma_1=(100, 500, 1000, 1500)$, we find $\Upsilon_{\rm inj}=(10.1, 2.0, 1, 0.7)$. 
} 
\label{fig:y1}
\end{figure*}

In \textbf{Fig.~\ref{fig:y1}a} we show the resulting electron spectra when $\gamma_1$ is varied, while leaving all the other parameters unchanged.
Clearly, at $\gamma=\gamma_1$ a break in the emitting electron spectrum generally occurs unless the injection rate is very steep (see Sect.~\ref{ssec:alpha}). 
The amount of the spectral index change
at the break energy can be understood as follows: For energies larger than $\gamma_1$, electrons 
are continuously injected and cooled at the same time. Hence, the produced power-law spectrum in this regime depends on the injected electron spectral index $s$.
At electron energies smaller than $\gamma_1$, only cooled particles accumulate. Thus the spectral shape depends on the injection rate index $\alpha$
(see Sect.\ref{ssec:alpha}). The combination of $s$ and $\alpha$ thus influences the broadness of the break in the emitting particle spectrum. In particular, 
breaks that are larger than typical cooling breaks, as implied, e.g., from observed GeV spectra of bright LAT FSRQs, become possible here. 

The corresponding photon spectra are shown in \textbf{Fig.~\ref{fig:y1}b}. The peak of the emitting electron spectrum 
translates into peaks in the corresponding components of the photon spectra. 
The different incident angles of the BLR target photon component, as compared to the accretion disk component, lead to a larger range of electron energies that 
can scatter photons 
up to the same energy. The geometric area from which the target photons can contribute to a certain scattered photon energy slowly decreases,
leading to a more gradual decrease, rather than a sharp cut-off towards high energies.
The BLR peak is shifted to higher energies with respect to the accretion disk peak, again the reason is the different incident angle distribution of the BLR 
target photons.
In the BLR component, head-on scattering with a higher energy transfer to the photon is common. Thus electrons with the same energy lead to higher energetic
scattered photons for the BLR target photons than for the accretion disk photons.
In \textbf{Fig.~\ref{fig:y1}b}, we also note the decreasing contribution of the Compton scattered BLR radiation to the total photon spectrum with 
increasing $\gamma_1$, a consequence of scattering in the Klein-Nishina regime of the cross-section with increasing energy.

\begin{figure*} %[!h]
        \centering
        \subfigure[Emitting electron spectra for different spectral injection indices $s$.]{\includegraphics[width=0.49\textwidth]{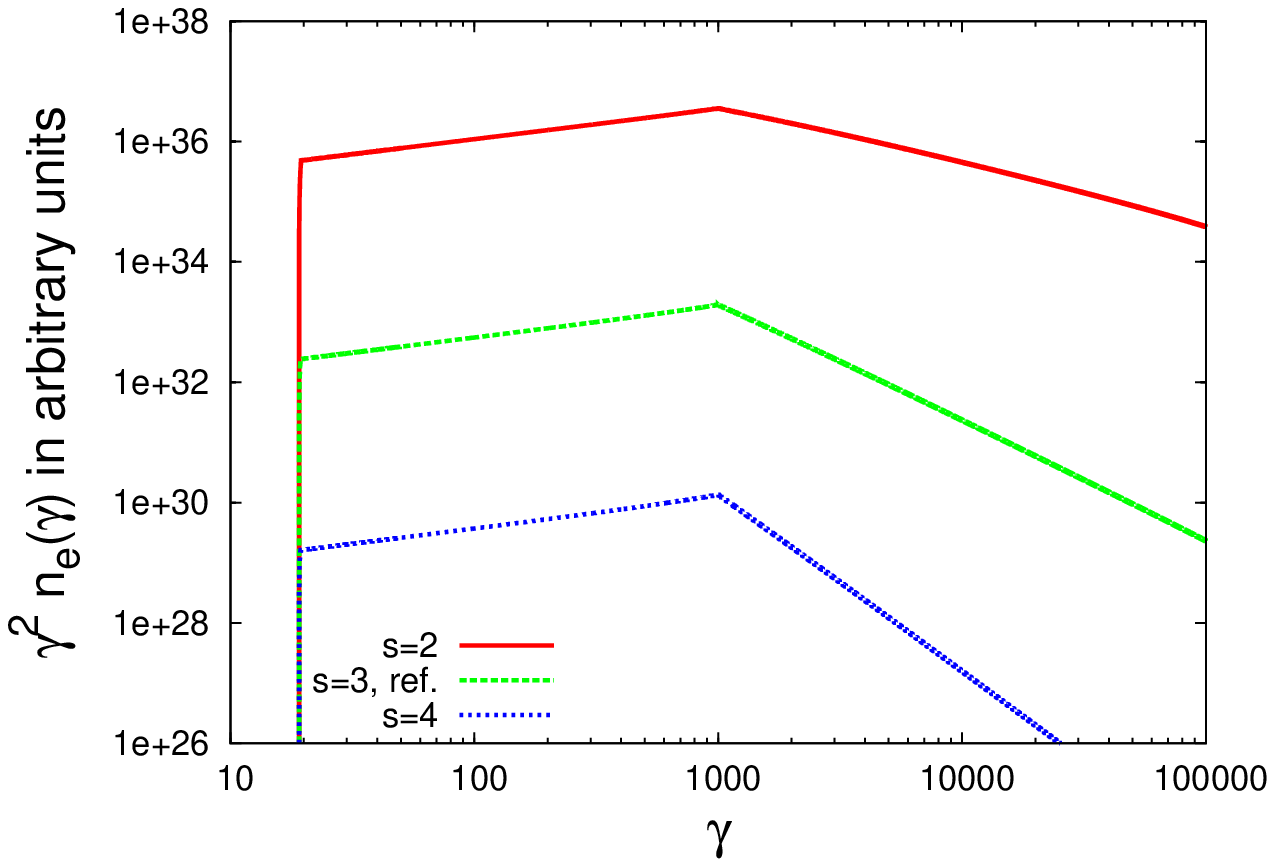}}\hfill
        \subfigure[Corresponding photon spectra to (a).]{\includegraphics[width=0.49\textwidth]{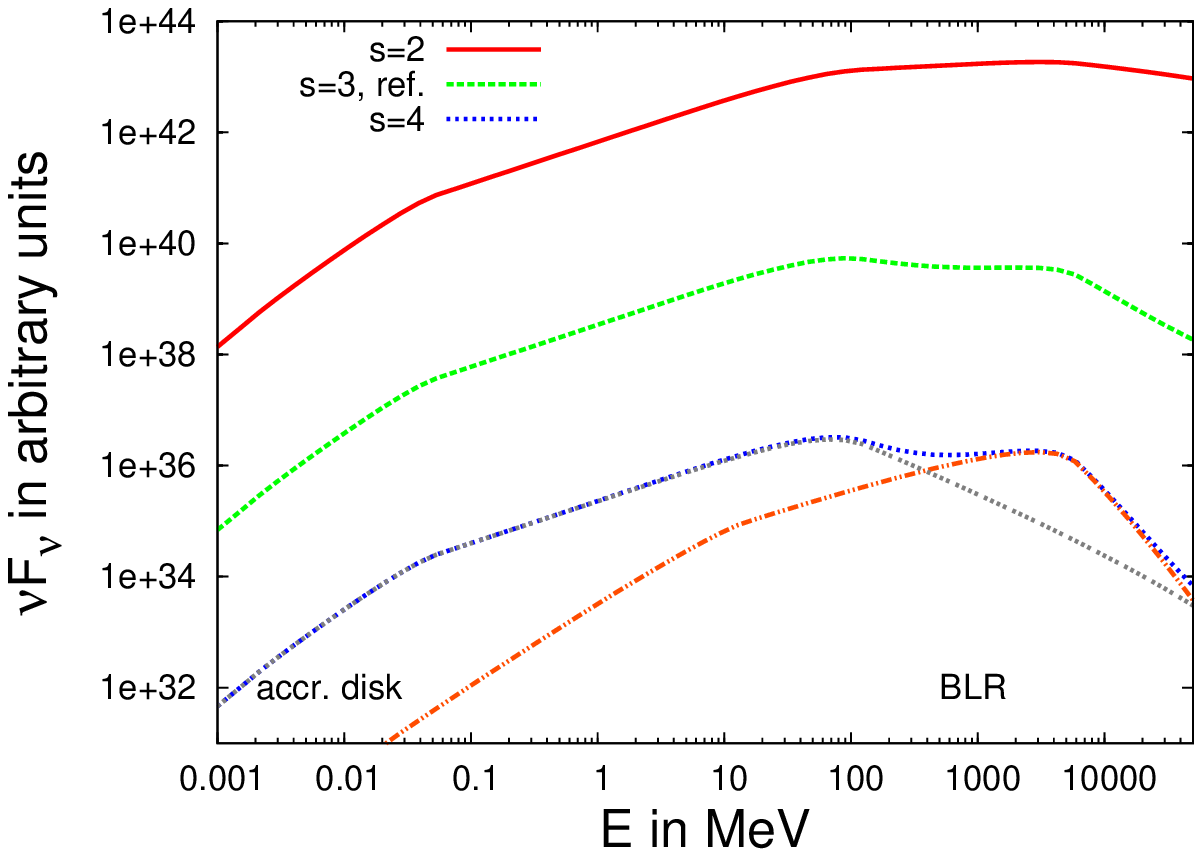}}\\
	\caption{Emitting electron and corresponding photon spectra for spectral injection indices $s$, varying with respect to the reference model. The total injected electron energies for 
$s=(2, 3, 4)$ are $\Upsilon_{\rm inj}=(4.7\cdot 10^3, 1, 5.1\cdot 10^{-4})$. In the case of $s=4$, the photon components produced by both target photon fields are shown separately 
as well.} 
\label{fig:svarBLR8}
\end{figure*}

As mentioned above, a change of the spectral injection index $s$ impacts the energy range above $\gamma_1$ in the emitting particle spectrum, becoming softer 
for softer injection spectra, and behaving as $n_e(\gamma)\propto \gamma^{-s-1}$.
Hence for a given injection rate index $\alpha$, the spectral break in the emitting particle spectrum becomes broader for harder injection spectra  
(see \textbf{Fig.~\ref{fig:svarBLR8}a}). For sufficient broad ambient electron spectra, the corresponding photon spectra from IC scattering the accretion disk 
and BLR radiation tend to merge 
into one component as is shown in \textbf{Fig.~\ref{fig:svarBLR8}b}. Again, we also note the impact of the IC scattering in the Klein-Nishina regime, which leads, 
for the same ambient electron 
spectrum, to softer photon spectra of the BLR-scattered component, as compared to the accretion disk scattered IC component beyond the respective peak energy.

\begin{figure*} %[!h]
        \centering
        \subfigure[Emitting electron spectrum for different $\gamma_2$.]{\includegraphics[width=0.49\textwidth]{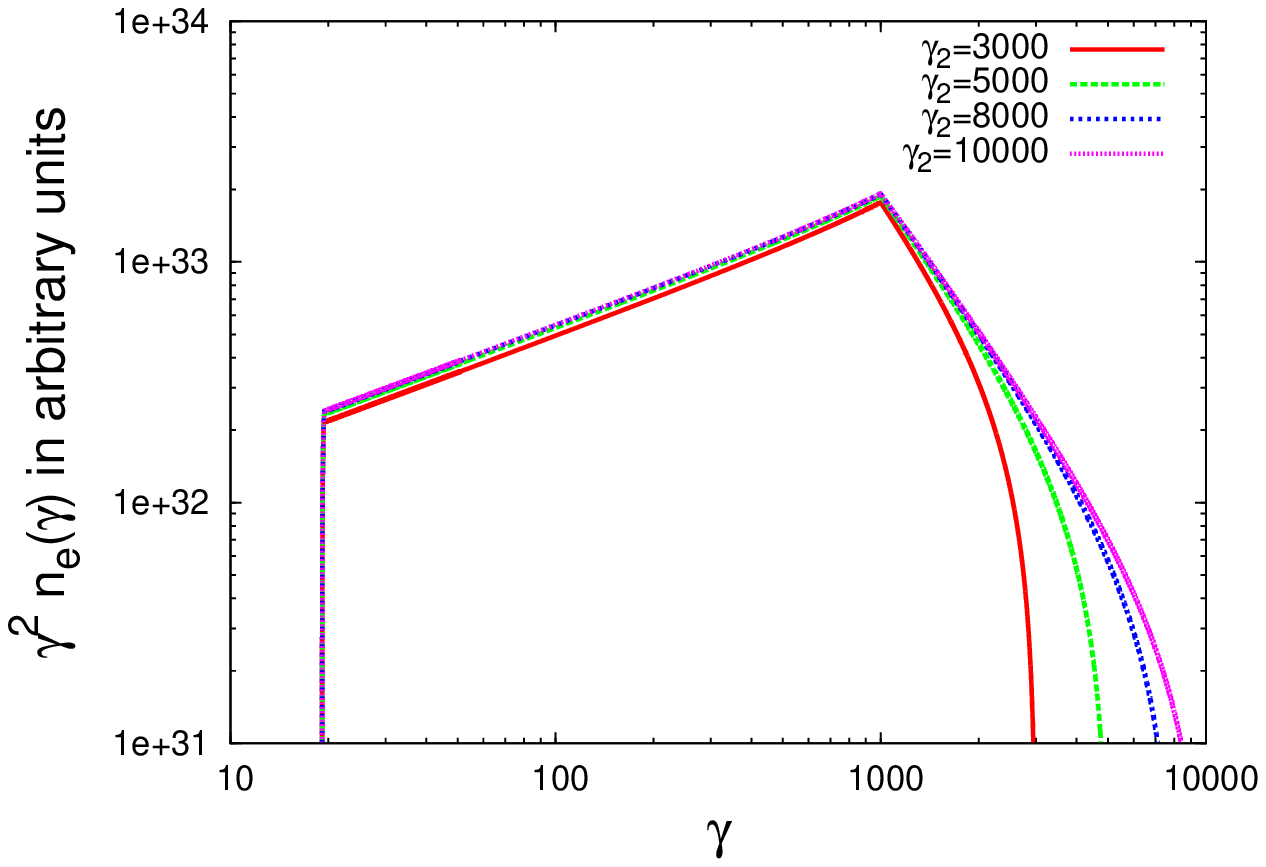}}\hfill
        \subfigure[Corresponding photon spectra to (a).]{\includegraphics[width=0.49\textwidth]{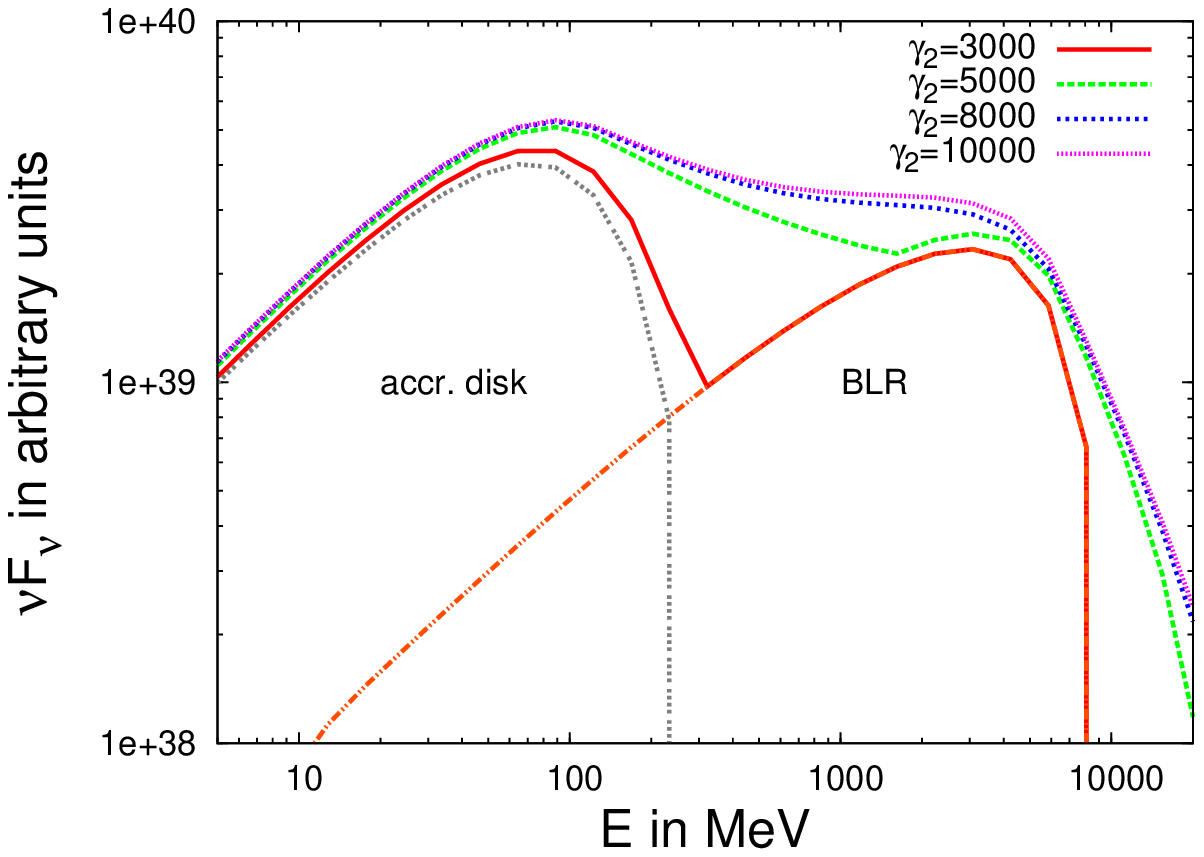}}\\
	\caption{Emitting electron spectra and corresponding photon spectra for $\gamma_2$ that vary with respect to the reference model.
The total injected electron energy is rather insensitive to $\gamma_2$. For $\gamma_2=(3000, 5000, 8000, 10000)$ we find $\Upsilon_{\rm inj}=(0.7, 0.8, 0.9, 0.9)$. In the case of $\gamma_2=3000$, we show the photon components produced by both target photon fields.} 
\label{fig:y2klein}

\end{figure*}

Very high energy electrons cool efficiently in the dense target radiation fields. Hence the particle spectrum above $\gamma_1$ quickly re-arranges to a 
state where injection rates balance the loss rates. For a sufficiently large high-energy cutoff of the injected electron spectrum, $\gamma_2$, where the 
radiative loss timescales at $\gamma_2$ are much smaller than the injection timescale, any variation of this parameter therefore only marginally impacts on
the resulting emitting electron and photon spectrum: The ambient particle spectra become very similar after only a short time owing to 
severe Compton losses.

For cases where the injected electron spectrum is bounded to an energy range where injection rates reach at least the radiative loss rates,
a variation of $\gamma_2$ results in a corresponding cut-off of the ambient particle and photon spectra,
as shown in \textbf{Fig.~\ref{fig:y2klein}}.
The values used in \textbf{Fig.~\ref{fig:y2klein}} for $\gamma_2$ within the reference model lead to a turn-over of the photon spectrum in the 
tens of MeV energy range when accretion disk photons are scattered, and in the GeV energy range when BLR target photons are scattered. The two photon 
components are visibly seperated only for sufficiently low $\gamma_2$.

\subsection{The particle injection mode}
\label{ssec:alpha}

The effects of different particle injection scenarios on the resulting emitting electron and photon spectra are explored in this section.
Particle injection starts within the jet at height $z_a$ above the accretion disk and continues with a rate $\propto z_i^{-\alpha}$ 
(where $\alpha$ is the injection rate index) up to a height $z_b$.

  \begin{figure*} %[!h]
        \centering
	\subfigure[Emitting electron spectra for various injection rate index $\alpha$.]{\includegraphics[width=0.49\textwidth]{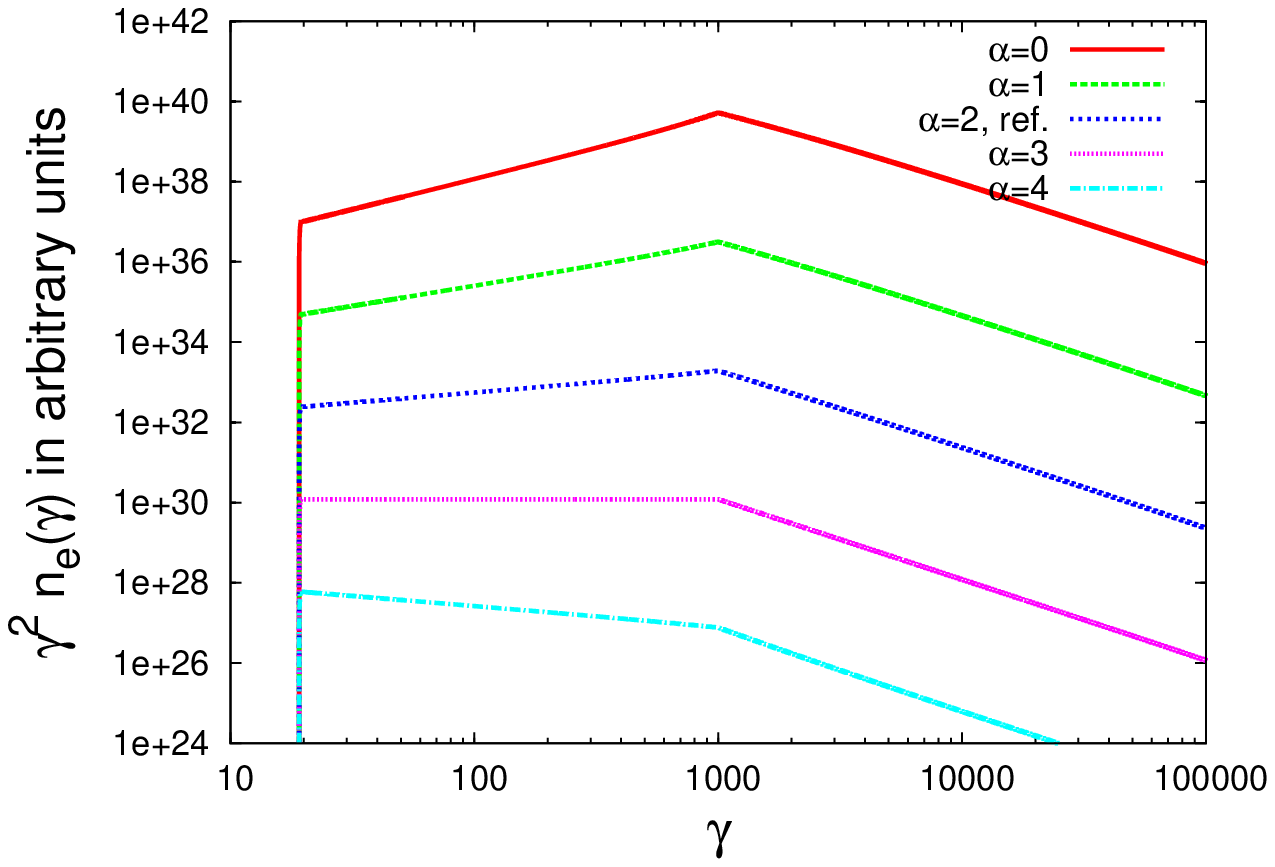}}
	\subfigure[Corresponding photon spectra to (a).]{\includegraphics[width=0.49\textwidth]{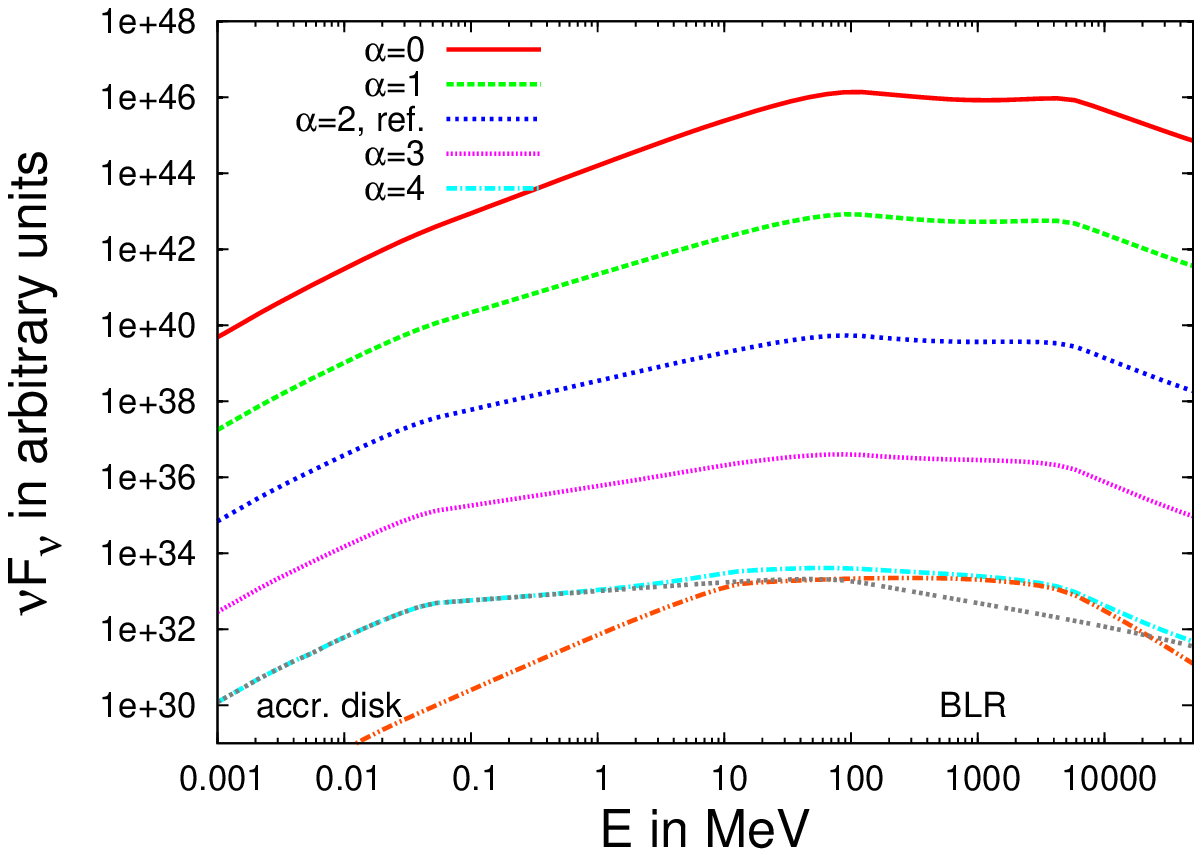}}
	\caption{Photon and ambient electron spectra with different injection rate parameters $\alpha$, with respect to the reference model. 
The total injected electron energies for $\alpha=(0, 1, 2, 3, 4)$ are calculated to $\Upsilon_{\rm inj}=(4\cdot 10^{5}, 512, 1, 0.3, 9\cdot 10^{-6})$.
In the case of $\alpha=4$, we also show the IC components produced by both target photon fields separately. } 
\label{fig:avarBLR6od001}
\end{figure*}

The injection rate index $\alpha$ controls the spectral shape of the ambient particle spectrum below $\gamma_1$ and (together with $s$) the spectral index 
change of the break.
\textbf{Fig.~\ref{fig:avarBLR6od001}} shows that harder pre-break spectral shapes develop for steeper injection rate distributions (large $\alpha$).
This can be understood as follows:
Particles in the pre-break regime $\gamma<\gamma_1$ are cooled electrons, which have suffered energy losses while propagating at least a distance that corresponds to $z_i^{-2} - z^{-2} < 2 K_n^{-1} (\gamma^{-1}-\gamma_1^{-1})$, where $z_i$ was the injection point. This implies a 
direct relation between $\gamma$ and $z_i$. Hence those particles injected at point $z_i$ contribute to the ambient spectrum at $\gamma<\gamma_1$. Since cooling rates are highest where the injection starts (at $z_a$), the number of particles injected there contribute to the lowest energies $\gamma\ll\gamma_1$ in the ambient spectrum. As a consequence, for very steep injection rate distributions, rather soft ambient spectra develop while, for shallow injection distributions, hard emitting particle spectra build up at $\gamma<\gamma_1$ (see \textbf{Fig. \ref{fig:avarBLR6od001}}).
Specifically, for constant injection rates ($\alpha=0$), we find an approximately $\propto \gamma^{-1/2}$ behaviour at $\gamma<\gamma_1$. For
$\alpha=3$, injection rates decrease with distance $z$ at the same rate as the cooling decreases. The ambient electron spectrum becomes independent of the blob location, and the well-known $n_e\propto \gamma^{-2}$ emitting electron spectrum develops for $\gamma<\gamma_1$. For well-developed electron distributions, $\alpha$ does not impact the spectral shape at $\gamma>\gamma_1$.

\begin{figure*} %[!h]
        \centering
        \subfigure[Emitting electron spectra when varying $z_a$ (in $R_g$)]{\includegraphics[width=0.49\textwidth]{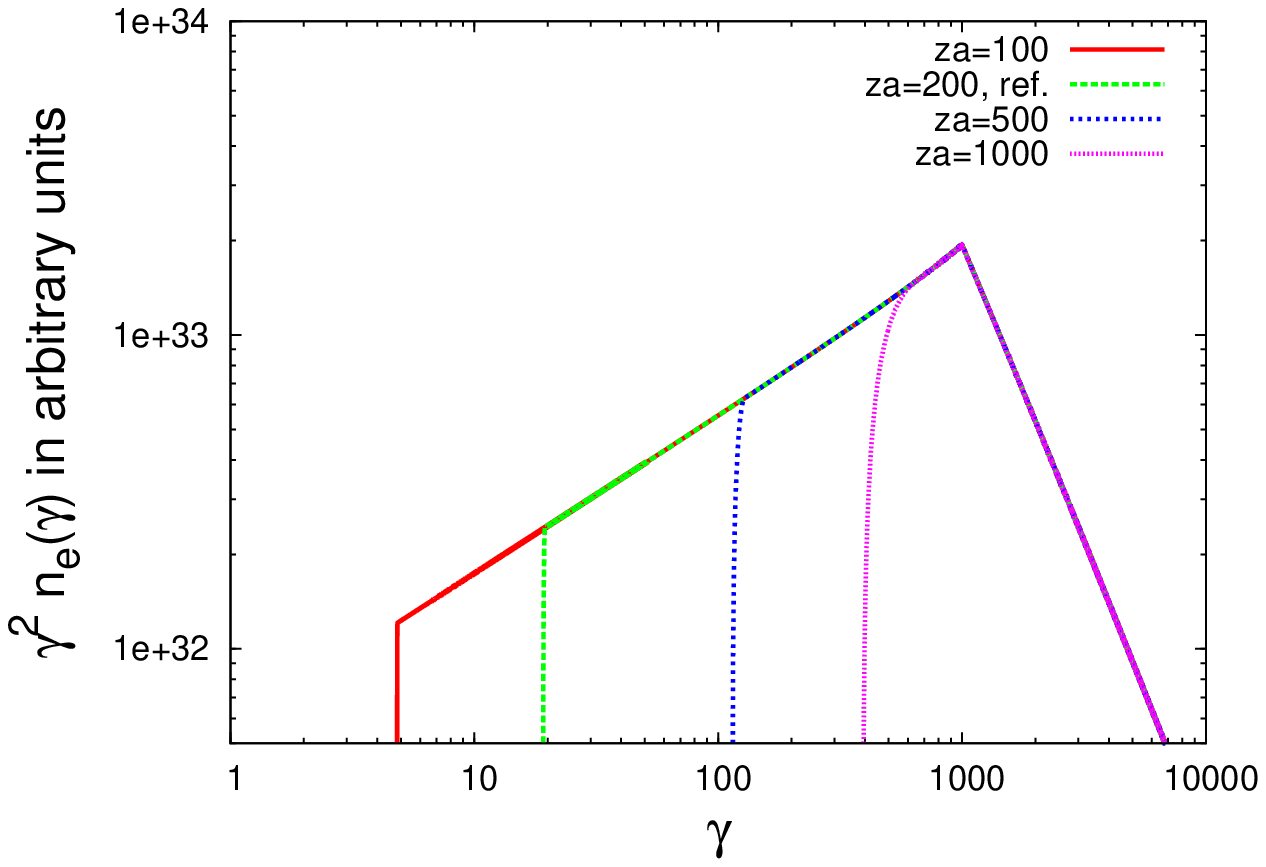}}\hfill
        \subfigure[Corresponding photon spectra to (a).]{\includegraphics[width=0.49\textwidth]{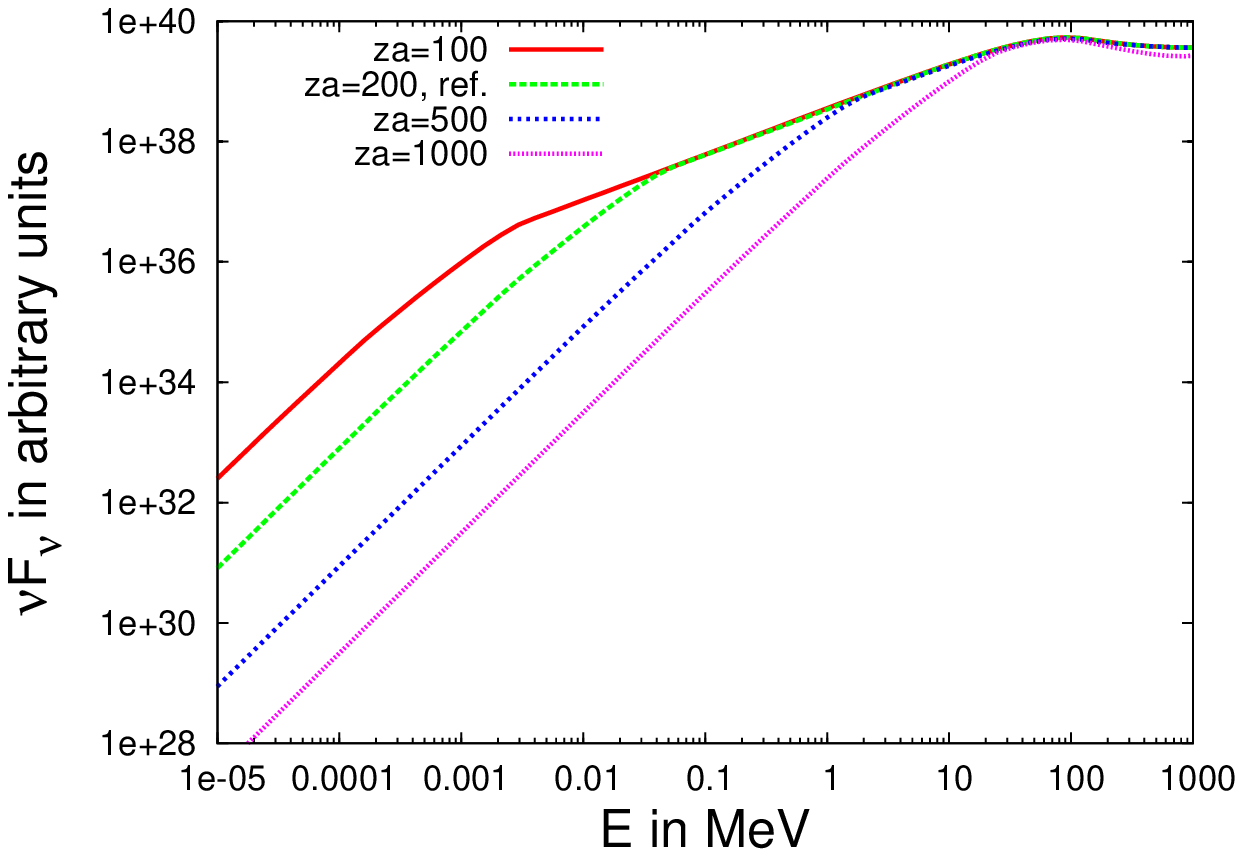}}\\
	\caption{Photon and ambient electron spectra when varying $z_a$ (in $R_g$) with respect to the reference model. The injected total electron energies for $z_a=(100, 200, 500, 1000)$  
are $\Upsilon_{\rm inj}=(2.1, 1, 0.33, 0.1)$. The total injected energy decreases with rising $z_a$. } 
\label{fig:zavarBLR6od001}
\end{figure*}

Particle injection starts within the jet at height $z_a$ above the accretion disk. \textbf{Figure~\ref{fig:zavarBLR6od001}} shows the resulting electron and photon spectra for different values of $z_a$, all other injection parameters are kept constant.

A higher cooling rate in the accretion disk radiation field, as well as a longer total cooling time (due to larger $z-z_a$ distance),
leads to lower electron energies with decreasing $z_a$ (see \textbf{Fig.~\ref{fig:zavarBLR6od001}}a). Hence, 
the corresponding photon spectrum below $\gamma_1$ (\textbf{Fig.~\ref{fig:zavarBLR6od001}}b) softens and extends to very low energies depending on $z_a$.

  \begin{figure*} %[!h]
        \centering
        \subfigure[Emitting electron spectra for various distances between the points $z$ and $z_b$ (in $R_g$)]{\includegraphics[width=0.49\textwidth]{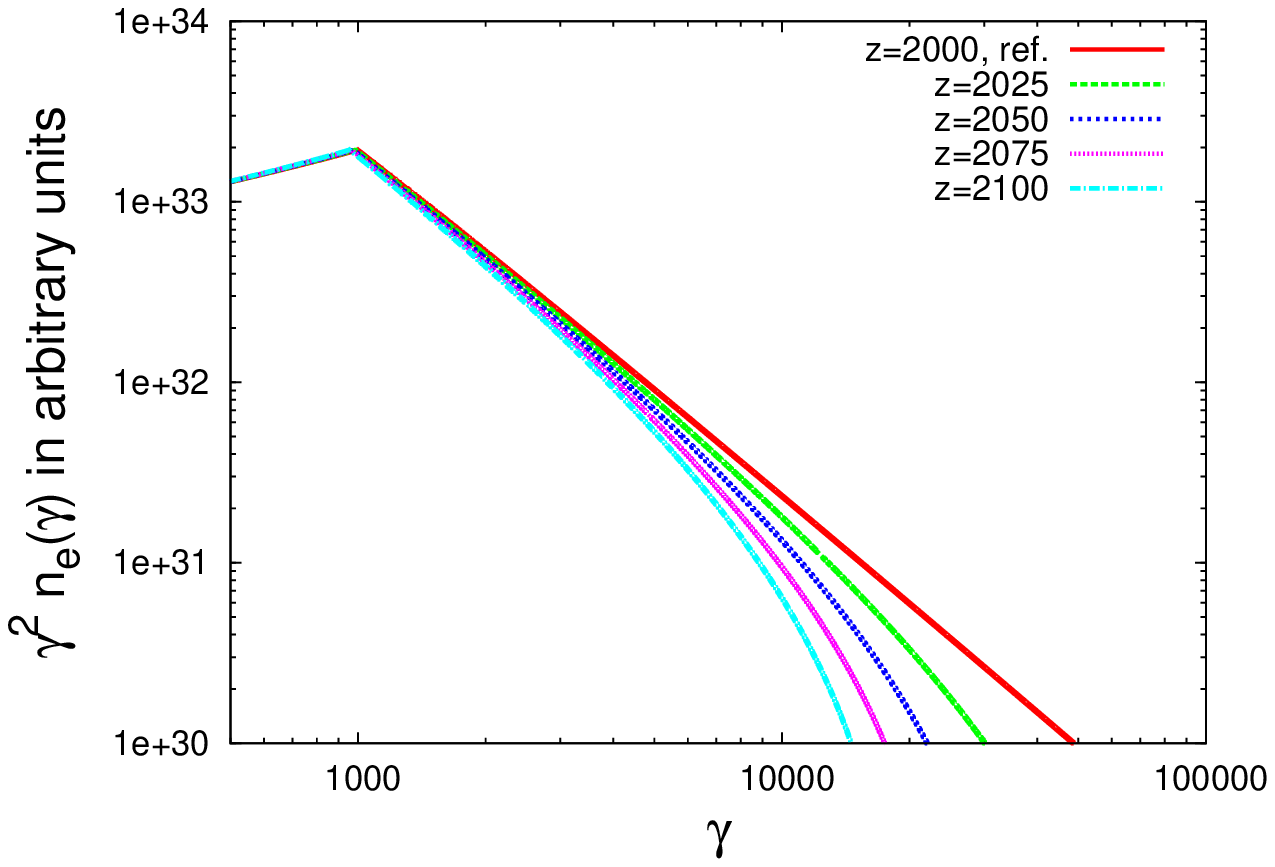}}\hfill
        \subfigure[Corresponding photon spectra to (a).]{\includegraphics[width=0.49\textwidth]{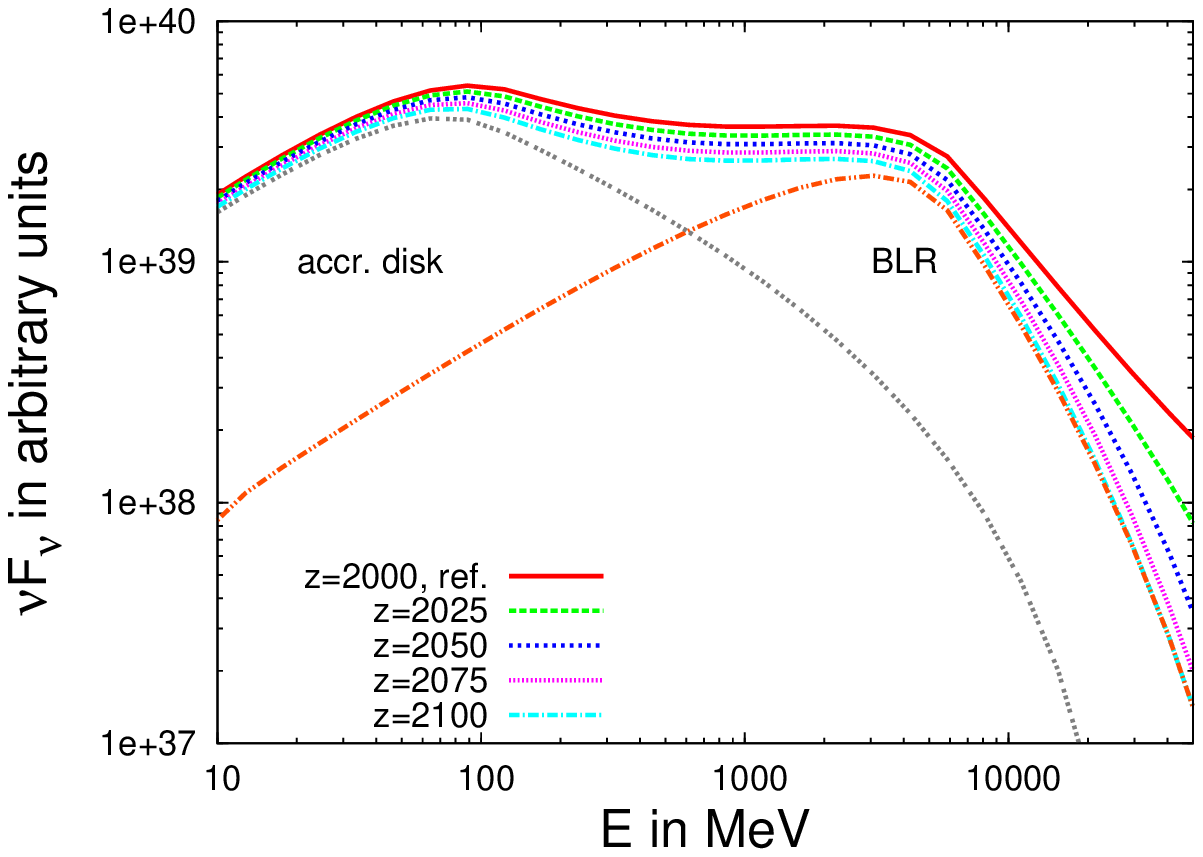}}\\
	\caption{Photon and ambient electron spectra when varying the distances between the points $z$ and $z_b$ (in $R_g$), with respect to the reference model. The total injected electron 
energy is $\Upsilon_{\rm inj}=1$ in all cases. For $z=2100$, we also show the IC photon components produced by the two target photon fields considered separately.} 
\label{fig:zvarBLR8od01}
\end{figure*}

Next, we explore the situation of a finite particle injection duration (corresponding to a distance $z_b-z_a$ in the jet) where the resulting spectra are observed at a given time after the last particle injection has occured. This time corresponds to a distance $z-z_b$ in the jet.
\textbf{Figure~\ref{fig:zvarBLR8od01}} shows the resulting electron and photon spectrum for various $z$ with fixed points $z_a$ and $z_b$. 
Because of the strong energy dependence of the electron cooling (with higher energy electrons cooling faster than lower energy electrons; see Sect.\ref{ssec:s}) we observe the exponential high energy cut-off 
in the electron spectrum moving to lower energies while, below  $\gamma_1$, the electron spectrum does not change significantly. 
 
The influence of the electron cut-off is reflected in the corresponding photon spectrum.
Although the electron injection spectral index $s$ is constant, the post-break photon spectrum becomes softer, owing to the lower energies of the electron energy cut-off.
In \textbf{Fig.~\ref{fig:zvarBLR8od01}b}, we also show the total photon spectrum decomposed into the IC-scattered accretion disk and BLR photon spectrum. We note that the turn-over in the photon spectrum at a few GeV has its origin in the break of the underlying electron spectrum.

  \begin{figure*} %[!h]
        \centering
        \subfigure[Emitting electron spectra when varying $z_b$ (in $R_g$)]{\includegraphics[width=0.49\textwidth]{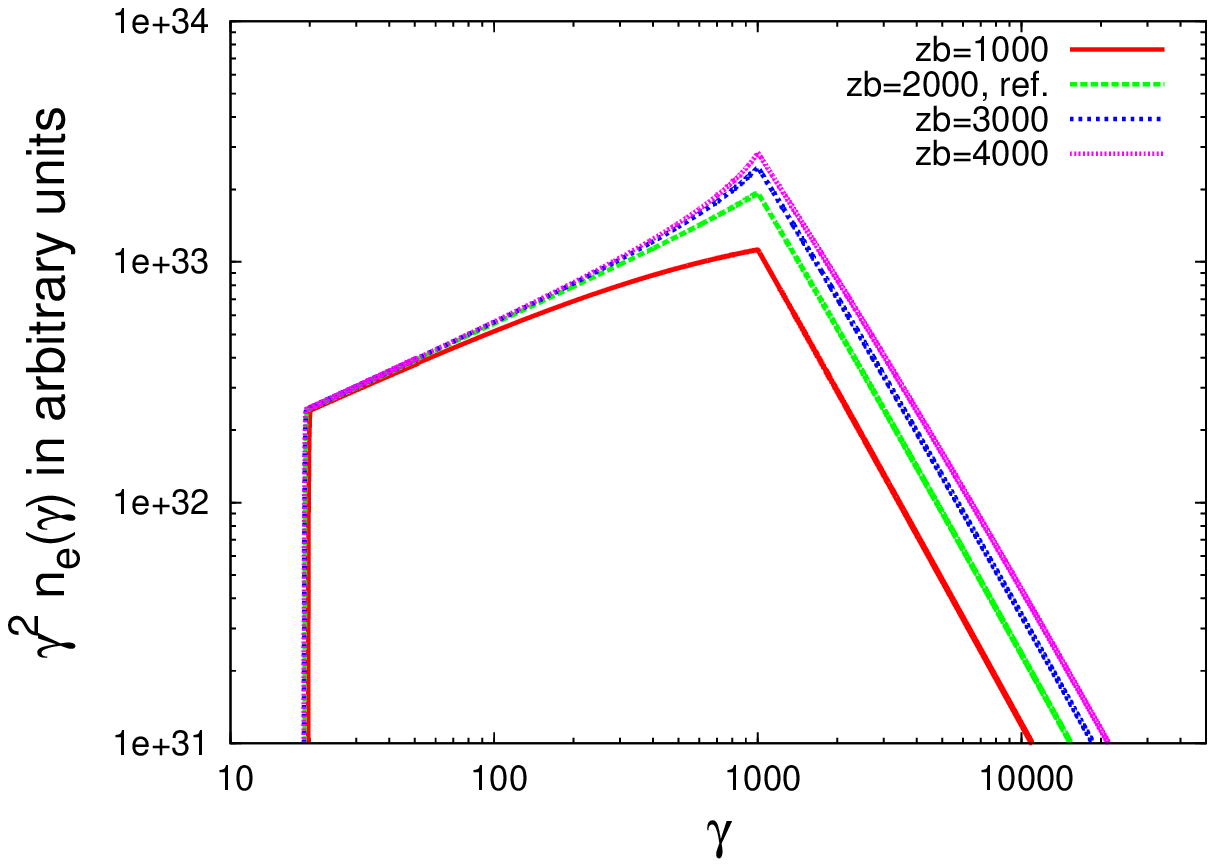}}\hfill
%        \subfigure[Normalized electron spectra for differing $z_b$ (in $R_g$)]{\includegraphics[width=0.49\textwidth]{Bilder/s4.0a0z=zbnorm.eps}}\\
	\subfigure[Corresponding photon spectra to (a).]{\includegraphics[width=0.49\textwidth]{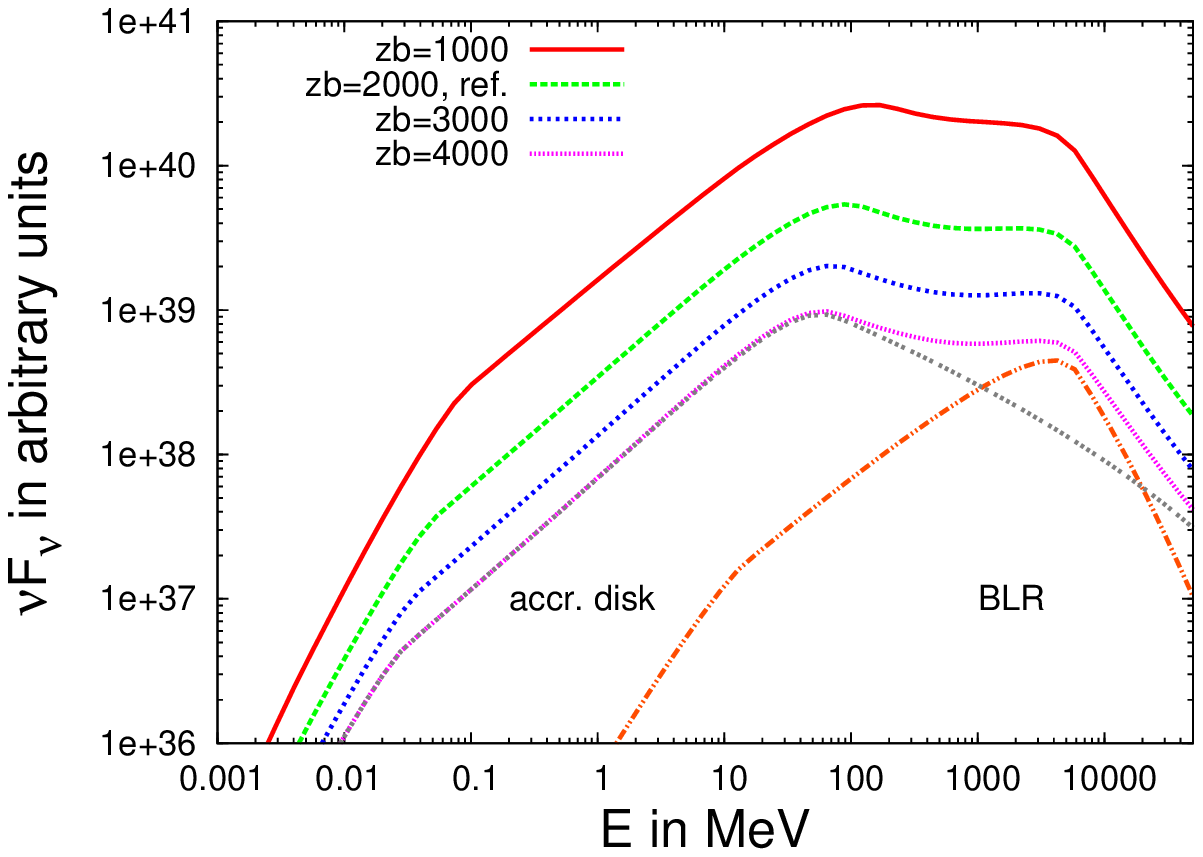}}\hfill
	\caption{Photon and ambient electron spectra for various $z_b$ (in $R_g$) with respect to the reference model. The total injected electron energies for 
$z_b=(1000, 2000, 3000, 4000)$ are $\Upsilon_{\rm inj}=(0.9, 1, 1.04, 1.06)$.
For $z_b=1000$, the IC photon components produced by the two target photon fields are also shown separately.} 
\label{fig:zbvara0}
\end{figure*}

  \begin{figure*} %[!h]
        \centering
        \subfigure[Normalized electron spectra for various $z_b$ (in $R_g$) and $\alpha=3$]{\includegraphics[width=0.49\textwidth]{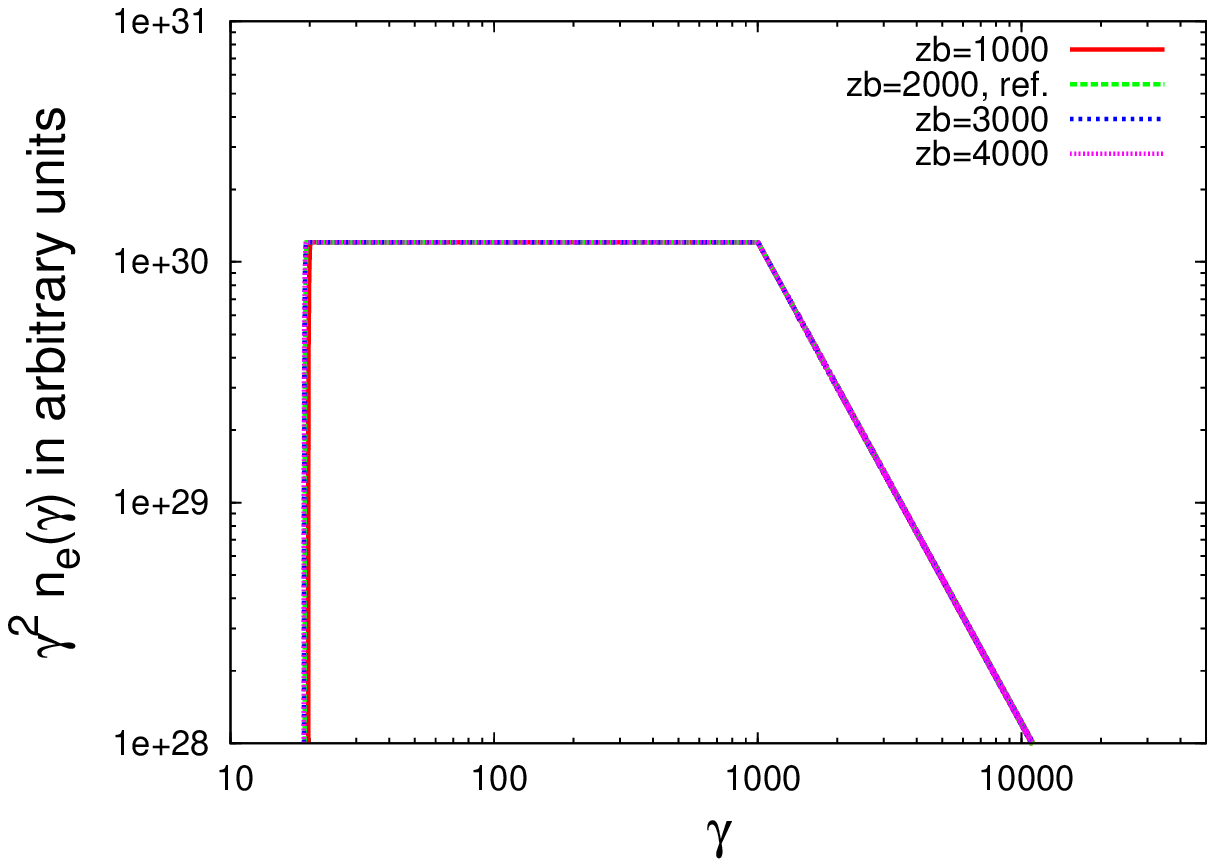}}\hfill
        \subfigure[Corresponding photon spectra for various $z_b$ (in $R_g$) and $\alpha=3$]{\includegraphics[width=0.49\textwidth]{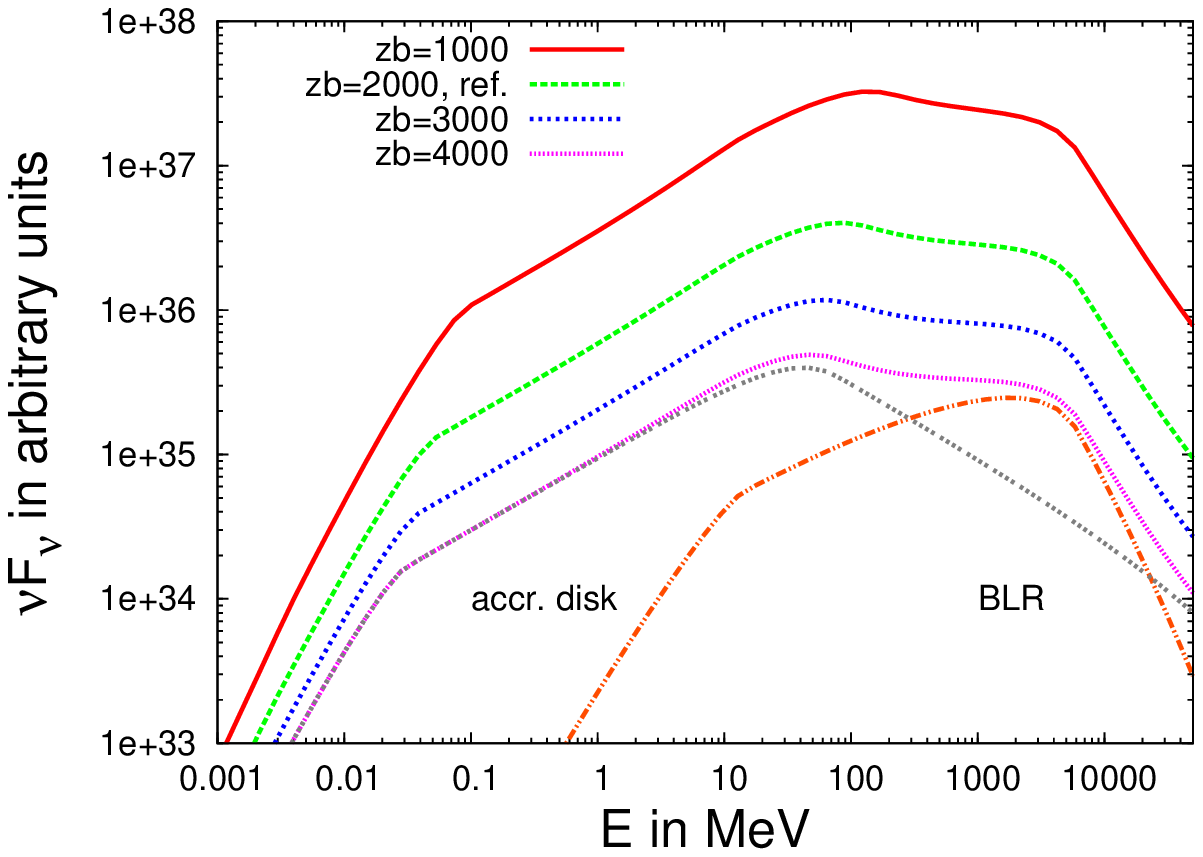}}\\
	\subfigure[Normalized electron spectra for various $z_b$ (in $R_g$) and $\alpha=7$]{\includegraphics[width=0.49\textwidth]{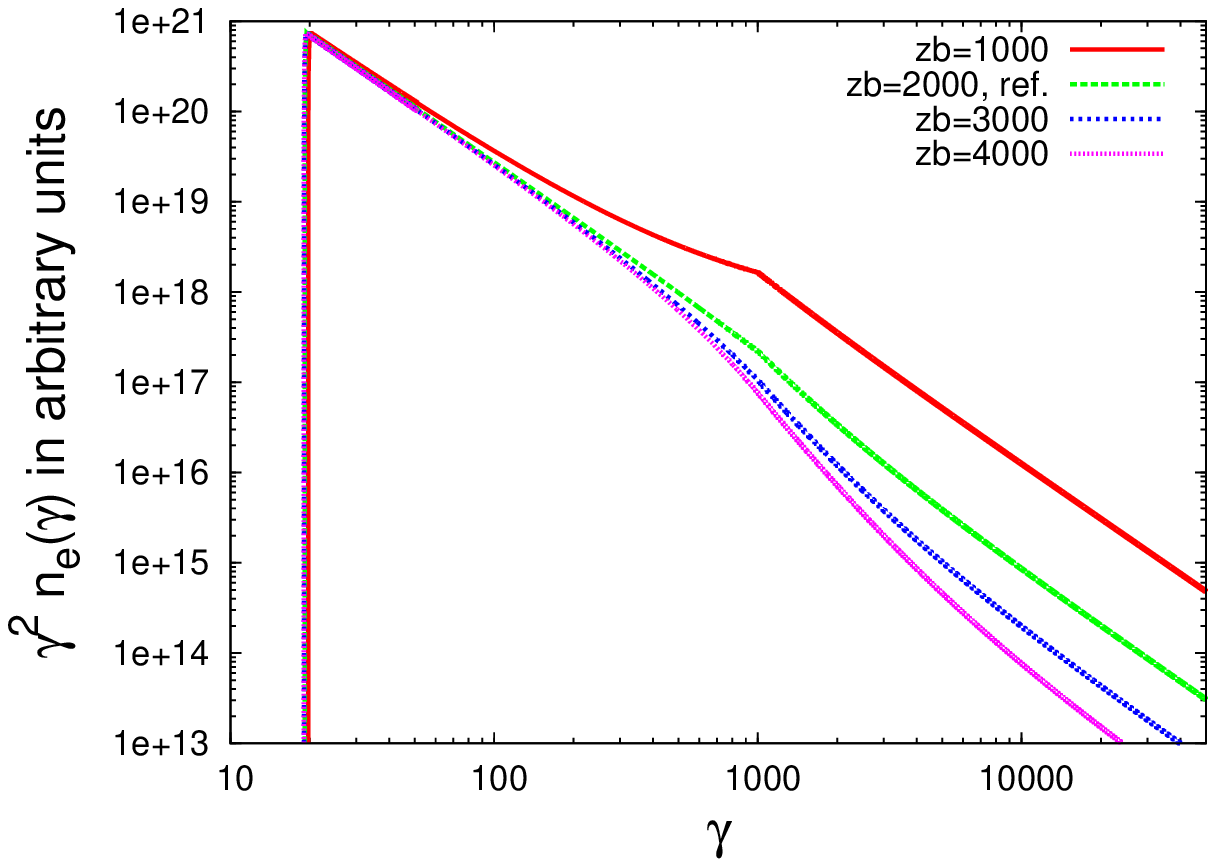}}\hfill
	\subfigure[Corresponding photon spectra for various $z_b$ (in $R_g$) and $\alpha=7$]{\includegraphics[width=0.49\textwidth]{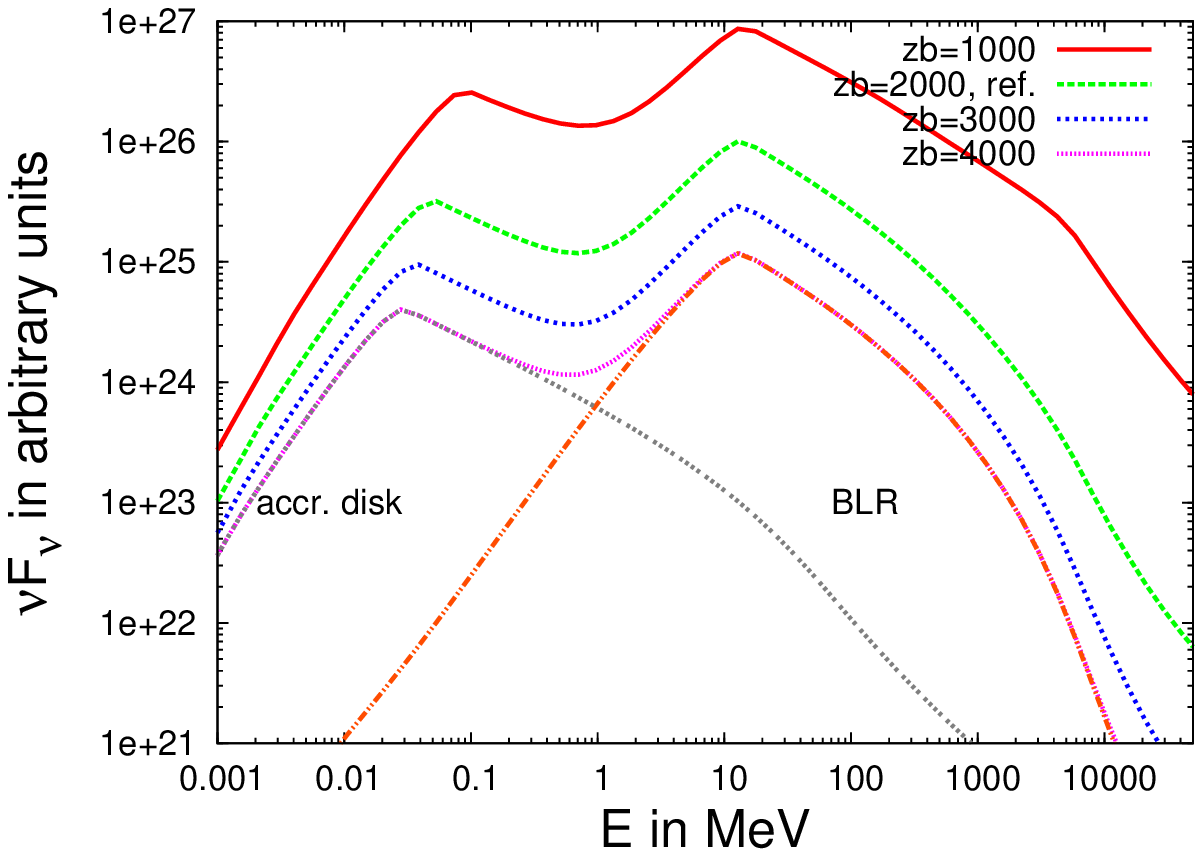}}\\
	\caption{Photon and ambient electron spectra for $z_b$ (in $R_g$) and $\alpha$ with respect to the reference model. The total injected electron energies for $\alpha=3$ and $z_b=(1000, 2000, 3000, 4000)$ are $\Upsilon_{\rm inj}=(2.7, 2.75, 2.8, 2.8) \cdot 10^{-3}$ while,
for $\alpha=7$, the total injected electron energies for $z_b=(1000, 2000, 3000, 4000)$ are $\Upsilon_{\rm inj}=(5.8, 5.8, 5.8, 5.8) \cdot 10^{-13}$. 
For $\alpha=3$ and $\alpha=7$ with $z_b=1000$, we also show the IC photon components produced by the two target photon fields separately.}
\label{fig:zbvara37}
\end{figure*}

We now explore the impact of a changing injection duration upon the resulting spectra. For this purpose, we vary $z_b$ while keeping $z_a$ fixed, and analyse the resulting particle and photon spectra at $z=z_b$. The longer the injection at a given rate, the higher the total injected particle energy. At the same time, the particles cool, at a, generally, decreasing rate with increasing distance from the black hole.
Hence, we find a strong dependence of the resulting spectral shapes upon the interplay between the injection and energy-loss rates. Therefore, we also present our results for various injection rate indices $\alpha$.

For a sufficiently shallow decrease of the injection rate, a pile-up forms in the emitting particle spectrum at $\gamma=\gamma_1$,
  with the spectral index below $\gamma_1$ becoming harder. This is a result of
the cooling rate of particles at low energies not being sufficiently large to be able to fully balance the injection rate at these energies. Long injection times at sufficiently large $z$
ensure an accumulation of particles at low energies, with reduced cooling rates leading to the pile-ups becoming more prominent with increasing $z_b$.
This is demonstrated in \textbf{Figs.~\ref{fig:zbvara0}a} and \textbf{~\ref{fig:zbvara0}b}, where the spectra are shown for $\alpha=2$ and various $z_b$.

Electrons pile up at a given particle energy $\gamma$ in a spectrum when, in momentum space, the net change of the number of electrons entering  this $\gamma$-bin (either from injection or from cooling down from a higher energy bin $\gamma+d\gamma$) and leaving this bin to $\gamma-d\gamma$, due to cooling, is positive. For an increasing cooling rate with energy, this occurs at the low-energy end of the injected spectrum, i.e., at $\gamma=\gamma_1$. 
For broad injection distributions $\alpha<3$ particle injection at large $z$, where cooling rates are diminished, is still significant. This facilitates accumulating particles within an energy bin. A pile-up forms.
As noted above, for $\alpha=3$, the emitting electron spectrum becomes independent of the blob location, and 
hence also of $z_b$. Thus, in this case, a pile-up does not form (see \textbf{Fig.~\ref{fig:zbvara37}a}). For $\alpha>3$, the
effects discussed above reverse and the spectral index becomes softer after the break $\gamma=\gamma_1$. In \textbf{Fig.~\ref{fig:zbvara37}c} we present the case of $\alpha=7$. Here the injection scenario becomes more ``instantaneous-like'', and we follow the evolution of the injected spectrum for a time that corresponds to $(z_b-z_a)/\beta_\Gamma c$. For increasing $z_b$, the emitting electron spectrum steepens while extending to low energies $\gamma<\gamma_1$ owing to cooling. In the limit of large $z_b$, all electrons have cooled and the injected power law is mapped into the $\gamma<\gamma_1$ energy regime.
The overall peak of the emitting electron spectrum for an instantaneous-like injection is therefore found at the low-energy end. Here, the emitting electrons correspond to the cooled particles that were injected at $\gamma_1$. Their energy depends on the cooling rate, which is highest at the injection point $z_a$. 

The corresponding photon spectra (see \textbf{Figs.~\ref{fig:zbvara0}b}, \textbf{~\ref{fig:zbvara37}b} and \textbf{~\ref{fig:zbvara37}d}) reflect the shapes of the respective ambient electron spectra, with some impact of the relative position of the 
emitting region with respect to the BLR (the latter has been discussed in more detail in {Sect.~\ref{ssec:BLR}}). 
%AFR: Seems to me not significant. 
%It seems that for higher $z_b$ the BLR peak and the accretion disk peak are further apart. That means that for a large distance $z_a$ to $z_b$ the photon spectrum
%has a more two peaked shape while a small distance leads to a more one peaked photon spectrum.
%This is a consequence of the more prominent pile-up in the electron spectrum for higher $z_b$. The differing spectral% indices of the photon spectra 
%in the high energy regime are a consequence of the differing spectral indices of the corresponding electron spectra. 
We note the shift of the IC peaks in \textbf{Fig.~\ref{fig:zbvara37}d} to lower energies which is due to the corresponding emitting electron spectra peaking in these cases at the very low-energy end.

\section{Modeling the 3C~454.3 and PKS~1510-089 multifrequency SEDs}
\label{sec:Data_fits}
To show the viability of our model, we apply it to quasi-simultaneous broadband data that was collected during the November 2010 outburst of 3C~454.3 (\citealt{2011ApJ...733L..26A}, \citealt{2012ApJ...758...72W}), with a particular focus on the {\it Fermi}-LAT spectrum that shows a distinct spectral decline at a few GeV.
Additionally, we use our model to also fit the quasi-simultaneous broadband SED of PKS~1510-089, as reported in \citealt{2010ApJ...716...30A}.
The low-energy photon data provides additional valuable constraints on the emitting electron distributions. Within our model, we calculate the synchrotron radiation from the ambient electron distribution following \citet{2009ApJ...692...32D}, assuming a tangled magnetic field of strength $B$, and including synchrotron-self absorption. % (following \citealt{2014ApJ...782...82D}). 

\subsection{Modeling the 3C~454.3 multifrequency SED}
\label{sec:Fit_to_data_3C4543}
\begin{figure*} %[!h]
        \centering
        \subfigure[Emitting electron spectrum]{\includegraphics[width=0.49\textwidth]{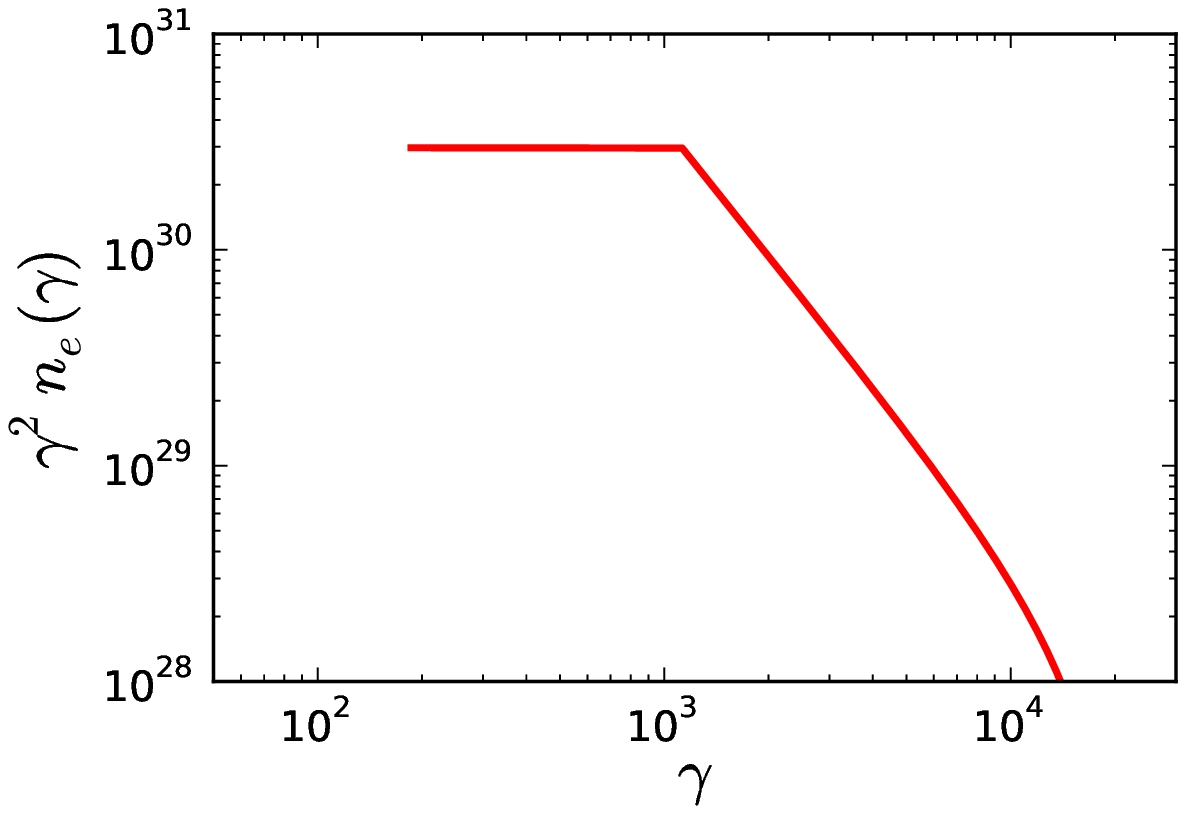}}\hfill
        \subfigure[Synchroton spectrum with \textit{Swift}, \textit{Herschel} and \textit{submillimeter array} data taken during the November 2010 flare (\citealt{2012ApJ...758...72W}).]{\includegraphics[width=0.49\textwidth]{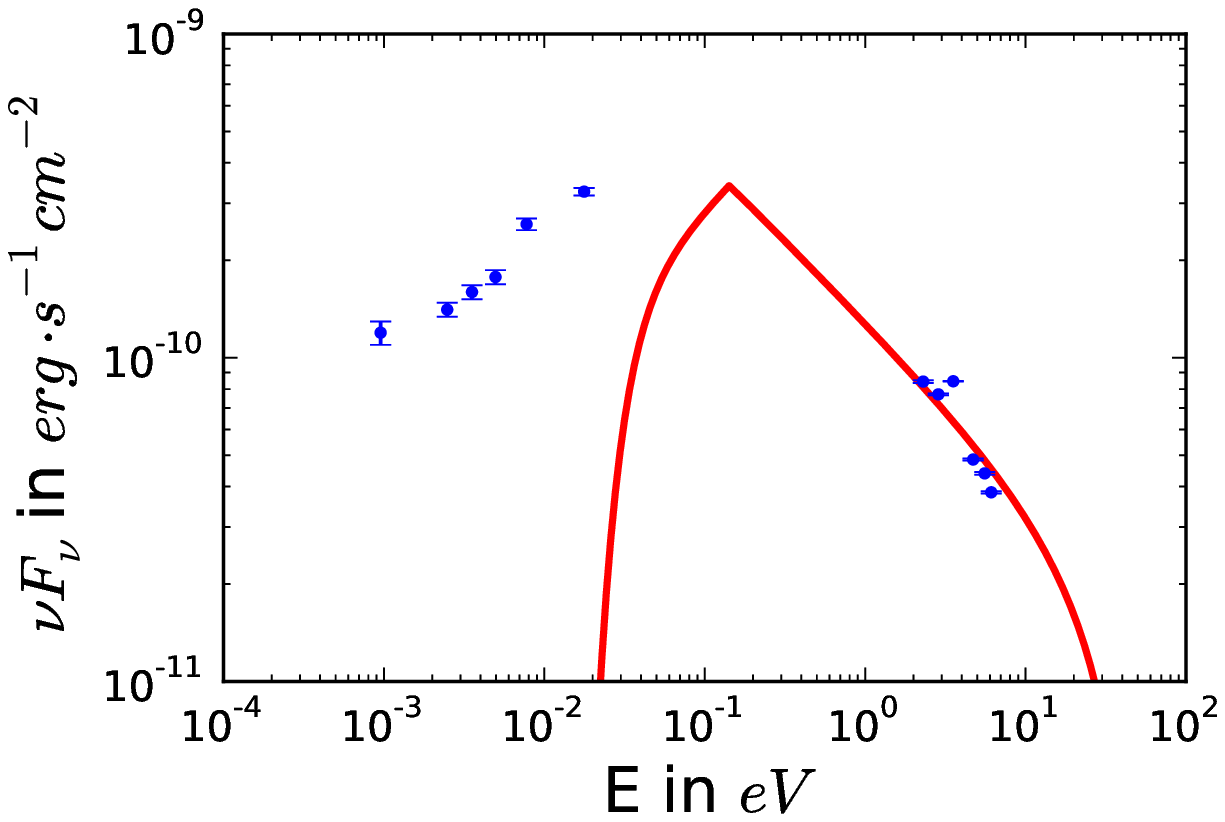}}\\

	\subfigure[Gamma-ray spectrum with Fermi-LAT data taken during the November 2010 flare (\citealt{2011ApJ...733L..26A}).]{\includegraphics[width=0.49\textwidth]{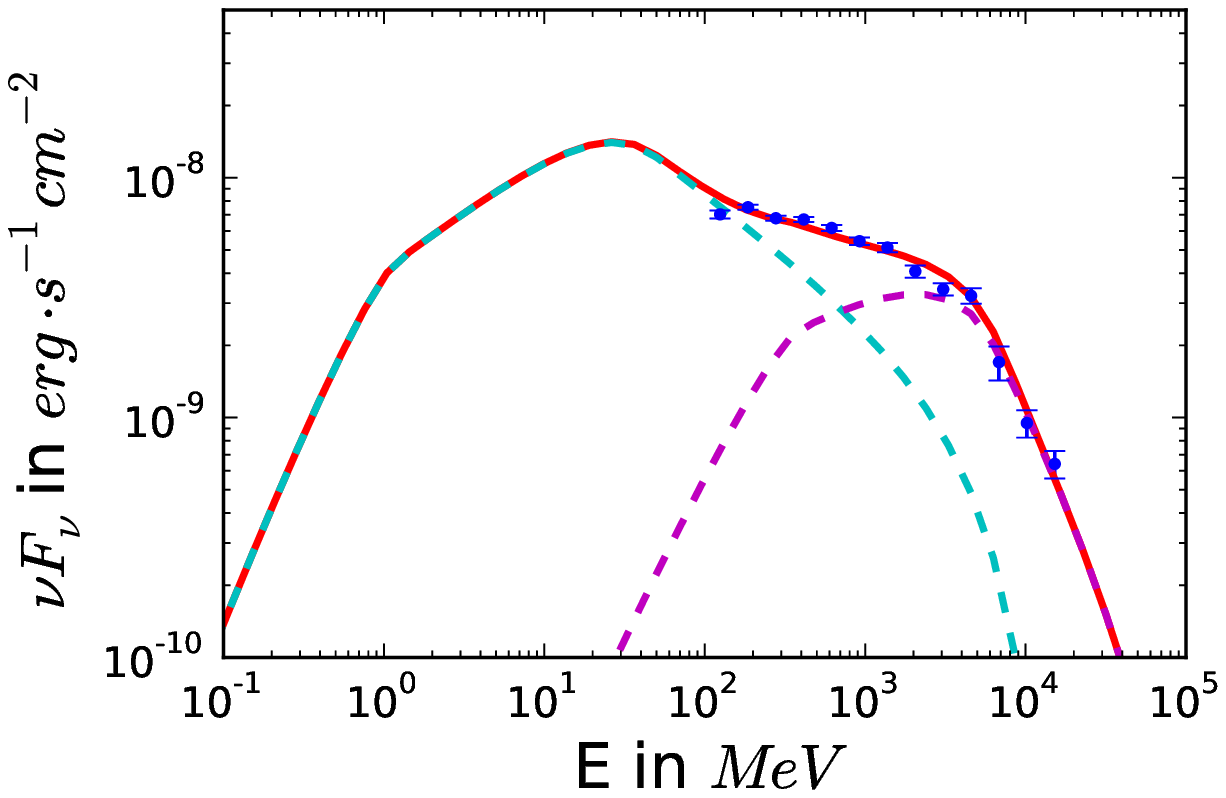}}\hfill
	\subfigure[Zooming into the $\gamma$-ray spectrum with Fermi-LAT data taken during the November 2010 flare (\citealt{2011ApJ...733L..26A}).]{\includegraphics[width=0.49\textwidth]{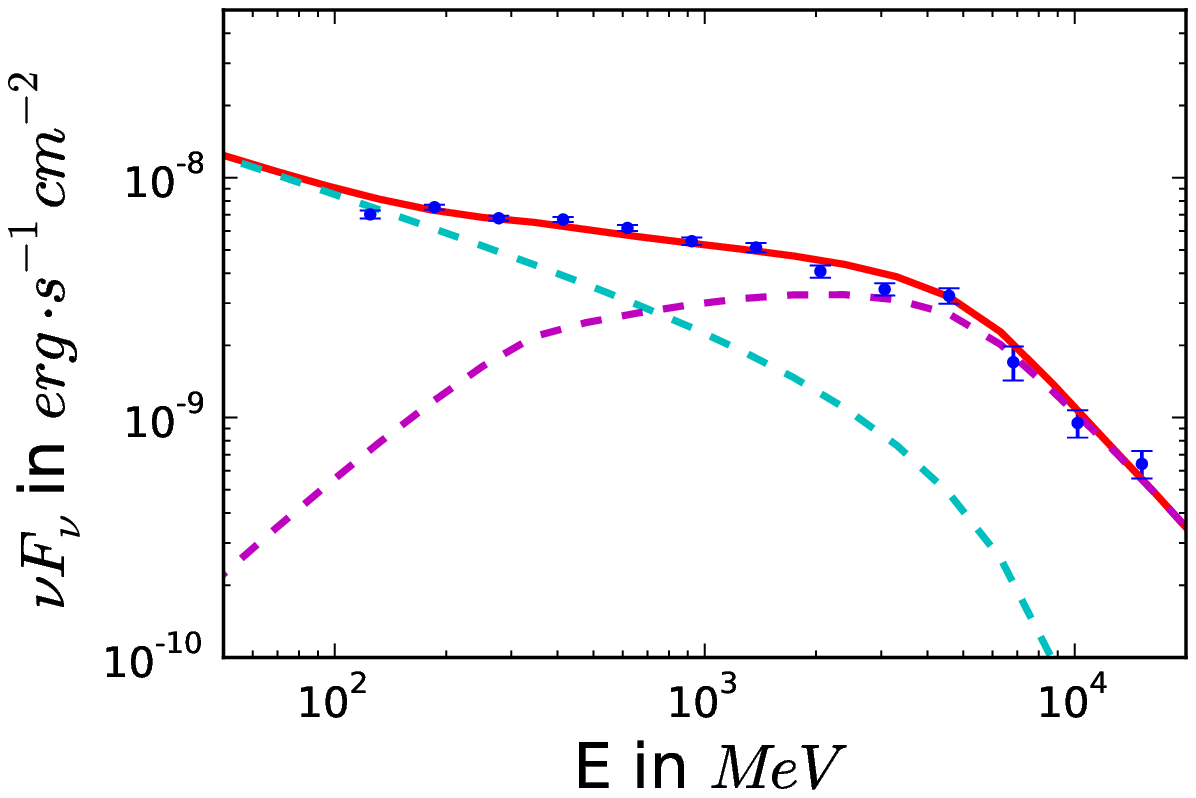}}\\
	\caption{Two-component IC model fit to multifrequency data of 3C~454.3. The model is shown in red with the data points shown in blue. 
The model parameter values are listed in \textbf{Table~\ref{tab:fit_param}}.} 
\label{fig:3C454.3_fit}
\end{figure*}

\begin{figure*} %[!h]
        \centering
        \subfigure[Emitting electron spectrum]{\includegraphics[width=0.49\textwidth]{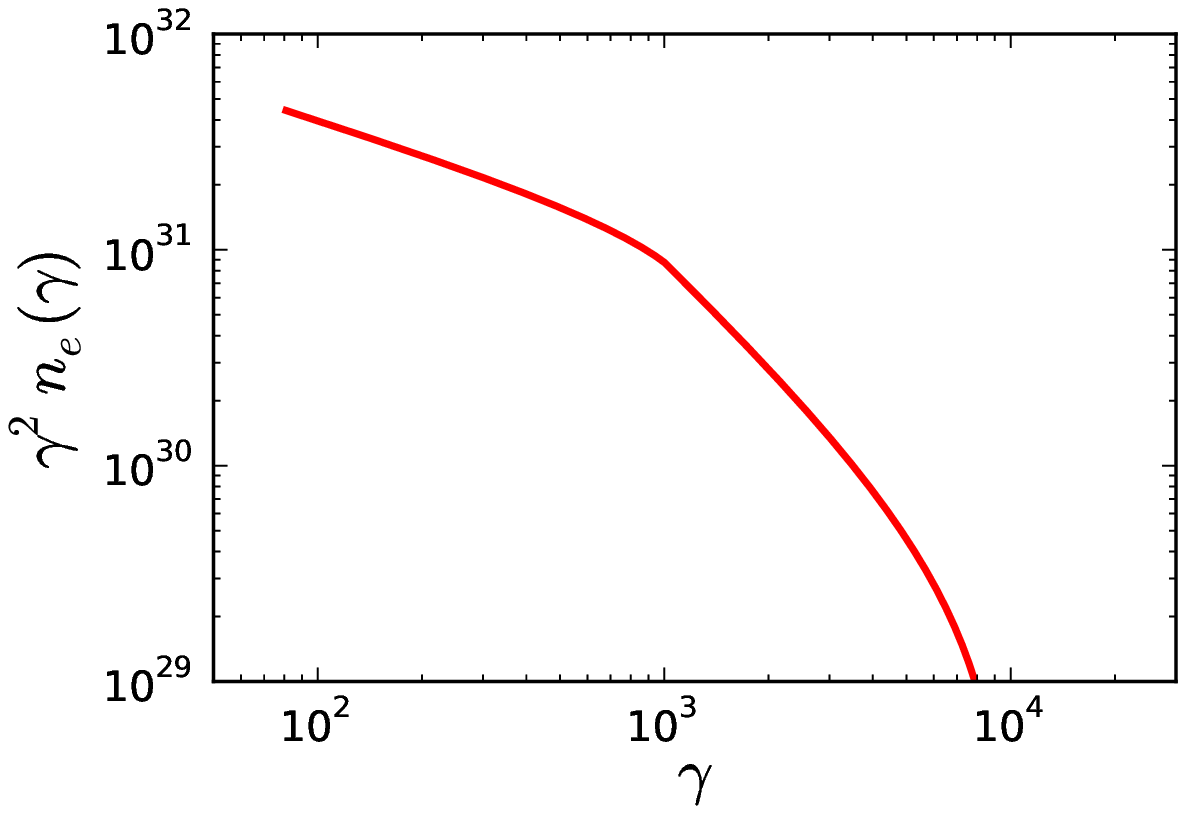}}\hfill
        \subfigure[Synchroton spectrum with \textit{Swift}, \textit{Herschel} and \textit{submillimeter array} data taken during the November 2010 flare (\citealt{2012ApJ...758...72W}).]{\includegraphics[width=0.49\textwidth]{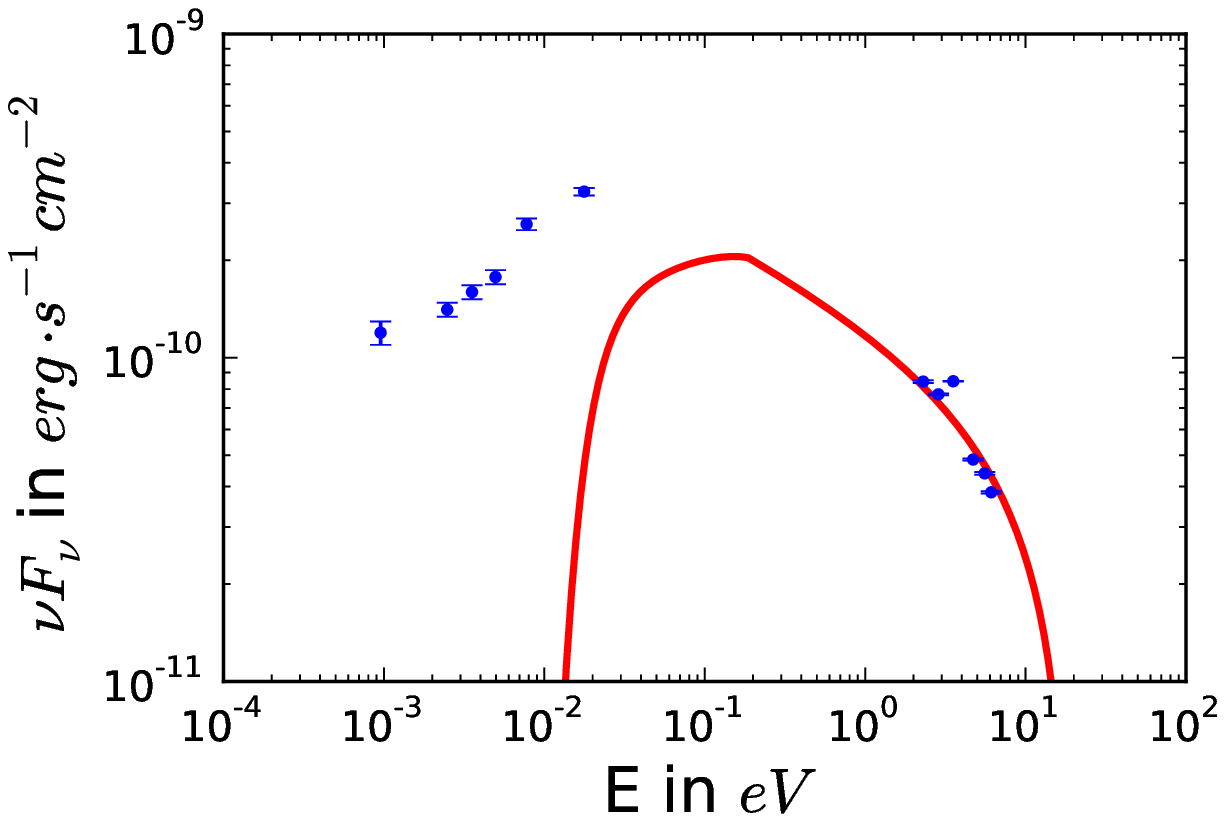}}\\
	\subfigure[Gamma-ray spectrum with Fermi-LAT data taken during the November 2010 flare (\citealt{2011ApJ...733L..26A}).]{\includegraphics[width=0.49\textwidth]{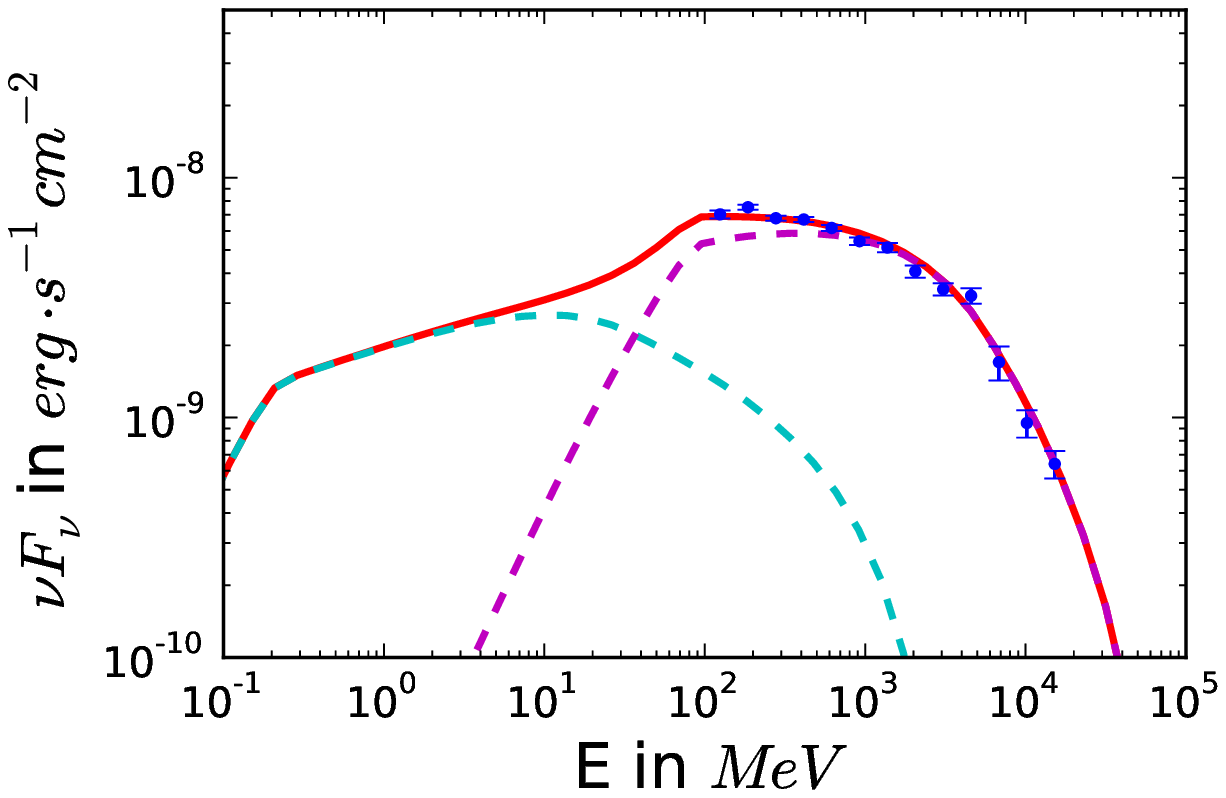}}\hfill
	\subfigure[Zooming into the $\gamma$-ray spectrum with Fermi-LAT data taken during the November 2010 flare (\citealt{2011ApJ...733L..26A}).]{\includegraphics[width=0.49\textwidth]{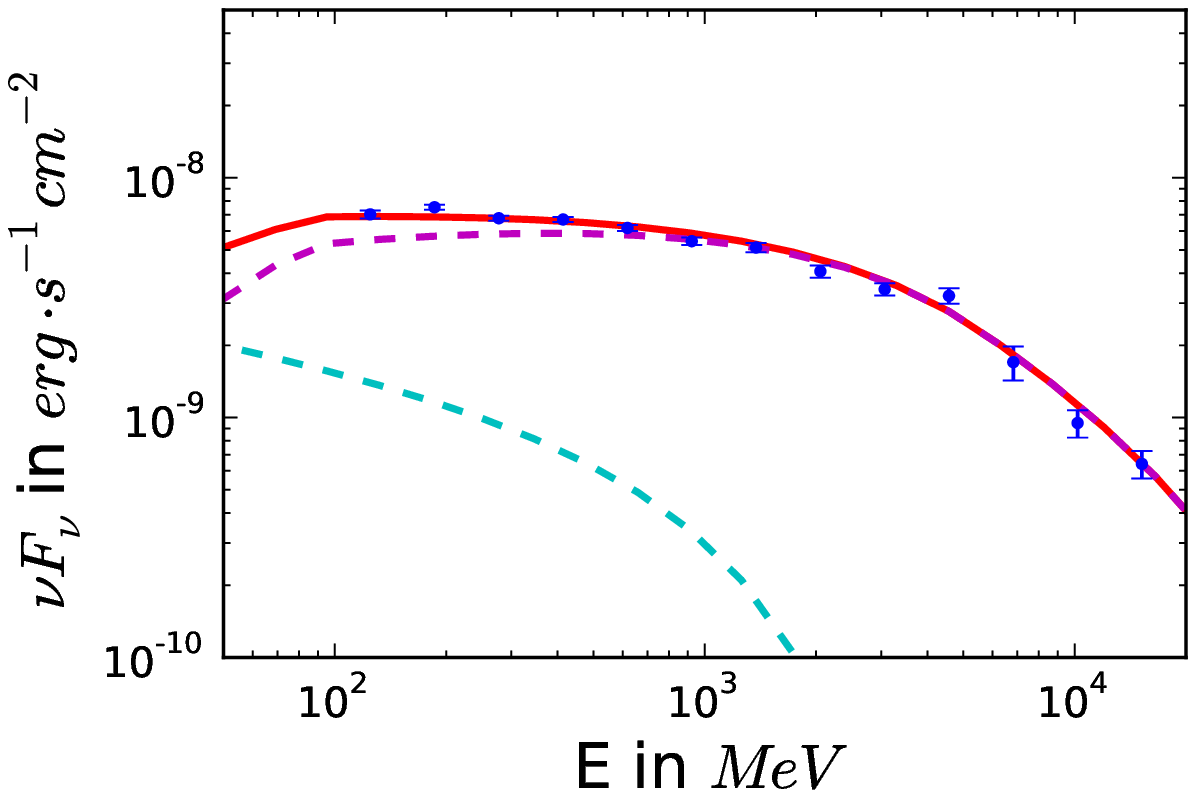}}\\
	\caption{Dominant-BLR IC model fit with multifrequency data of 3C~454.3. The model is shown in red with the data points shown in blue.
The model parameter values are listed in \textbf{Table~\ref{tab:fit_param}}. } 
\label{fig:3C454.3_fit_BLR}
\end{figure*}

In the following, we discuss two possible model fits to the data. The parameter values used for these fits are listed in \textbf{Table~\ref{tab:fit_param}}, including the required particle power $P_{\rm par}$ of the complete injection event, which was calculated by assuming that the total power in particles (electrons and cold protons) is tenfold the power in nonthermal electrons, the magnetic field power $P_B$, and the equipartition field $B_{\rm eq}$.

In \textbf{Fig.~\ref{fig:3C454.3_fit}} we present a model representation of the broadband data with parameter values similar to those used in \citet{2010ApJ...714L.303F}.
The size of the BLR used in our model agrees well with \citealt{Bonnoli01012011}, but is different from the one used in \citet{2010ApJ...714L.303F}.
Other slight differences in the fit parameter values are due to the different data set we used. 
We note that the (observer frame) sizes of the emitting region, which are used in the model fits we present, are roughly around the size limits inferred by \citet{2011AdSpR..48..998S}.
Here, we searched for particle injection parameter values that lead, in the framework of our model, to an ambient snapshot electron spectrum similar to the ad hoc ambient electron spectrum used in the aforementioned paper. 
We find that continuous electron injection is indeed able to generate a broken power-law emitting electron spectrum that is similar to the one assumed in \citet{2010ApJ...714L.303F}. 
For this model fit, the $\gamma$-ray photon spectrum is explained by the combination of a BLR and an accretion disk target photon component in the same way as in \citet{2010ApJ...714L.303F}.
We found reasonable fits for a range of particle injection parameter values. The injection spectral index $s$ of the electron spectrum is required to lie within the range of $2$ and $3$. 
The model fit is somewhat insensitive to changes in $\alpha$, the injection rate index, providing $\alpha \leq 4$. 
For $\alpha > 4$, the ambient electron spectrum would become too soft to be able to fit the low energy part ($<1$~eV) of the optically thin synchrotron radiation component. 
The parameter $\gamma_1$ controls the position of the break in the emitting electron spectrum. Hence the model fit is quite sensitive to changes in this parameter. 
We found reasonable fits with $\gamma_1$ between $800$ and $1300$. 
Furthermore, sensible fits were found when $z=z_b$ was between $1500$ and $3000$. For $\gamma_2 > 20000$ the post-break ($>3$ GeV) spectral index does not adequately represent the data. 
For $\gamma_2<10000$ the GeV-spectrum becomes two-peak shaped, which does not fit the data well. The first injection point 
$z_a$ is not well constrained here by the data since it merely influences the pre-break photon spectrum. Reasonable fits can be achieved for values between $z_a=100$ and $800$. 
%The observed synchroton spectrum further constrains the injection index $\alpha$ to values smaller than 4.5.   

In the previous section, we demonstrated that broken power-law ambient electron spectra with changes in the spectral index, whihc are greater than the cooling break, can develop as a result of the continuously varying injection of non-thermal particles along the jet. We apply this property to the broadband data of 3C~454.3 to demonstrate that the observed decline of the GeV-spectrum in bright FSRQs can be consistently modeled by this kind of scenario. Unlike in the previous fit, where two IC components are required, this scenario requires only one, but dominating IC component. In \textbf{Fig.~\ref{fig:3C454.3_fit_BLR}}, we show a model fit to the data with a higher optical thickness ($\tau_{\rm BLR}=0.05$) of the BLR. As a consequence, the BLR IC component dominates at $\gamma$-ray energies $>10$~MeV. The GeV spectral turn-over here is due to a corresponding break in the emitting electron spectrum, i.e., not the result of a combination of the accretion disk and BLR IC component. This different behavior also leads to 
differences with respect to the sensitivity of the model parameter values. The parameter value of $z_a$ here is more constrained, since the pre-break emission of the BLR IC component dominates in the sub-GeV energy range (around 100 MeV). We require $z_a<200$ to adequately fit the data. Similarly, changes in the last injection point $z_b$ also impact the shape of the resulting ambient spectrum in the pre- as well as post-break regime. Choosing $z_b$ between $2000$ and $3000$ leads to reasonable model fits. The injection rate index $\alpha$ influences the pre-break spectrum, and is therefore rather tightly constrained, as well, to values between $\alpha=4$ and $4.5$, with electron injection spectral indices lying in the range $s=1.5-2.5$. This choice also leads to a satisfying model representation of the observed synchrotron spectrum. The parameter values for $\gamma_1$ and $\gamma_2$ are similarly constrained, as in the previous model fit.

In summary, both model fits to the flare 3C~454.3 data set point towards
gradually decreasing injection rates that start close to the inner
boundary of the BLR but do not extend beyond the outer boundary of the BLR at the time of observations.
\citet{2014ApJ...789..161N} explored the efficiency of several constraints
for the location of the blazar zone, such as the collimation, SSC, cooling and internal $\gamma$-ray opacity constraints.
Although, in our modeling, we adopted a somewhat weaker accretion disk and BLR luminosity
than \citet{2014ApJ...789..161N} for the November 2010 flare of 3C~454.3,
we find that the blazar zone location in our model fits is in agreement with \citet{2014ApJ...789..161N}.

The injection spectrum is found to be surprisingly narrow, between
$\gamma \sim 10^3-10^4$. The
Thomson optical thickness $\tau_{\rm BLR}$ of the diffuse BLR gas
influences the relative importance of the BLR and accretion disk IC
components. Accordingly, for
a very low $\tau_{\rm BLR}$, the spectral decline observed at a few GeV from
3C~454.3 can be explained by the combination of both IC components while,
for sufficiently
larger $\tau_{\rm BLR}$, the GeV-break in the photon spectrum is due to an
intrinsic break in the emitting electron spectrum where the injection
properties play a decisive
role. The break energy is determined by the low-energy cut-off $\gamma_1$ of
the injected electron spectrum.
The observed, approximately constant, break energies that are independent of flux state
(e.g., \citealt{2011ApJ...733L..26A}) imply a correspondingly constant $\gamma_1$
for a constant Doppler factor, if there is no rapid change from a continuous to
impulsive-like injection mode (or vice versa).
The energy density of the magnetic field for these model fits is found below the jet-frame disk energy density up to distances $\sim 2000-3000R_g$ above the black hole, and therefore justifies the neglect of synchrotron losses up to these distances. %The last particle injection point in our model fits lies slightly beyond this point. Hence, the expected source-frame VHE spectrum of 3C~454.3 from our modeling has to be taken with caution.

Potentially, $\gamma$-ray absorption, followed by pair production
in the BLR radiation field, may impact the emerging photon spectrum
(e.g., \citealt{2003APh....18..377D}). 
Here, not only absolute luminosities of the BLR are important, but also its geometrical structure. For example, for a flat BLR geometry in blazars with typical BLR parameters, \citet{2014PASJ...66...92L} demonstrate that high energy photons up to tens of GeV can escape this region.

We therefore investigate the importance of this process for the above
parameter values.
For a smooth broadband target spectrum, most of the photon-photon interactions take place in a small energy 
interval that is centered on $\epsilon_{\rm target}\approx 3.1 \epsilon_s^{*-1}$  (e.g., \citealt{2007ApJ...665.1023R}) in the galaxy frame
owing to the strongly peaked cross-section near the threshold.
Gamma-gamma absorption in the BLR radiation field therefore modifies the observed $\gamma$-ray spectrum beyond $\epsilon\approx 3.1 (\epsilon_{0*}(1+z_r))^{-1}$,
which is above the highest photon energy detected from this source for our parameter choice of $\epsilon_{0*}=10(m_e c^2)^{-1}$eV.

During the April and September 2013 outbursts of this source, \citet{2014ApJ...790...45P} noted a very different flaring behavior: hard LAT-flares with exceptional hard power-law spectra, as opposed to the soft flaring events from 3C~454.3 that are considered in this work, which typically show strong declines or even cut-offs in the GeV range. From $\gamma\gamma$-pair production arguments and the non-detection of any KN-curvature during these hard flare events, \citet{2014ApJ...790...45P} argue that the emission region must be beyond or at the outer rim of the BLR, where the bulk of the gamma-rays are produced by Comptonization of photons from the IR torus.

\begin{table}[!htb]

 \centering
  \begin{tabular}{p{0.14\textwidth}|p{0.14\textwidth}|p{0.14\textwidth}}
  \hline
  General Parameters &  Two-component model & BLR-dominant model \\ \hline
  Redshift  & 0.859	& 0.859 \\
  $\Gamma_{\rm bulk}$	 &20	& 20\\
  $D$& 30	& 30	\\
  $R_b$& $ 10^{16}$cm	&  $7\cdot 10^{15}$cm	\\
  $B$& 0.6 G	& 1.0 G	\\   \hline \hline	
  Electron injection &       \\ \hline
  $\gamma_1$ & $1125$	 & $1000$	\\
  $\gamma_2$ & $20000$	 & $10000$	 \\
  $z_a$ & $350 R_g$	& $200 R_g$\\
  $s$ & $3$	& $2$\\
  $\alpha$ & $3$	& $4$\\
  $z$ & $3000 R_g$ & $3000 R_g$	\\
  $z_b$ & $3000 R_g$	& $3000 R_g$\\ \hline \hline
  Accretion and black hole & \\ \hline
  $M_8$ & 20 & 20\\
  $l_{\rm edd}$ & 0.04 & 0.03 \\
  $\epsilon_f$ & 1/12 & 1/12\\ \hline \hline
  BLR parameters & \\	\hline
  $\tau_{\rm BLR}$ & 0.0084 & 0.05\\
  $R_i$ & $50 R_g$ & $50 R_g$\\
  $R_o$ & $10000 R_g$ & $10000 R_g$  \\	
  $\zeta$ & -2 & -2 \\ \hline \hline
  Energetics & \\	\hline
  $B/B_{eq}$ & 0.31 & 0.06 \\
  $P_{\rm par}$ & $2.9 \cdot 10^{47}$ erg/s & $4.6 \cdot 10^{48}$ erg/s \\ 
  $P_{B}$ & $2.8 \cdot 10^{46}$ erg/s & $1.9 \cdot 10^{46}$ erg/s\\ \hline \hline
  \end{tabular}

\caption{Model parameters fitting the November 2010 flare state SED of 3C~454.3.}
  \label{tab:fit_param}
\end{table}

\subsection{Modeling the PKS~1510-089 multifrequency SED}

The FSRQ PKS~1510-089 at redshift $\sim 0.361$ is among the best observed $\gamma$-ray blazars with a well-constrained SED from radio to $\gamma$ rays.  We apply our model to
the quasi-simultanuous broadband data from the January 2009 flare, as published in \citet{2010ApJ...716...30A}, the first flare in a series of outbursts from this source in early 2009. These observations indicate variability on timescales of a day down to a few hours (e.g., \citealt{2010ApJ...721.1425A}). Though PKS~1510-089 is a known VHE FSRQ \citep{2010HEAD...11.0705W}, it has not been detected at VHEs during the considered outburst.
Apart from $\gamma$-ray observations, this source has also been monitored during the 2009 activity at radio, optical (including polarization measurements) and X-rays \citep{2010ApJ...710L.126M}. The appearance of two new knots on VLBA 43 GHz radio maps around August 2008 and April 2009, which move with superluminal speeds of $\sim 24c$ and $\sim 22c$ along the jet, seem to indicate a relation between these flaring events and their appearance \citep{2010ApJ...710L.126M}. The optical polarization vector started rotating by $\sim 720^o$ just after the flare that we consider here.
The {\it{Fermi}}-LAT spectrum, during the January 2009 flare, indicates a pronounced steep decline at a few GeV, which has been observationally described by a log-parabolic shaped spectrum \citep{2010ApJ...721.1425A}.

We present two sets of parameter values (see \textbf{Table~\ref{tab:fit_param_PKS1510}}) that have been found to represent the data from the January 2009 flare well. Here again, the listed
particle power $P_{par}$ (including electrons and cold protons) is given as tenfold the total injected power of the nonthermal electrons. 
To quantify the black hole mass and accretion disk luminosity, we use $M_8=5$ and $l_{edd}=0.16$, which agrees well with data presented in \citet{2008A&A...491L..21P}.

\begin{figure*} %[!h]
        \centering
        \subfigure[ Emitting electron spectrum. ]{\includegraphics[width=0.49\textwidth]{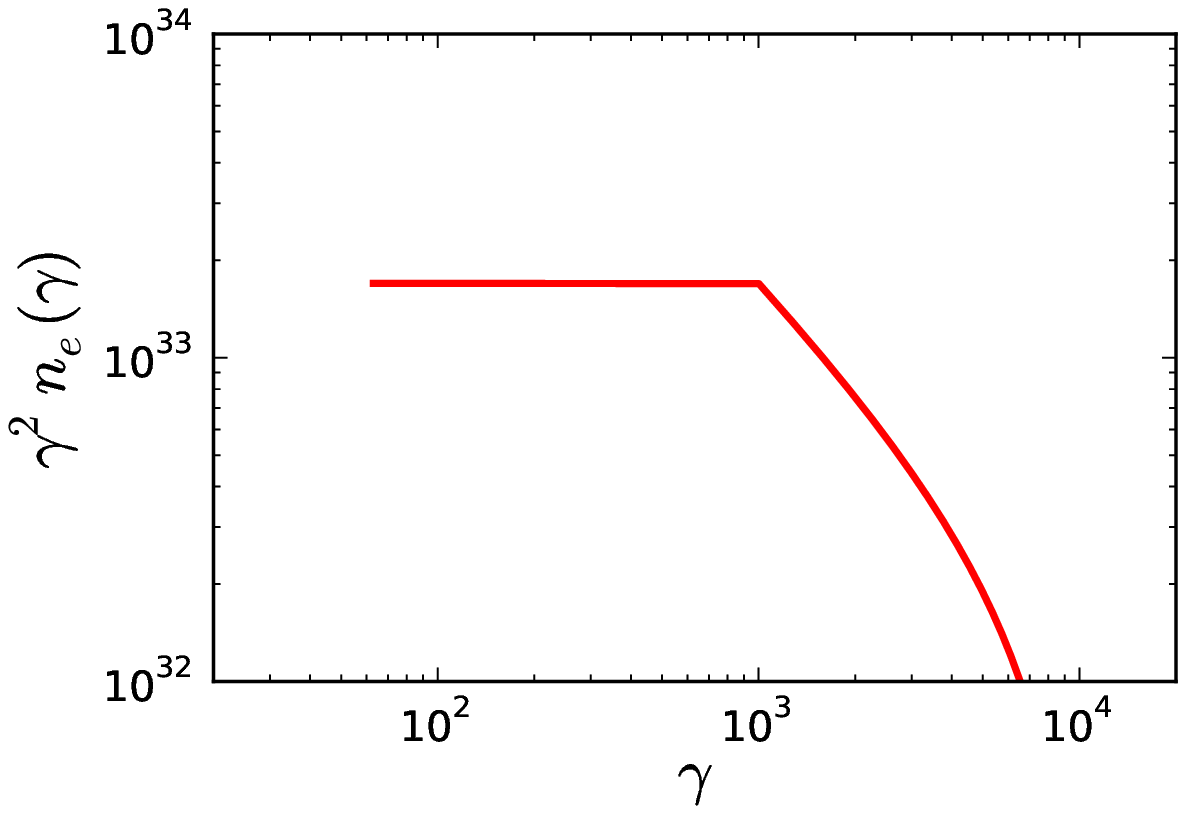}}\hfill
        \subfigure[ Synchroton spectrum fit to data from \citealt{2010ApJ...716...30A}.]{\includegraphics[width=0.49\textwidth]{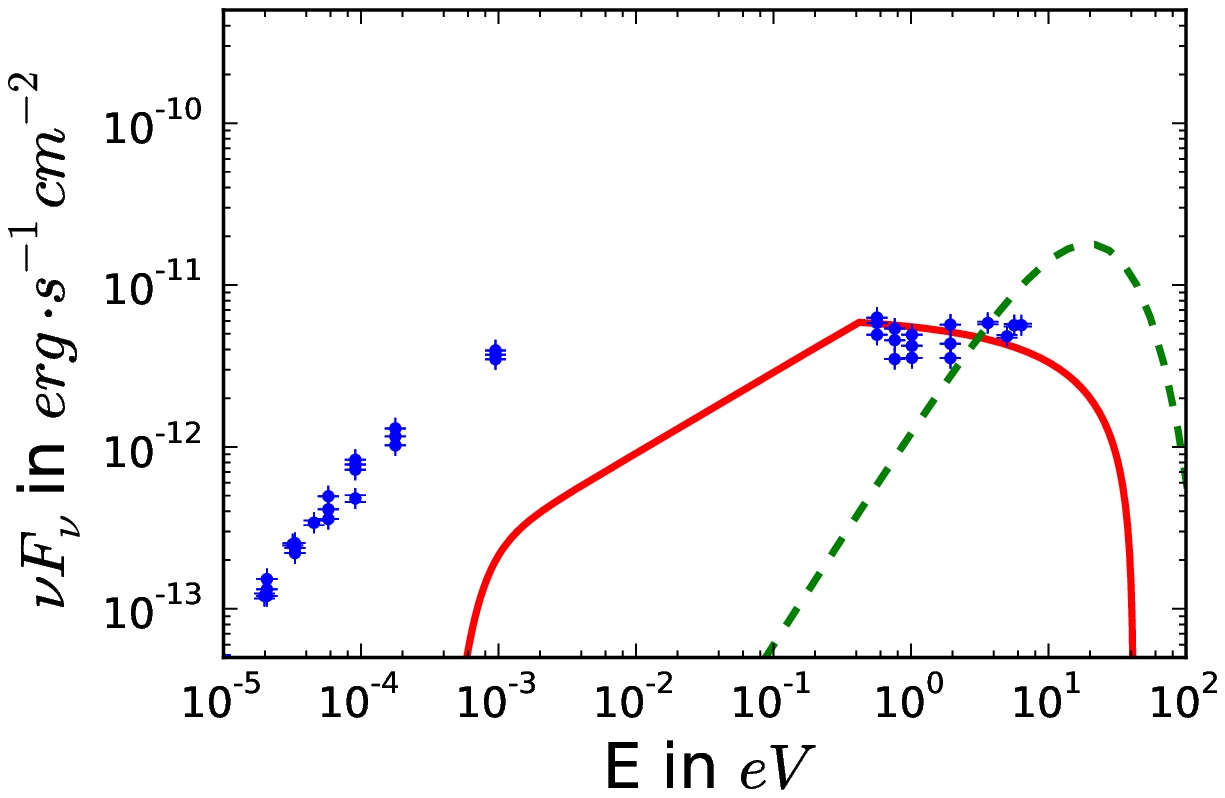}}\\

	\subfigure[  Gamma-ray spectrum with Fermi-LAT data \citealt{2010ApJ...716...30A}.]{\includegraphics[width=0.49\textwidth]{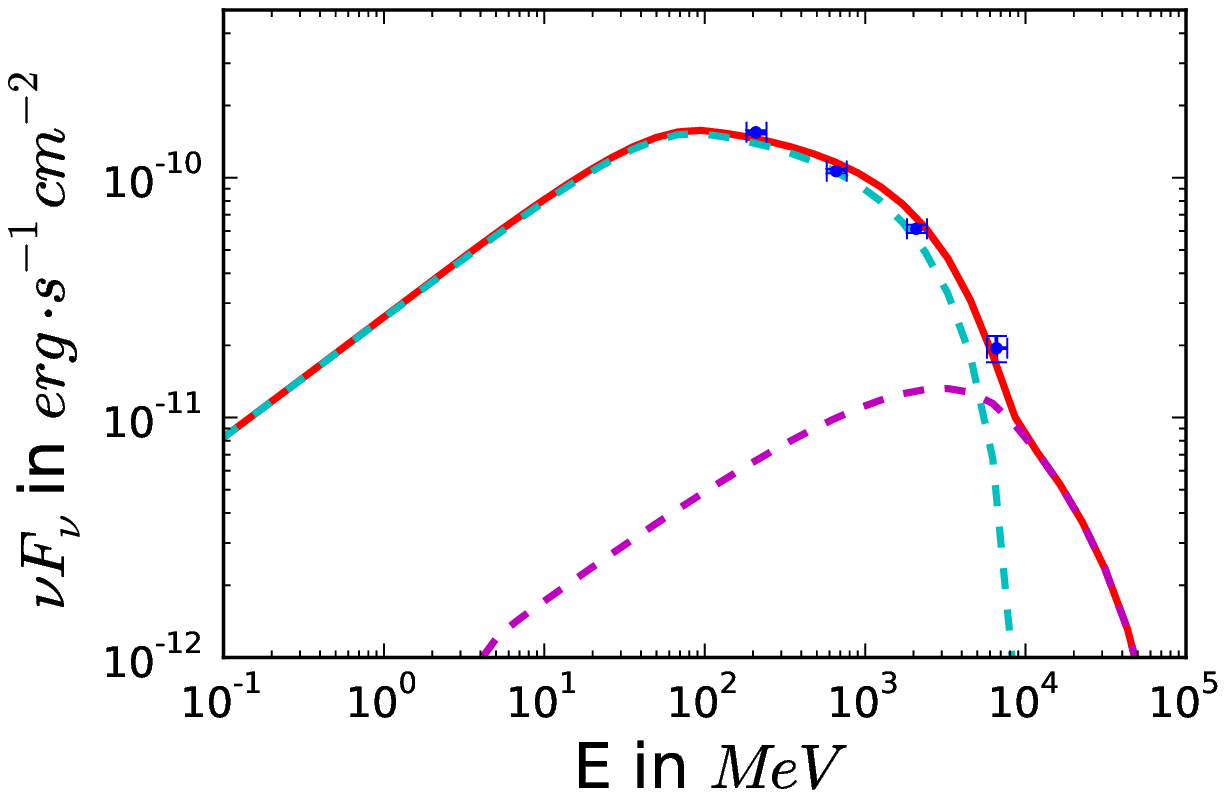}}\hfill
	\subfigure[  Zooming into the $\gamma$-ray spectrum with Fermi-LAT data \citealt{2010ApJ...716...30A}.]{\includegraphics[width=0.49\textwidth]{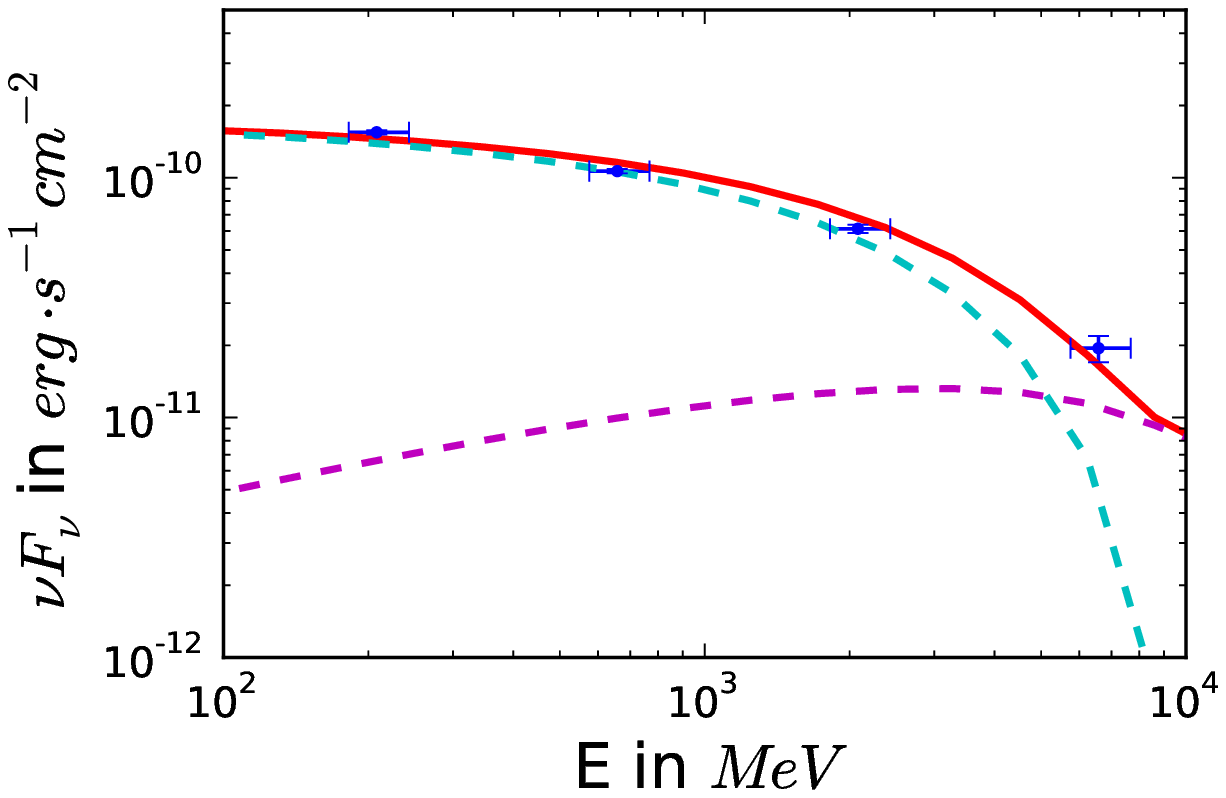}}\\
	\caption{ Two-component IC model fit to multifrequency data of PKS~1510-089. The model is shown in red with the data points shown in blue. 
The model parameter values are listed in \textbf{Table~\ref{tab:fit_param_PKS1510}}.} 
\label{fig:PKS1510_fit}
\end{figure*}

In \textbf{Fig.~\ref{fig:PKS1510_fit}}, we show a model fit where the joint Compton-scattered disk and BLR radiation explains the observed LAT-spectrum.
In this fit, we keep the BLR geometry unchanged with respect to the 3C~454.3 fits, and use a (observer-frame) size of the emitting region of $R_b=1 \cdot 10^{15}$cm, in agreement with \citet{2010ApJ...721.1425A} and the inferred limits of \citet{2011AdSpR..48..998S}.
We now discuss the robustness of the parameter values used for our model fit. 
As in the fit to 3C~454.3, the model is quite sensitive to changes in $\gamma_1$ since it controls the position of the breaks in the electron spectrum 
and, in turn, also controls the position of the respective breaks in the photon spectra. The injection spectral index $s$ for this fit is required to lie between 1.5 and 2.5 for a reasonable fit. 
The model fit is again rather insensitive to changes in the injection parameter $\alpha$ since changes in this parameter mostly influence the spectral index before the break in the respective photon spectra. 
We get the best fit for $z=z_b=3000R_g$, but values from $2000R_g$ to $4000R_g$ also lead to reasonable fits. 
For $\gamma_2$, we can achieve reasonable fits up to a value of $2\cdot 10^4$, otherwise the model does not describe the high energy part ($>1$ GeV) of the data adequately.
As in the 3C~454.3 case, the first injection point $z_a$ is not well constrained. Values between $z_a=200R_g$ and $z_a=800R_g$ lead to reasonable fits.
We note that the parameter values for this fit are quite similar to the ones used for the 3C~454.3 data set. 

\begin{figure*} %[!h]
        \centering
        \subfigure[Emitting electron spectrum.]{\includegraphics[width=0.49\textwidth]{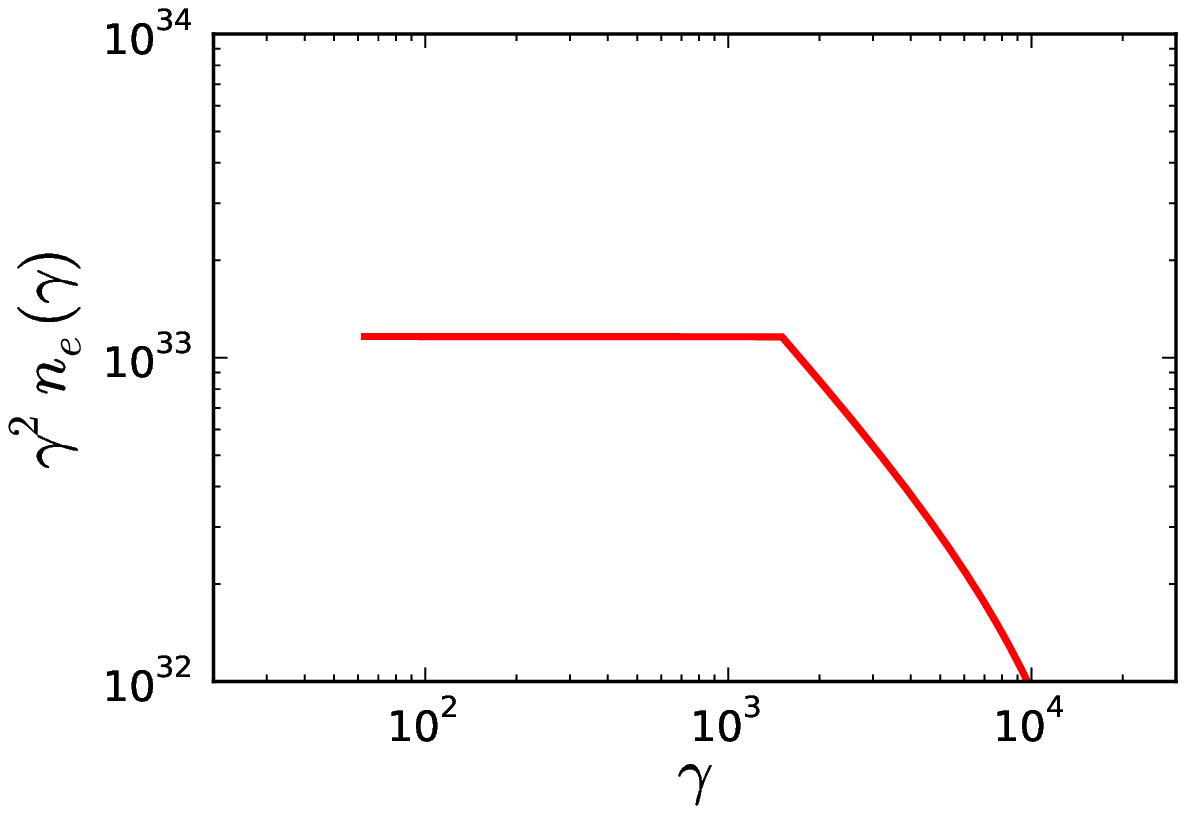}}\hfill
        \subfigure[ Synchroton spectrum fit to data from \citealt{2010ApJ...716...30A}.]{\includegraphics[width=0.49\textwidth]{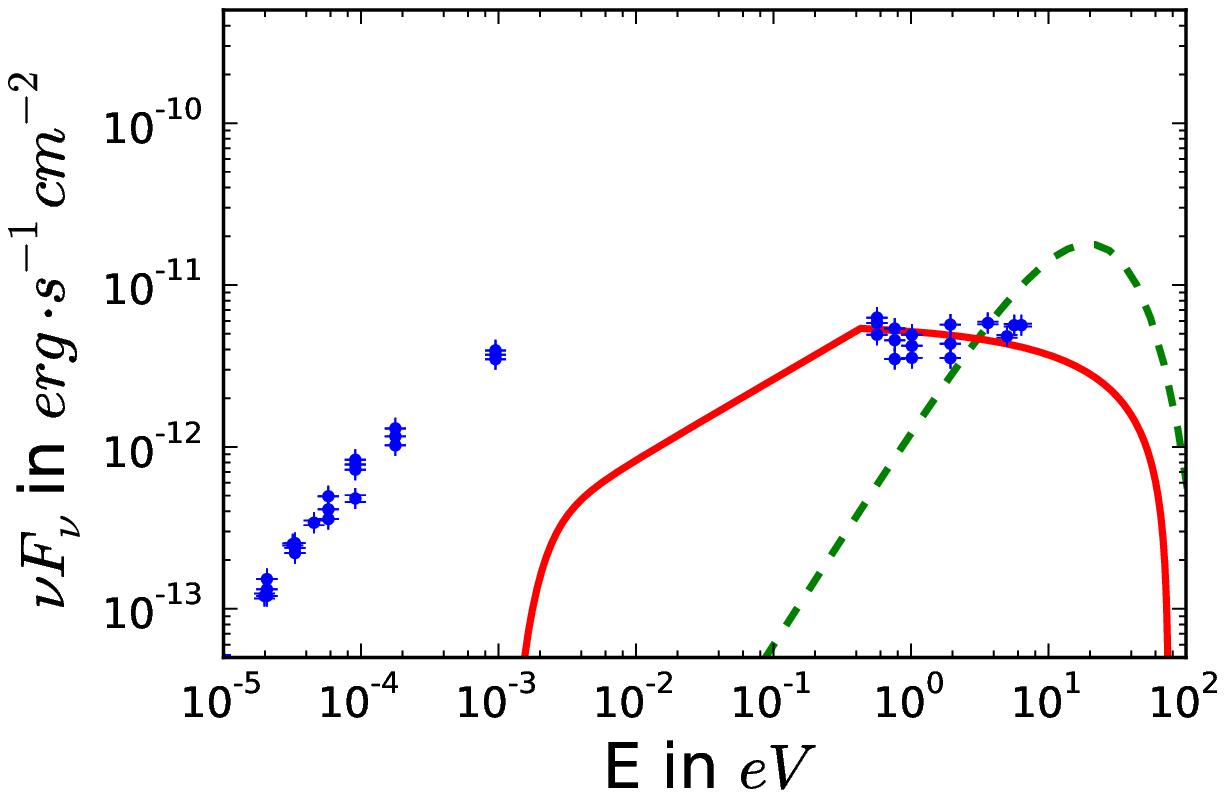}}\\
	\subfigure[ Gamma-ray spectrum with Fermi-LAT data \citealt{2010ApJ...716...30A}.]{\includegraphics[width=0.49\textwidth]{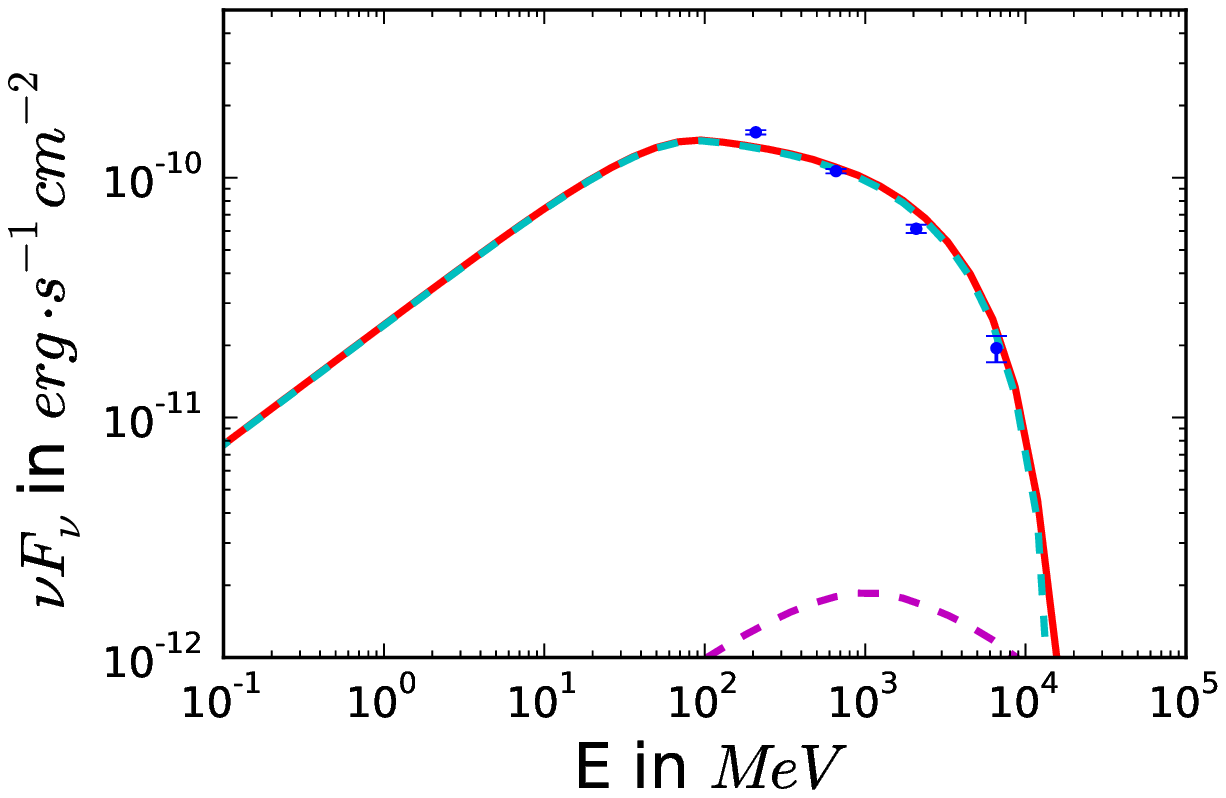}}\hfill
	\subfigure[ Zooming into the $\gamma$-ray spectrum with Fermi-LAT data \citealt{2010ApJ...716...30A}.]{\includegraphics[width=0.49\textwidth]{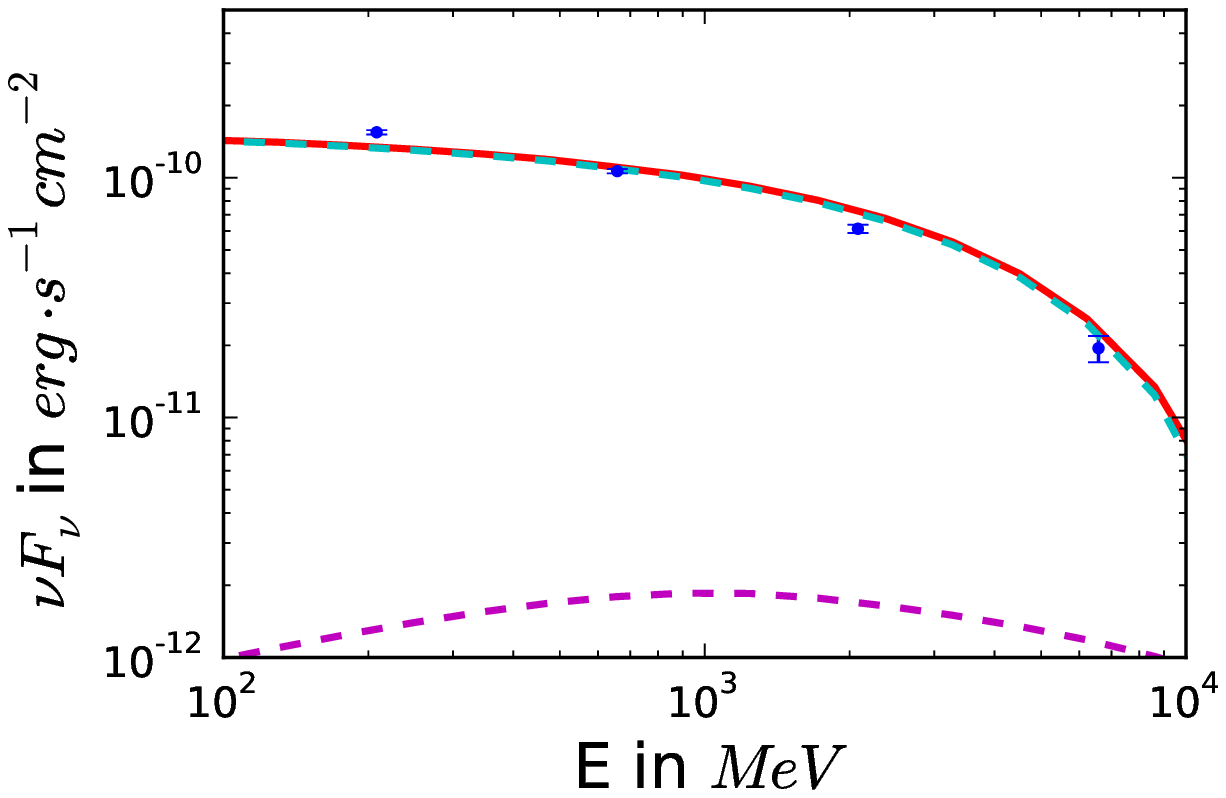}}\\
	\caption{ Dominant accretion-disk IC model fit with multifrequency data of PKS~1510-089. The model is shown in red with the data points shown in blue.
The model parameter values are listed in \textbf{Table~\ref{tab:fit_param_PKS1510}}. } 
\label{fig:PKS1510_fit_Accr}
\end{figure*}

In \textbf{Fig.~\ref{fig:PKS1510_fit_Accr}} we present a model that is able to explain the $\gamma$-ray data with Compton-scattered accretion disk radiation only.
Here, the break in the GeV-spectrum is generated by the break in the corresponding emitting electron spectrum. 
By reducing the geometric thickness of the BLR and keeping a realistic BLR optical depth, the importance of the BLR as a target photon field is diminished
for the emitting region, since it is located at snapshot time just beyond the BLR border. Particle injection started when the emission region was inside the BLR and stopped just beyond the border of the BLR. We found that the accretion disk radiation dominated the target photon field energy density in this case.
Other than the change of the BLR geometry, the fit parameter values are similar to the ones fitting the two-component model with only slight changes in $\gamma_1$ and the bulk Doppler factor. Our fit is found to be sensitive again to changes in $\gamma_1$ since it controls the position of the break. The injected electron spectrum is as narrow as in the 3C~454.3 fit, extending from $\gamma \sim 10^3-10^4$. For the other parameters, we reached reasonable fits at roughly the same parameter ranges that we used for the two-component fit.

%(-> IS PUT TO SUMMARY SECTION.....This fit shows that the geometry of the BLR is as important as the optical depth for the importance of the BLR IC component.) 

 Both model fits to the PKS~1510-089 data require a continuous injection scenario with gradually decreasing injection rates. Thus, impulsive injection scenarios (i.e., $\alpha>5$) are again disfavored.
In both of our model representations of the PKS~1510-089 data set, the accretion disk IC component is more prominent than in the 3C~454.3 fits.  
The energy density of the magnetic field for these model fits is found below the jet-frame disk energy density up to distances $\sim 2000-3000R_g$ above the black hole, and therefore justifies the neglect of synchrotron losses up to these distances.

\citet{2010ApJ...721.1425A} modelled this flare by also using a stationary leptonic blazar model, however, they were using parameter values that put more emphasis on the (assumed non-stratified) BLR and less on the accretion disk radiation field than our model parameter set values (and also the modeling of \citet{2013ApJ...768...54B} of this source). As a consequence the LAT-data are explained rather by dominantly Compton-scattered BLR-photons from lower energetic electrons than our model envisions. The corresponding SSC component is found to be negligible.

Interestingly, our spectral modeling of this flaring event being due to the Compton scattered accretion disk and BLR photons from a $\sim 1$G magnetized jet emission region, is in qualitative agreement with the scenario \citet{2010ApJ...710L.126M} proposed for this activity period.

\citet{2014ApJ...789..161N} place the emission region of the April 2009 flare of this source outside the BLR, while no constraints were
put on the January 2009 outburst. Our model fit parameters for this January 2009 flare point towards an emission region that is located inside the BLR at the time of this flare.
If the series of flares in 2009 of PKS~1510-089 were causally connected, possibly originating from the same so-called blob that was moving along the jet, then the location of this blob that caused the January 2009 flare would be well outside the BLR at the time of the April 2009 flare, assuming no strong blob-braking. This would be 
compatible with the results of \citet{2014ApJ...789..161N}.

We note that the {\it HERSCHEL} PACS \& SPIRE data from observations of this source in late 2011 (\citealt{2012ApJ...760...69N})
lie above our model curves. Hence, if this IR flux were comparable during the here considered 2009 event, these IR photons would have to be produced by a further region within the jet with possibly some notable contribution from the dusty torus.

Similar to 3C~454.3 (see Sect.~4.1), we find $\gamma\gamma$-pair production being negligible for the parameter set used for the above model fits and below the last LAT data point (\citealt{2013arXiv1307.4050B}).

\citet{2012MNRAS.424..789C}  modelled the last flare in the extended early 2009 flaring episode of this source by using a time-dependent Monte Carlo Fokker-Planck code, where the emission region is treated as an extended cylindrical volume within the jet. In addition to a particle ``pick-up'' term in the Fokker-Planck equation, they also included a stochastic acceleration term and took into account all external and internal target photon fields. Although, in their modeling, they did not discuss pronounced GeV spectral turn-over, but rather focused on spectral variability, they note the impact of not only internal but also external light-travel time effects causing complex spectral shapes and variations in this March 2009 flaring event. In our approach to explain complex high-energy spectral shapes, we found that the injection history already has a strong impact, and is able to reproduce the typically observed pronounced turn-overs at GeV energies.

\label{Fit_to_data_PKS}

\begin{table}[!htb]

 \centering
  \begin{tabular}{p{0.14\textwidth}|p{0.14\textwidth}|p{0.14\textwidth}}
  \hline
  General parameters &  Two-component model & Accretion-disk dominated model \\ \hline
  Redshift  & 0.361	& 0.361 \\
  $\Gamma_{\rm bulk}$	 &20	& 20\\
  $D$& 30	& 20	\\
  $R_b$& $ 10^{15}$cm	&  $ 10^{15}$cm	\\
  $B$& 1.6 G	& 1.1 G	\\   \hline \hline	
  Electron injection &       \\ \hline
  $\gamma_1$ & $1000 $	 & $1500$	\\
  $\gamma_2$ & $10000$	 & $20000$	 \\
  $z_a$ & $200 R_g$	& $200 R_g$\\
  $s$ & $2$	& $2$\\
  $\alpha$ & $3$	& $3$\\
  $z$ & $3000 R_g$ & $3000 R_g$	\\
  $z_b$ & $3000 R_g$	& $3000 R_g$\\ \hline \hline
  Accretion and black hole & \\ \hline
  $M_8$ & 5 & 5\\
  $l_{\rm edd}$ & 0.16 & 0.16 \\
  $\epsilon_f$ & 1/12 & 1/12\\ \hline \hline
  BLR parameters & \\	\hline
  $\tau_{\rm BLR}$ & 0.003 & 0.003\\
  $R_i$ & $50R_g$ & $100R_g$\\
  $R_o$ & $10000 R_g$ & $1750 R_g$  \\	
  $\zeta$ & -2 & -2 \\ \hline \hline
  Energetics & \\	\hline
  $B/B_{eq}$ & 1.21 & 0.17 \\
  $P_{\rm par}$ & $1.3 \cdot 10^{45}$ erg/s & $1.4 \cdot 10^{46}$ erg/s \\ 
  $P_{B}$ & $1.9 \cdot 10^{45}$ erg/s & $3.9 \cdot 10^{44}$ erg/s\\ \hline \hline
  \end{tabular}

\caption{Model parameters fitting the SED of PKS~1510-089.}
  \label{tab:fit_param_PKS1510}
\end{table}

\section{Summary and conclusion}
\label{sec:Discussion}

Within the framework of a leptonic emission model for blazars, where
inverse Compton scattering on external radiation fields dominates, we
considered the case of continuous particle injection of a given simple power
law along the jet with various injection rates, durations, and locations of the
emission region within the jet.
We calculated the emitting particle spectrum that results from this kind of setup
at a snapshot time $t_{\rm obs}$, taking into account radiative cooling
(in the Thomson regime) in the accretion disk and BLR radiation field during injection
self-consistently. This prescription offers the possibility to explain a
broad range of ambient electron spectral shapes, including spectral breaks that are
significantly larger than simple cooling breaks. The injection properties may provide valuable
information on the acceleration mode at work. Hence, our simple
phenomenological prescription has the potential to find interesting clues in this aspect
by rigorously analyzing the ambient electron spectrum that is required for the
broadband fit of a given quasi-simultaneous multifrequency data set,
in conjunction with constraints that are inferred from correlated variability
studies and $\gamma$-ray opacity arguments.

For example, the peak of the emitting electron spectrum can, in general,
be linked to the minimum injection energy $\gamma_1 m_e c^2$ if the
injection rate distribution is not too steep. For impulsive-like injections (i.e.,
large $\alpha$ within our model) the peak in the ambient particle
spectrum is, instead, associated with the smallest cooled electron energy.
The peak in the electron spectrum results in corresponding spectral turn-overs in the 
Compton-scattered photon components. 
Thus $\gamma_1$ is a parameter that may influence the $\gamma$-ray break energy. 

Our model has the potential to describe broad high-energy components, as well as double-peaked ones (with one peak corresponding to
Compton-scattered accretion disk photons, and the second one from IC-scattered BLR radiation field).  
The former case arises for not too steep injection spectra, while very steep ones, or injections where $\gamma_2$ is suitably small, can give SEDs with double-peaked $\gamma$-ray spectra. In general the injection spectral index $s$, together with $\gamma_2$ and the time since last injection (given the accretion disk luminosity), mostly determine the post-break shape of the spectrum. For example, fitting the flare state SEDs of 3C~454.3 and PKS~1510-089 required rather conservative injection indices ($s=2-3$), as expected from, for example, Fermi acceleration.

The injection rate index $\alpha$ strongly influences the ambient electron spectrum ahead of the peak. 
Observationally, this may be visible below the photon turnover frequency in case of one Compton-scattered component that dominates the high-energy regime, or in the
synchrotron component in the optically thin pre-break energy range. The accretion disk IC component would mask the
low-energy part of the BLR IC component in a two-component IC scenario. 
%By applying our model to the flare state SED of 3C~454.3 we found both the one- as well as the two-component IC scenarios possible.
Accordingly, the rate of injection might be directly reflected in the pre-break shape of the GeV-spectrum in the former case.

Limits in the initial injection point $z_a$ of the flare can be derived if the smallest particle energy in the cooled electron spectrum can be estimated,
given the accretion disk luminosity.

In summary, if the ambient electron spectrum is unambiguously determined (e.g., from the optically thin synchrotron spectrum), 
it is possible to set limits on the injection rate index $\alpha$. Together with analyzing the high energy component with sufficient
broadband coverage and a fair knowledge of the accretion disk radiation, the remaining particle injection properties can be constrained.

We have demonstrated this for the SEDs of two FSRQs, 3C~454.3 and PKS~1510-089 where the flare state broadband SEDs have been analyzed to derive information on the
required injection scenario. Two principally different scenarios per source were found to be viable: a one-component IC model, where either the Compton-scattered BLR or accretion disk radiation field dominates, 
and a two-component model as proposed in \citet{2010ApJ...714L.303F} for 3C~454.3. Accordingly, the origin of the observed spectral turn-over at a few GeV is either found in the corresponding 
break in the ambient electron spectrum, or is due to the combination of Compton-scattered accretion disk and BLR radiation field.

The observed turnover energy in the GeV range is, in both scenarios, approximately determined by $\propto \gamma_1^2 \Gamma \epsilon_{*0}D$.

A constant GeV break energy that is independent of flux state has been reported for 3C~454.3 (\citealt{2011ApJ...733L..26A}), which implies that $\gamma_1$ is broadly independent of the flux level
for not significantly changing beaming parameters during flux variations.
The required injection parameters for both cases of the two sources studied disfavor impulsive-like/instantaneous electron injection,
but point towards rather narrow particle spectra injected continuously with a gradually decreasing rate into the moving plasma blob.
For all of our model fits, the emitting region is either inside or just beyond the BLR, which is compatible with the results of \citet{2014ApJ...789..161N}, thus making the BLR an integral part of the model. In this context, we also note that the BLR geometry impacts the resulting $\gamma$-ray spectra. A good knowledge of
the BLR structure therefore improves accurate and unambiguous modeling of the $\gamma$-ray spectrum. 
In all scenarios, a further spectral break would be expected at a few tens of MeV. In the one-component IC scenario for 3C~454.3, this break arises from the 
combination of IC-scattered accretion disk and BLR photons. The break energy is determined by the smallest energy in the 
ambient electron spectrum, which in turn is influenced by the initial injection point $z_a$. In the two-component IC scenario, this energy 
corresponds to the IC scattered (in the accretion disk) ambient particle break energy or the smallest energy of this 
spectrum.
Potentially, these scenarios could, observationally, be distinguished by an improved coverage of the 0.1-1~eV (12.4-1.3 $\mu m$) energy range
of the optically thin jet emission quasi-simultaneously to the $\gamma$-ray observations, and/or
precision measurements at MeV energies, in conjunction with GeV observations, which might become possible with future
projects like {\it AstroMeV}\footnote{http://astromev.in2p3.fr} or the proposed {\it GAMMA-LIGHT} (\citealt{2013NuPhS.239..193M}) instrument. Extending the GeV energy range of the \textit{Fermi}-LAT instrument down to few tens of MeV and using the LAT Pass 8 event-level analysis \citep{2013arXiv1303.3514A} may already provide a possibility of distinguishing between models that show notably different spectral signatures in that energy range, as is the case, of the presented 3C~454.3 models.

\begin{acknowledgements}
We thank Beno\^{i}t Lott for valuable discussions and providing tables of the 3C~454.3 data. We also acknowledge particularly valuable discussions with Chuck Dermer and Ann Wehrle on this work. This project was funded by the FWF Doctoral School CIM Computational 
Interdisciplinary Modelling under contract W 1227-N16 (DK-plus CIM), and by a Marie Curie International Reintegration Grant (grant 24803) within the 7th European Community Framework Programme. 
The publication is supported by the Austrian Science Fund (FWF). This work was performed on the computational resource bwUniCluster funded by the Ministry of Science, Research and the Arts Baden-Württemberg and the Universities of the State of Baden-Württemberg, Germany, within the framework program bwHPC.
\end{acknowledgements}

\bibliography{Manuscript_AA_2014_24738} 
\end{document}